\documentclass[12pt,preprint]{aastex631}
\usepackage{natbib,natbibspacing}
\usepackage{graphicx}

\newcommand{\kms}{km s$^{-1}$}

\newcommand{\lya}{Ly$\alpha$}

\begin{document}

\title{Identifying Circumgalactic Medium Absorption in QSO 
Spectra: A  Bayesian Approach}
\author{Jennifer E. Scott}
\affil{Department of Physics, Astronomy, and Geosciences, 8000 York Road,
Towson University, Towson, MD 21252}

\author{Emileigh S. Shoemaker}
\altaffiliation{Lunar and Planetary Laboratory, University of Arizona, 1629 East University Boulevard, Tucson, AZ 85721}
\affil{Department of Physics, Astronomy, and Geosciences, 8000 York Road,
Towson University, Towson, MD 21252}

\author{Colin D. Hamill}
\altaffiliation{Department of Earth, Atmospheric and Planetary Sciences, 550
Stadium Mall Drive, Purdue University, West Lafayette, IN 47907}
\affil{Department of Physics, Astronomy, and Geosciences, 8000 York Road,
Towson University, Towson, MD 21252}

\section{Abstract}
We present a study of candidate galaxy-absorber pairs for 43 low redshift QSO sightlines ($0.06 < z < 0.85$) observed with  the
{\it Hubble Space Telescope}/Cosmic Origins Spectrograph 
that lie within the footprint of the Sloan Digital Sky Survey with a statistical approach to match absorbers 
with galaxies near the QSO lines of sight using only the SDSS Data Release 12 photometric data for the galaxies, including estimates of their redshifts.
Our Bayesian methods combine the SDSS photometric information with measured properties of the circumgalactic medium
to find the most probable galaxy match, if any, for each
absorber in the line of sight QSO spectrum. We find $\sim$630 candidate galaxy-absorber pairs using two different statistics.
The methods are able to reproduce pairs reported in the targeted spectroscopic studies 
upon which we base the 
statistics at a rate of 72\%.
The properties of the galaxies comprising the candidate pairs 
have median redshift, luminosity, and stellar mass, all estimated from the photometric data, 
$z=0.13$, $L=0.1L^*$, and $\log(M_*/M_\sun) = 9.7$.
The median impact parameter of the candidate pairs is $\sim$430~kpc, or  $\sim 3.5$ times the galaxy virial radius.
The results are broadly consistent with the high \lya\ covering fraction out to this radius found in previous studies.
This method of matching absorbers and galaxies can be used to prioritize targets for spectroscopic studies, and we present 
specific examples of promising systems for such follow-up.

\section{Introduction}
Since the discovery of QSO absorption lines, their origin in both
the gaseous material in galaxies and in the truly intervening intergalactic medium (IGM) has been debated 
\citep{GunnPeterson1965,BahcallSalpeter1965,BahcallSpitzer1969}.
Statistical studies of QSO absorption lines have long been the primary avenue for characterizing the unvirialized structures in the cosmic web.
See reviews by \cite{Meiksin2009} and \cite{McQuinn2016}. The connection of the strongest of these absorption features, the damped \lya\ systems with
$\log{N_{\rm HI}} \gtrsim 20$,
to galaxy disks has also long been recognized and leveraged to measure the physical properties of galaxy gas phases at high redshift and their 
evolution over cosmic time \citep{Wolfe2005}.

Absorption features with lower column densities
offer a very effective probe of the diffuse gas in galaxy halos, provided that absorption features can be attributed to a 
particular galaxy.
Early studies of low redshift galaxy-absorber relationships
\citep{Morris1993,Lanzetta1995,Bowen1995,Bowen1996,Bowen1997,Tripp1998,Chen1998,Chen2001a,Chen2001b,Bowen2002}
established that galaxies over a wide range of luminosity and morphological type possess
extended \ion{H}{1} halos that can be probed using the line of sight spectra of QSOs at
impact parameters $\lesssim200$~kpc.
The studies showed that the \ion{H}{1} halos have $\sim70$\% covering fractions and 
extents that scale with galaxy luminosity.

The one-to-one association of absorption features with $\log{N_{\rm HI}} < 14$ with galaxies is more ambiguous \citep{Chen2005}.
Statistical studies show associations of galaxies and weaker absorbers to 1~Mpc or more \citep{Morris1993, Tripp1998, Tejos2014}, consistent with
the positions of galaxies within the large scale structure of the cosmic web \citep{Prochaska2011b,Burchett2020,Wilde2020}.

In the last decade, understanding of the structure of the gaseous halos within the virial radii of galaxies, 
i.e. the circumgalactic medium or CGM, and its role in the physical
relationship between the IGM and galactic processes 
has expanded, thanks to
targeted spectroscopic surveys of QSOs with close projected distances to regular galaxies at low 
redshift \citep{WakkerSavage2009,Prochaska2011b,Tumlinson2011,Tumlinson2013,Stocke2013,Werk2013,Werk2014,Keeney2017} 
and high \citep{Rudie2012,Rudie2013,Turner2014,Turner2015,Rudie2019}
or specific populations such as dwarf galaxies \citep{Bordoloi2014,Johnson2017,Borthakur2016},
starbursts \citep{Borthakur2013,Heckman2017}, early type galaxies \citep{Thom2012}, luminous red galaxies \citep{Smailagic2018,Zahedy2019} or quasar host galaxies \citep{Johnson2015}.
Comparisons with hydrodynamic simulations \citep{Oppenheimer2018, Peeples2019, Nelson2019}
enable detailed examinations of the physical
interplay between IGM accretion, metal production in stars, and outflows from supernovae.

The results of the targeted 
studies of the low redshift CGM, e.g. from the COS-Halos Project, demonstrate the presence of a substantial gas reservoir within $\sim300$~kpc of galaxies, with an
\ion{H}{1} covering fraction of nearly 100\% for
star forming galaxies and $\sim$75\% for passive galaxies \citep{Tumlinson2013} to $\sim150$~kpc and an average covering fraction of 60\% out
as far as 500~kpc \citep{Liang2014}.
The Keck Baryonic Structure Survey 
finds galaxy-absorber correlations are particularly
tight for high column density absorbers, $\log N(\rm HI) > 14.5$,
and present
at lower column densities out to scales of 2~Mpc at $z\sim2.3$ \cite{Rudie2012}.
The \lya\ covering fraction in this survey is found to be $\sim$80\% within $\sim200$~kpc for $\log N(\rm HI)=15-17$.

A declining CGM metal surface density with increasing
distance from galaxies is manifested as a distinct anticorrelation between the equivalent width of absorption features measured along lines
of sight at different impact parameters. These relationships were shown in early studies of \ion{H}{1}  and \ion{C}{4} \citep{Chen2001a,Chen2001b}
and have now been measured in a variety of other metal line transitions including 
\ion{O}{6}, \ion{N}{5}, \ion{Si}{4}, \ion{Si}{3}, \ion{C}{3}, and \ion{C}{2},
allowing for detailed ionization modeling, indicating a multiphase structure of a hot $T\sim10^{5.5}$ K corona with smaller, cooler clouds
embedded within \citep{Werk2014,Werk2016,Burchett2015,Keeney2017}.

Several observed properties of the CGM indicate the influence of galactic outflows from supernovae.
\ion{H}{1} is common for sightlines intersecting the halos of both passive and star forming galaxies, with modest 
dependence on
star formation rate in the sense that the strongest features are typically observed near star forming galaxies \citep{Tumlinson2011,Thom2012,Borthakur2016}.
There is a strong association of \ion{O}{6} with galaxies, particularly 
massive galaxies with high specific star formation rates, $\gtrsim10^{-11}$ yr$^{-1}$ \citep{ChenMulchaey2009,Prochaska2011b,Mathes2014,Werk2016,Prochaska2019},
extending to $\sim330-350$~kpc \citep{Muzahid2014,Bielby2019}.
On the other hand, \ion{C}{4} is estimated to extend to $\sim100$~kpc in galactic halos at both low \citep{Chen2001b} and high redshift \cite{Steidel2011}.
Dwarf galaxies show a similar gaseous extent and a correlation with star formation rate similar to \ion{O}{6},
though weaker than found for more massive galaxies \cite{Bordoloi2014}.
In a blind survey of  \ion{C}{4} with $z<0.16$, 
\cite{Burchett2016} find that \ion{C}{4} absorption is dependent upon both stellar mass and
galaxy environment, with high mass ($M_* > 10^{9.5} M_{\sun}$) galaxies in low density environments showing the
highest rates.
All of these studies lead to the model of an extended warm, metal enriched CGM bound to galaxies.
The general conclusion is that galaxies contain significant reservoirs of \ion{H}{1} and metals within
$\sim$300~kpc and that these reservoirs play a key role in the star formation processes within the galaxy \citep{Tumlinson2017}.

Because establishing the physical relationship between specific absorption features and galaxies relies on measuring a coincidence in redshift, the targeted surveys 
make use of
extensive spectroscopic observations of galaxies in QSO fields \citep{ChenMulchaey2009,Prochaska2011a,Werk2012,Keeney2018,Prochaska2019}
or archival data of galaxy redshifts and other properties \citep{WakkerSavage2009,Liang2014}.
We define a process by which likely galaxy-absorber pairs may be identified so that
these galaxies may be targeted and follow-up spectroscopic observing resources optimally placed, 
for example by determining whether fainter galaxies closer to a particular QSO sightline versus brighter galaxies at
larger impact parameter may be most likely to result in specific absorption features in the QSO line of sight spectrum.

The methods presented here leverage the abundance of photometric data in the Sloan Digital Sky Survey, 
and the ability to use photometry alone to estimate photometric redshifts and other galaxy properties.
A photometric redshift alone is not sufficiently accurate to establish galaxy-absorber relationships, but 
in combination with other measured or estimated galaxy properties such as color, virial radius, stellar mass; absorber properties
such as equivalent width and the presence of metals; and characteristics of the potential galaxy-absorber pair, e.g. impact parameter,  
we may draw upon the wealth of targeted CGM studies to modify a set of priors
and  obtain a probability that a particular galaxy and absorber are physically associated with one another. 
We use this probability to examine trends in the galaxy-absorber pairs versus galaxies within the same
impact parameter that are not associated with absorbers.
Our approach is similar to \cite{Liang2014} in that we are mining the SDSS datasets  and comparing to QSO spectra from the Cosmic Origins Spectrograph (COS)
on the {\it Hubble Space Telescope}.
However, because we do not require a spectroscopically confirmed galaxy redshift, we have the ability to investigate 
more possibilities for associations between galaxies near the QSO lines of sight and absorption features in the QSO spectra.
In this way, the method compares to photometric redshift surveys of $z < 1$ DLAs and Lyman limit systems \citep{Chen2003, Rao2011}. 
We compare our results to known galaxy-absorber pairs in our fields as a check on the method.

Throughout this paper, we assume the following values of cosmological parameters: 
$H_0 =69.6$, $\Omega_M =  0.286$, $\Omega_{\lambda}= 0.714$ \citep{Bennet2014}.

\section{Data}
Of the 82 QSOs with COS spectra in the sample of \cite{Danforth2016} and archived in the Mikulski Archive for Space Telescopes (MAST), 
43 of them, and their surrounding fields, overlap with the footprint 
of the Sloan Digital Sky Survey (SDSS). Table~\ref{tab:data} lists these QSOs, which constitute our sample.
We use the \cite{Danforth2016} absorption linelists downloaded from MAST, 
with identifications and redshifts, equivalent widths and significance values
for the absorption features in the COS spectra.
\begin{deluxetable*}{lllrl}
\tablecaption{HST/COS QSO Sample in SDSS Footprint\label{tab:data}}
\tablewidth{0pt}
\tablehead{
\colhead{Quasar} &\colhead{Short name\tablenotemark{1}} &\colhead{RA (J2000.0)} &\colhead{Dec (J2000.0)} &\colhead{$z_{\rm em}$} }
\startdata
PG0003+158   &pg0003     &00:05:59.24      &+16:09:49.0      &0.4509     \\
PG0026+129   &pg0026    &00:29:13.71      &+13:16:04.0      &0.1420     \\
QSO0045+3926 &q0045    &00:48:18.99      &+39:41:11.6      &0.1340     \\
PG0157+001   &pg0157   &01:59:50.25      &+00:23:41.3      &0.1631     \\
SDSSJ080908.13+461925.6	&s080908    &08:09:08.14      &+46:19:25.7     &0.6563      \\
PG0832+251   &pg0832   &08:35:35.80      &+24:59:41.0      &0.3298     \\
PG0844+349   &pg0844   &08:47:42.45      &+34:45:04.4      &0.0640     \\
MRK106       &mrk106   &09:19:55.36      &+55:21:37.4      &0.1230     \\
SDSSJ092554.43+453544  &s09255b    &09:23:54.43      &+45:35:44.3      &0.3295      \\
SDSSJ092909.79+464424  &s092909    &09:29:09.79      &+46:44:24.0      &0.2400      \\
SDSSJ094952.91+390203  &s094952    &09:49:52.91      &+39:02:03.9      &0.3656      \\
PG0953+414   &pg0953   &09:56:52.41      &+41:15:22.1      &0.2341     \\
PG1001+291   &pg1001   &10:04:02.59      &+28:55:35.2      &0.3297     \\
FBQS1010+3003 &f1010   &10:10:00.70      &+30:03:22.0      &0.2558     \\
TON1187       &ton1187  &10:13:03.20      &+35:51:23.0      &0.0789     \\
1ES1028+511   &1es1028  &10:31:18.50      &+50:53:36.0      &0.3604     \\
1SAXJ1032.3+5051 &1sj1032     &10:32:16.10       &+50:51:20.0      &0.1731      \\
PG1048+342    &pg1048  &10:51:43.90      &+33:59:26.7      &0.1671     \\
PG1049-005    &pg1049  &10:51:51.48      &-00:51:17.6      &0.3599     \\
HS1102+3441   &hs1102  &11:05:39.80      &+34:25:34.4      &0.5088     \\
SBS1108+560   &sbs1108 &11:11:32.20      &+55:47:26.0      &0.7666     \\
PG1115+407    &pg1115  &11:18:30.30      &+40:25:54.0      &0.1546     \\
PG1116+215    &pg1116  &11:19:08.60      &+21:19:18.0      &0.1763     \\
PG1121+422    &pg1121  &11:24:39.18      &+42:01:45.0      &0.2250     \\
SBS1122+594   &sbs1122 &11:25:53.79      &+59:10:21.6      &0.8520     \\
TON580        &ton580 &11:31:09.50      &+31:14:05.0      &0.2902     \\
3C263         &3c263  &11:39:56.99      &+65:47:49.2      &0.6460     \\
PG1216+069    &pg121  &12:19:20.93      &+06:38:38.5      &0.3313     \\
PG1222+216    &pg122  &12:24:54.40      &+21:22:46.0      &0.4320     \\
3C273         &3c273  &12:29:06.70      &+02:03:08.7      &0.1583     \\
Q1230+0115    &q1230  &12:30:50.00      &+01:15:21.5      &0.1170     \\
PG1229+204    &pg1229  &12:32:03.61      &+20:09:29.4      &0.0630     \\
PG1259+593    &pg1259  &13:01:12.90      &+59:02:07.0      &0.4778     \\
PG1307+085    &pg1307  &13:09:47.01      &+08:19:48.3      &0.1550     \\
PG1309+355    &pg1309  &13:12:17.80      &+35:15:21.0      &0.1550     \\
SDSSJ135712.61+170444  &s135712  &13:57:12.61      &+17:04:44.1     &0.1500      \\
PG1424+240    &pg1424  &14:27:00.39      &+23:48:00.0      &0.6035     \\
MRK478        &mrk478  &14:42:07.47      &+35:26:23.0      &0.0791     \\
TON236        &ton236  &15:28:40.60      &+28:25:29.7      &0.4500     \\
1ES1553+113   &1es1553  &15:55:43.04      &+11:11:24.4      &0.4140     \\
PG1626+554    &pg1626  &16:27:56.12      &+55:22:31.5      &0.1330     \\
MRK1513       &mrk1513  &21:32:27.92      &+10:08:18.7      &0.0630     \\
PG2349-014    &pg2349  &23:51:56.12      &-01:09:13.1      &0.1737   
\enddata
\tablenotetext{1}{Reference in Tables~\ref{tab:rvir} and \ref{tab:mstar}}
\end{deluxetable*}

We queried the SDSS DR12 database \citep{Alam2015} to gather all galaxies within three degrees of each QSO sightline. 
This search radius corresponds to a radius of $\sim$0.16(83)~Mpc for $z=0.00073(0.85)$ the 
lowest(highest) redshift absorption lines in the QSO sample. These queries returned $u, g, r, i,$ and $z$ magnitudes of 
all galaxies with clean photometry and $r > 14$ within the search radius as well as their
photometric redshift and position angle and the Galactic extinction. We corrected galaxy magnitudes for 
extinction in each band and applied small offsets in $u$ and $i$ to convert to AB magnitudes. 
For all calculations based on these magnitudes, we use K-corrections from \cite{Chilingarian2010} and \cite{Chilingarian2012}, 
and thus restrict our sample to $z_{\rm phot}< 0.5$, the applicable range for their parameterizations.
To generate the final galaxy catalogs for each QSO field, we also filtered the results to include only those with the highest
quality photometric redshifts, SDSS photoErrorClass = 1, rms=0.043
\citep{Beck2016}. This limits our ability to treat the lowest redshift galaxy-absorber pairs with $z \lesssim 0.01$.

\begin{figure}
\plottwo{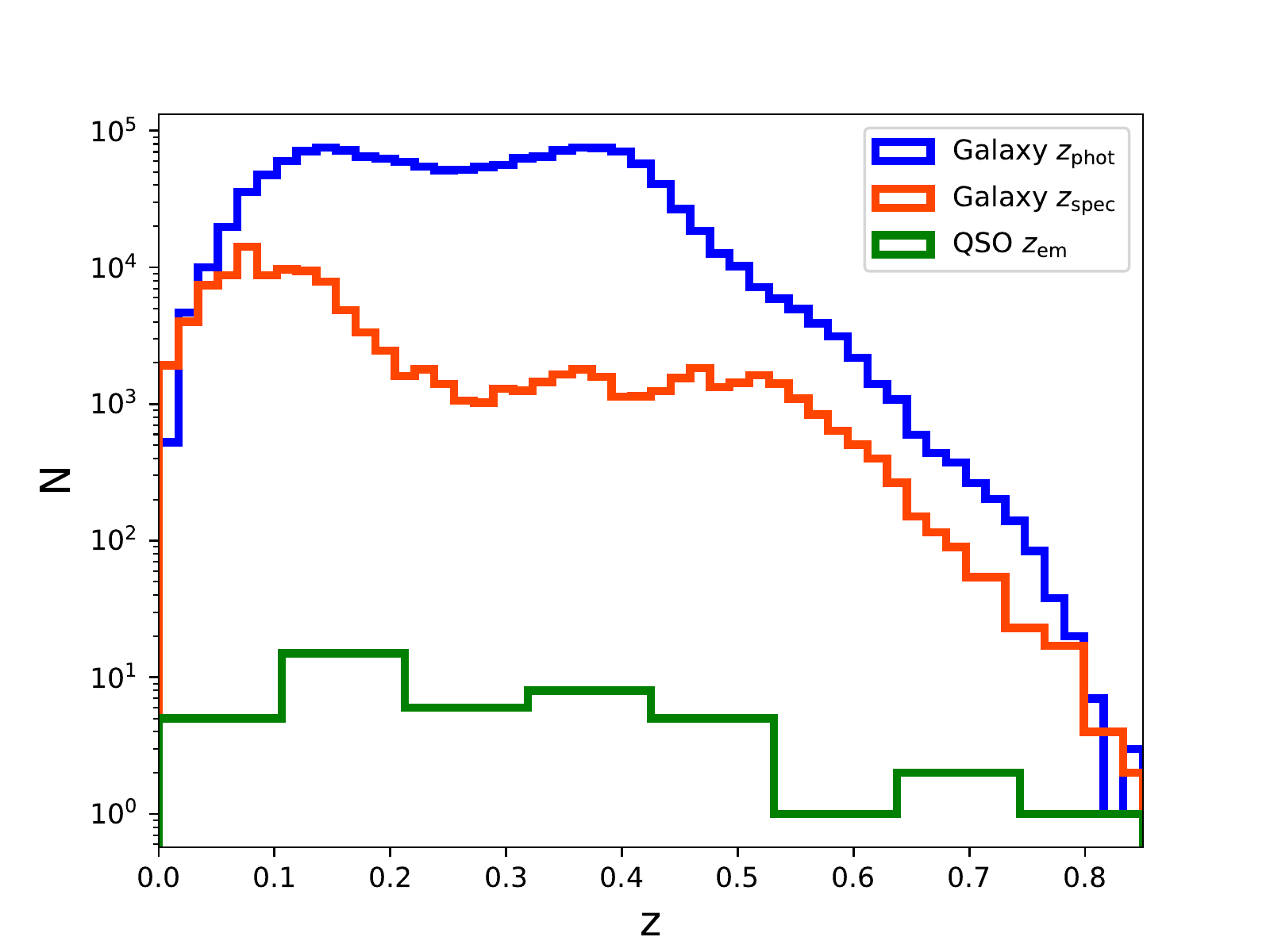}{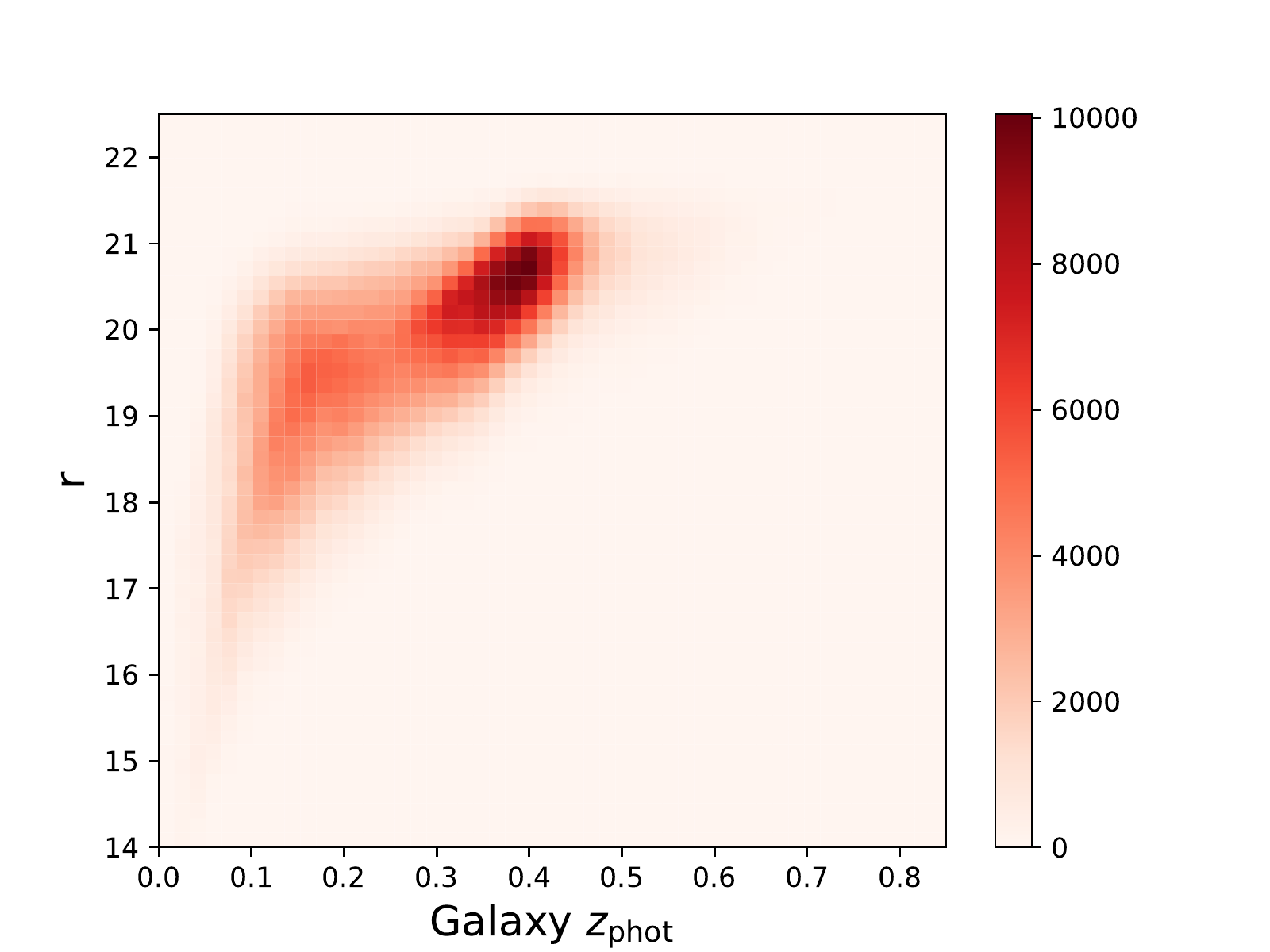}
\caption{{\it Left:} Distribution of sample QSO emission redshifts (green); SDSS galaxy photometric redshifts (blue); 
and SDSS galaxy spectroscopic redshifts (red)
{\it Right:} Distribution of sample SDSS galaxy $r$ magnitudes and photometric redshifts}~\label{fig:zhist}
\end{figure}

The resulting galaxy catalogs for the 43 QSO fields contain $\sim$93K galaxies each.
Distributions of the redshifts of the sample QSOs, as well as the SDSS photometric and spectroscopic 
redshifts where available
of all the galaxies in the SDSS sample are shown in Figure~\ref{fig:zhist}, along with
the photometric redshifts and SDSS $r$ magnitudes of all sample galaxies.
In order to evaluate our methods, we discuss the agreement between a galaxy's $z_{\rm phot}$ and its more secure $z_{\rm spec}$ in particular cases in
Appendix~\ref{sec:sightlines},
but we do not use the $z_{\rm spec}$ values in our analysis.
Given the completeness of the SDSS photometry, $\sim$90\% for galaxies to $r \sim 21.2$ \citep{Rykoff2015}, we estimate that the catalogs achieve the same completeness to 
$L \sim 0.3-0.5L^*$ for $z=0.5$ and to $L \sim 0.01L^*$ for $z=0.1$. The median redshift of the sample absorbers is 0.11.

\section{Identifying Galaxy-Absorber Pairs}
We use the linelists generated from the COS spectra of the QSO sightlines and provided as high-level science products through MAST 
and the SDSS photometry of the galaxies in those QSO fields as input into an algorithm that calculates
a Bayesian probability that a particular absorption system is associated with each galaxy in the QSO field.
The likelihoods 
draw upon previous work characterizing the distribution of absorber equivalent widths in various ionic species with  galaxy impact 
parameter and intrinsic galaxy 
properties estimated from the SDSS photometry.
Because we have only photometric redshifts for our sample galaxies, our objective is to define
a statistic that will incorporate other available data such as absorber rest equivalent width, $W$, galaxy-absorber impact parameter, $\rho$, and galaxy properties
such as virial radius $r_{\rm vir}$ or stellar mass $M_{*}$ using empirical relationships derived from
galaxy-absorber pairs identified by previous spectroscopic surveys.

The Bayesian statistic we define \citep{Trotta2017} is the posterior probability of a true match between an absorber and galaxy $i$,
given the combination of observed parameters: absorber $W$, 
galaxy-absorber $\rho$, and galaxy $r_{\rm vir}$ or $M_*$, expressed here as $\bf p$. The probability is multiplied over all ionic species, $j$:
\begin{equation}
P_i (m_z | {\bf p} ) = \frac{P_i(m_z)}{P({\bf p})} \prod\limits_{j}^{} P_{ij}({\bf p}|m_z)
\label{equ:bayes}
\end{equation}
This statistic uses the prior $P_i(m_z)$ that galaxy $i$
is a match for the absorber 
based on its photometric redshift alone, assuming Gaussian errors and integrated over $z_{\rm abs} \pm 500$ \kms. 

The additional information about each potential galaxy-absorber pair, $W$, $\rho$, $r_{\rm vir}$, $M_{*}$ is contained in the likelihood, $P({\bf p}| m_z)$, 
the conditional probability of observing $\bf p$ given a galaxy-absorber relationship.
This likelihood is calculated from a probability distribution that is assumed 
to be Gaussian, with parameters defined by the empirical relationships found for galaxy-absorber pairs 
in species $j$. Two definitions of $P({\bf p}|m_z)$ are described and investigated below.
The posterior probability is normalized by summing over the possible outcomes, i.e. the absorber is matched to any one of the SDSS galaxies:
\begin{equation}
P({\bf p}) = \sum_i \Bigl[ P_i(m_z) \prod\limits_{j}^{} P_{ij}({\bf p}|m_z) \Bigr]
\label{equ:norm}
\end{equation}
This construction makes the {\it a priori} assumption that each absorber in the QSO spectra can be attributed to the CGM of one of the SDSS galaxies
in the field, i.e. it does not include a term accounting for the probability that the absorber is caused by the CGM of an 
undetected galaxy or is intergalactic in origin. 
However, in practice, the methodology described here does leave $\sim$40\% of \lya\ lines unpaired, indicating 
that they arise in the IGM or 
a galaxy fainter than the SDSS detection limit. 

Our algorithms calculate the probability described in Equation~\ref{equ:bayes} 
in a pairwise fashion. For each galaxy within a minimum impact parameter of the QSO sightline and judged on the basis of the photometric 
redshift to be in the QSO foreground, we calculate the probability of its association with each absorption system in the QSO spectrum, where
absorption systems are defined here to be any group of features within 300~\kms\ of one another.
The calculated galaxy-absorber impact parameter based on the photometric redshifts are necessarily uncertain, 
so we allow for this impact parameter to be as large as 2~Mpc. 
If the galaxy and absorber are truly ``matched", they can be assumed to lie at approximately the same redshift, and so the impact parameter can be calculated instead from 
$z_{\rm abs}$. This $\rho(z_{\rm abs})$ is capped at 500~kpc.
The addition of the empirical CGM $\rho - W$ relationships provides further constraints on the statistical method.

The output of the algorithm is a list of up to ten most probable galaxy matches to each absorption system.
We preserve galaxy-absorber pairs with $P(m_z | {\bf p} ) > 0.1$.
Because the probability defined in Equation~\ref{equ:bayes} is a product over all ionic species in a system, the resulting value can be
smaller than the probability calculated for individual ions. For these multicomponent systems, we preserve pairs for which the
value of the probability for at least one \lya\ component exceeds the 10\% threshold.
The algorithm often results in double matches, i.e. galaxies that are paired with more than one absorber in the top ten list.
To determine our final list of candidate galaxy-absorber pairs, we impose a uniqueness criterion that selects the highest probability absorber paired with each galaxy, and that
allows for lower ranked galaxies to be associated with a particular absorption system if the galaxy with the highest probability is a better match for a different absorber.
For this, we must compare different absorption system probabilities with one another, so again, we use the top \lya\ component probability for multicomponent systems.
This uniqueness criterion results in an absorber being paired with a galaxy that is first in its top ten list about 61\% of the time. In another $\sim$22(9)\%  
of cases, the galaxy selected as the
unique match is the second(third) in the top ten list.  

We devised two different statistics based on Equation~\ref{equ:bayes}, each using a different construction of 
$P({\bf p} | m_z)$. 
One is based on the established empirical relationships between absorber equivalent width and galaxy-absorber impact parameter as a ratio with the galaxy's virial radius. 
We refer to this as the virial radius method. Our second statistic is based on the proposed CGM fundamental plane \citep{Bordoloi2018} which also incorporates the
galaxy stellar mass, and so this is labeled the stellar mass method.

\subsection{Virial Radius Method}
We use Equation~\ref{equ:bayes} and linear fits to absorber rest equivalent width $W$ versus galaxy-absorber impact parameter 
normalized by galaxy virial radius, $\rho/r_{\rm vir}$
to define the statistic $P(r_{\rm vir})$. The
parameters of the fits for each ion are listed in Table~\ref{tab:fits}.
We calculate $r_{\rm vir}$ using the \cite{Richter2016} parameterization of the \cite{Stocke2013} halo abundance matching curves.
\begin{deluxetable}{lllllrr}
\tablecaption{Galaxy-Absorber Pair Fits \label{tab:fits}}
\tablewidth{0pt}
\tablehead{
\colhead{Ion} &\multicolumn{2}{c}{$W$ vs. $\rho/r_{\rm vir}$} &\multicolumn{2}{c}{$W$ vs. $\rho$}  &\multicolumn{2}{c}{$\Delta W$ vs. $\log{M_*/M_{\odot}}$} \\
\cline{2-7}
\colhead{}    &\colhead{slope} &\colhead{intercept}  &\colhead{slope} &\colhead{intercept}  &\colhead{slope} &\colhead{intercept}
}
\startdata
\ion{H}{1} &-0.553  &-0.268 &$-1.02\times10^{-2}$ &2.996  &0.129   &-1.492 \\
\ion{O}{6} &-0.0114 &-0.427 &$-1.68\times10^{-2}$ &2.806  &-0.199  &1.702 \\
\ion{C}{2} &-0.663  &-0.830 &$-4.75\times10^{-2}$ &2.948  &0.493   &-0.348 \\
\ion{C}{3} &-0.523  &-0.467 &$-1.30\times10^{-2}$ &2.861  &0.0459  &-0.592 \\
\ion{C}{4} &-0.358  &-0.625 &$-3.28\times10^{-2}$ &2.689  &-0.121  &1.0621 \\
\ion{Si}{2} &-0.718  &-1.016 &$-4.85\times10^{-2}$ &2.655  &-0.0342  &0.00864 \\
\ion{Si}{3} &-1.015  &-1.048 &$-4.23\times10^{-2}$ &2.841  &0.0765   &-1.170 \\
\ion{Si}{4} &-1.016  &-1.382 &$-4.50\times10^{-2}$ &2.474  &-0.169   &1.517 \\
\enddata
\end{deluxetable}

\begin{figure}
\gridline{\fig{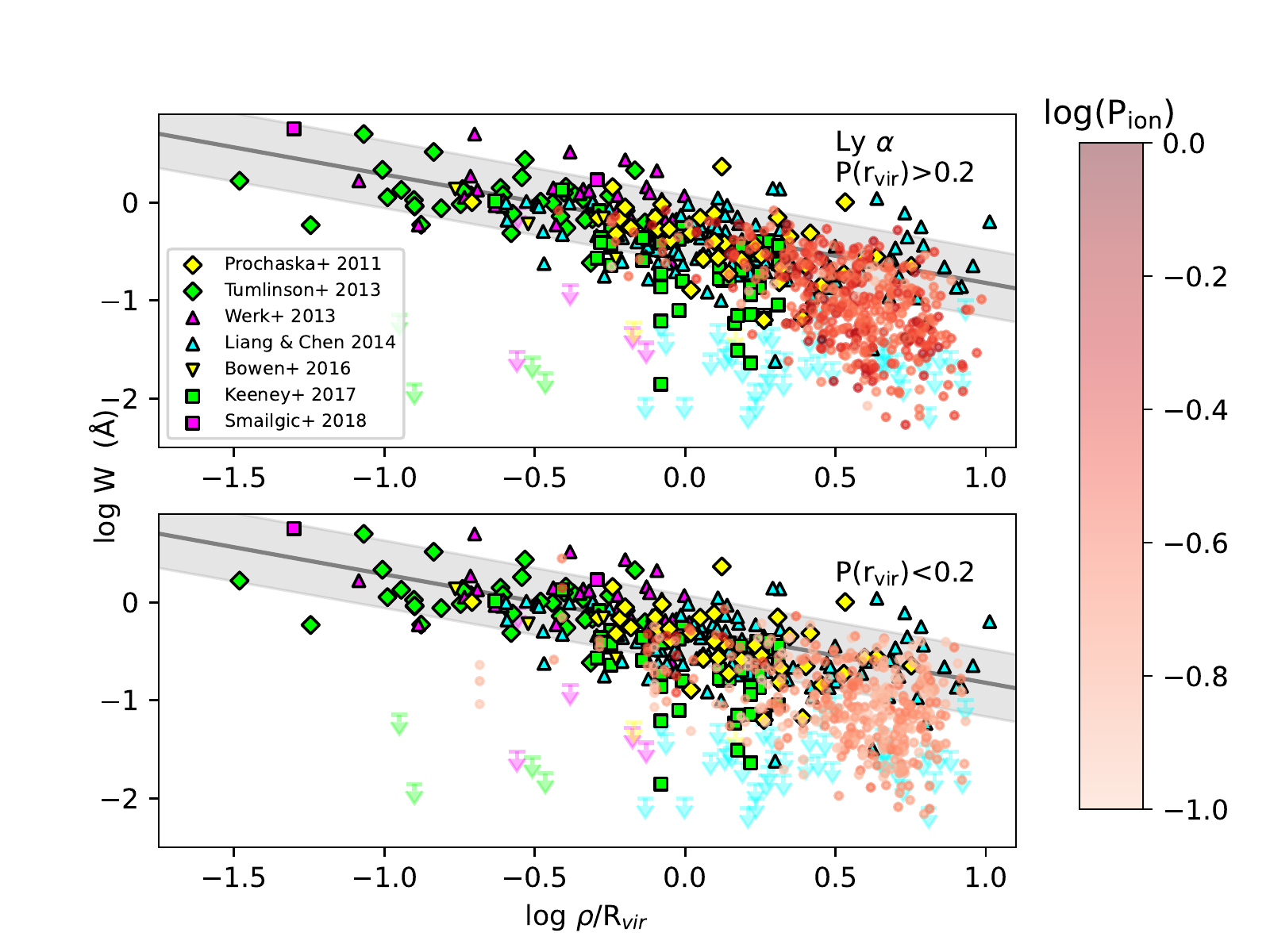}{1.\textwidth}{(a)} }
\caption{Absorber equivalent width versus ratio of impact parameter
to virial radius for all unique galaxy-absorber pairs for (a) \lya\, (b) \ion{C}{2}
(c) \ion{Si}{2}, (d) \ion{C}{3}, (e) \ion{Si}{3}, (f) \ion{C}{4} and (g) \ion{Si}{4}, and (h) \ion{O}{6}.
Black lines and gray shaded
regions show the fits to the literature points and  $1\sigma$ confidence intervals.
Red points are the unique galaxy-absorber
pairs identified by the virial radius method, with shading
indicating the value of $P_{\rm ion}$ ($=P(r_{\rm vir})$ for single component \lya\ absorbers).
Upper limits from \cite{Prochaska2011b} (yellow), \cite{Werk2013} (magenta), \cite{Tumlinson2013} (green), and \cite{Liang2014} (cyan).
In panel (a), pairs with $P(r_{\rm vir})> 0.2$ are shown on the top,
and those with $P(r_{\rm vir}) < 0.2$ are on the bottom.}~\label{fig:rvir}
\end{figure}

Figure~\ref{fig:rvir} shows the literature measurements for  \lya, \ion{O}{6}, \ion{C}{4}, \ion{C}{3}, \ion{C}{2}, \ion{Si}{4}, \ion{Si}{3}, and \ion{Si}{2}
absorption paired with galaxies on the basis of spectroscopic redshifts. The fits to these points
determine the parameters of the 
probability distribution for a given galaxy-absorber pair for each of the primary ionic species observed in the COS spectra.
Fit parameters are listed in Table~\ref{tab:fits}.
The mean of the Gaussian probability density is the linear fit value for $W$ given $\rho/r_{\rm vir}$ of the galaxy-absorber pair and its variance is determined by the
$1\sigma$ confidence region about the fit. 
In Figure~\ref{fig:rvir}, we also show upper equivalent width limits reported in the literature but these points are not included in the fit.
For \lya,
including these upper limits had no significant effect on the slope or the overall galaxy-absorber probability calculations. 
The metal line fits were more sensitive to these upper limits but resulted in an physically implausible positive slope for \ion{Si}{4}.
The value of $P({\bf p} | m_z)$ is the integral, evaluated over $\pm 5\sigma_W$, of the product of
this Gaussian probability density about the measured absorber equivalent width and a second Gaussian probability density 
associated with the measurement itself, i.e. with
mean equal to $W_{\rm obs}$ and $\sigma$ equal to the reported equivalent width error, $\sigma_W$.

The red points in Figure~\ref{fig:rvir} 
mark the unique galaxy-absorber candidate pairs identified by the virial radius method, with shading indicating the value of the probability
used in the uniqueness criterion, $P_{\rm ion}$. This is the same as the overall probability, $P(r_{\rm vir})$, for absorbers consisting of a single \lya\ line, but
is the value of the largest \lya\ component probability for multicomponent systems.
For \lya, the pairs shown in the top panel correspond to those with overall probability $P(r_{\rm vir})> 0.1$. Pairs marked by shaded red points 
in the bottom panel have
$P(r_{\rm vir}) < 0.1$. Due to the imposition of the 10\% threshold to select only the most reasonable candidate pairs, the points 
in this bottom panel
correspond to 
multicomponent absorption systems. In other words, no galaxies paired with single \lya\ absorbers with $P(r_{\rm vir}) < 0.1$ are part of our final candidate lists.

\renewcommand{\thefigure}{\arabic{figure} (Cont.)}
\addtocounter{figure}{-1}

\begin{figure}
\gridline{\fig{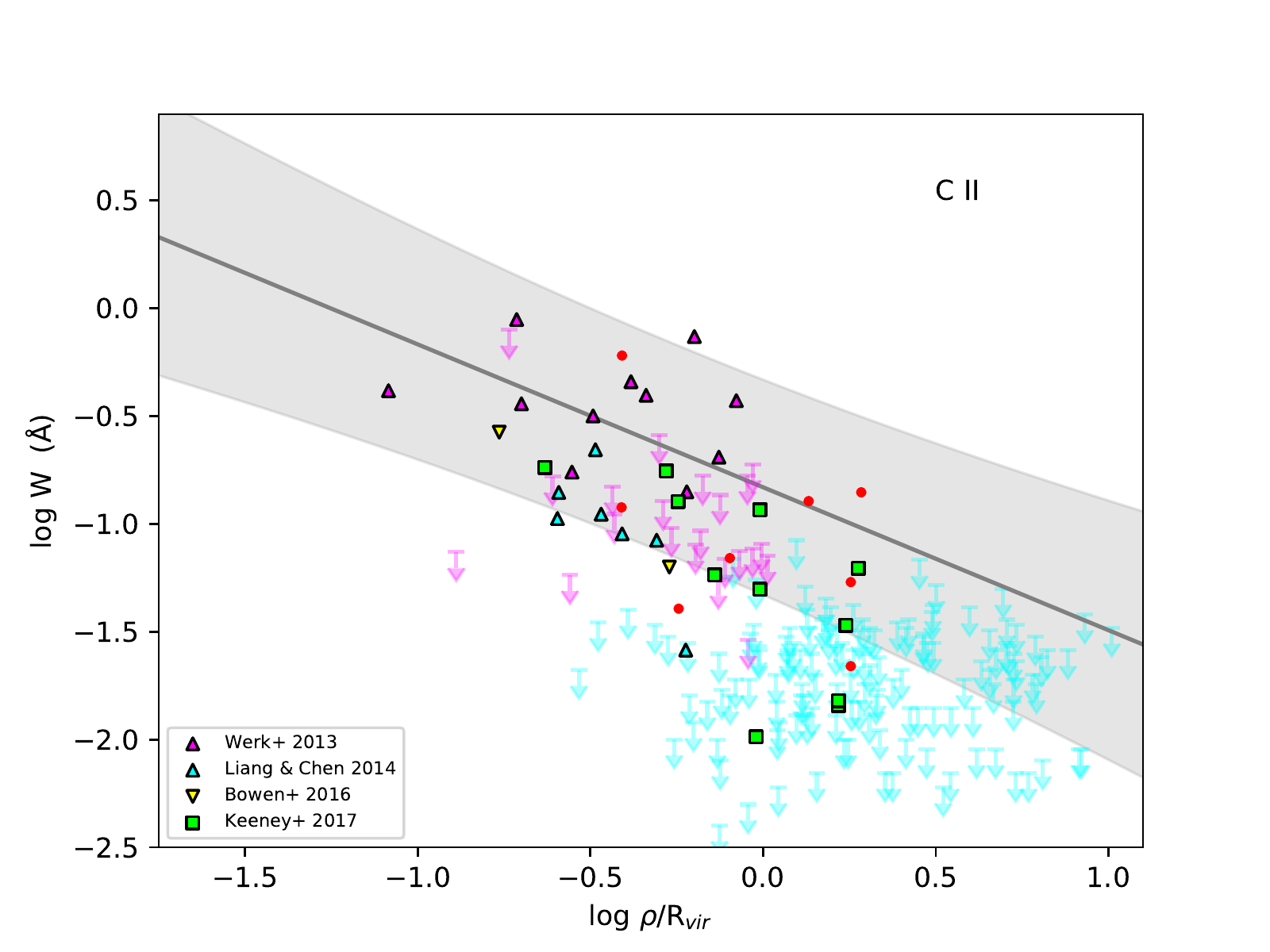}{0.4\textwidth}{(b)}
          \fig{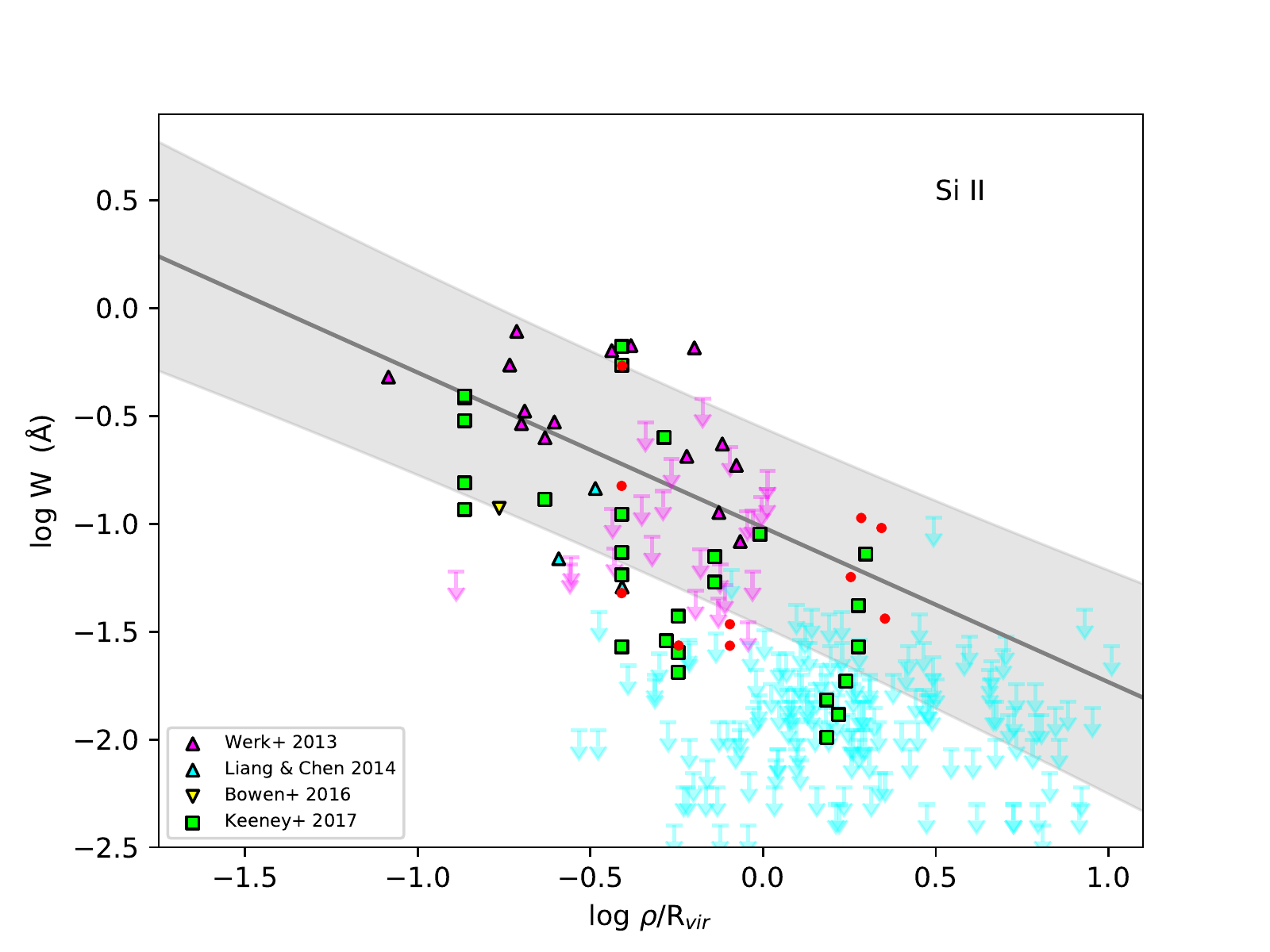}{0.4\textwidth}{(c)}}
\gridline{\fig{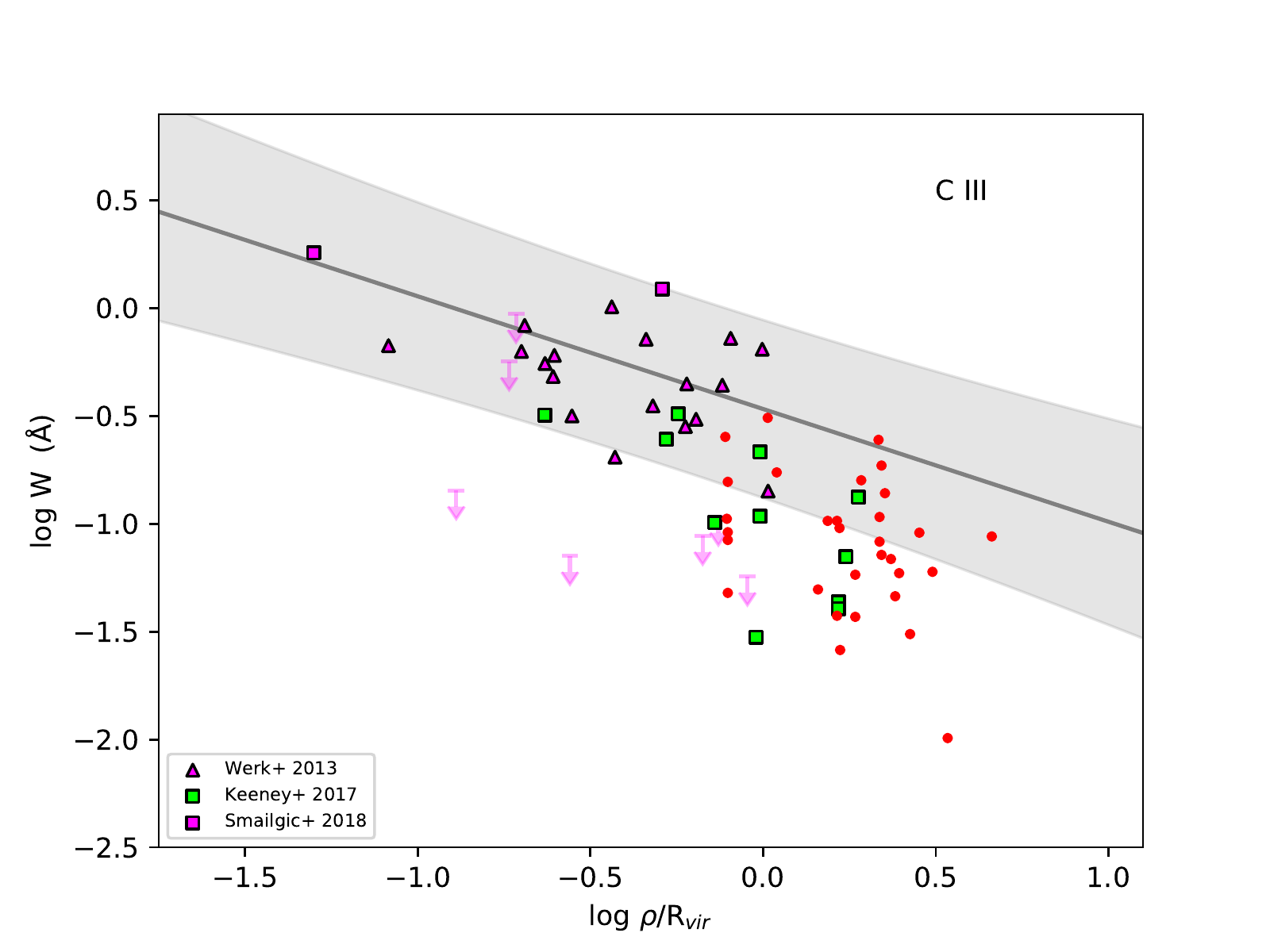}{0.4\textwidth}{(d)}
          \fig{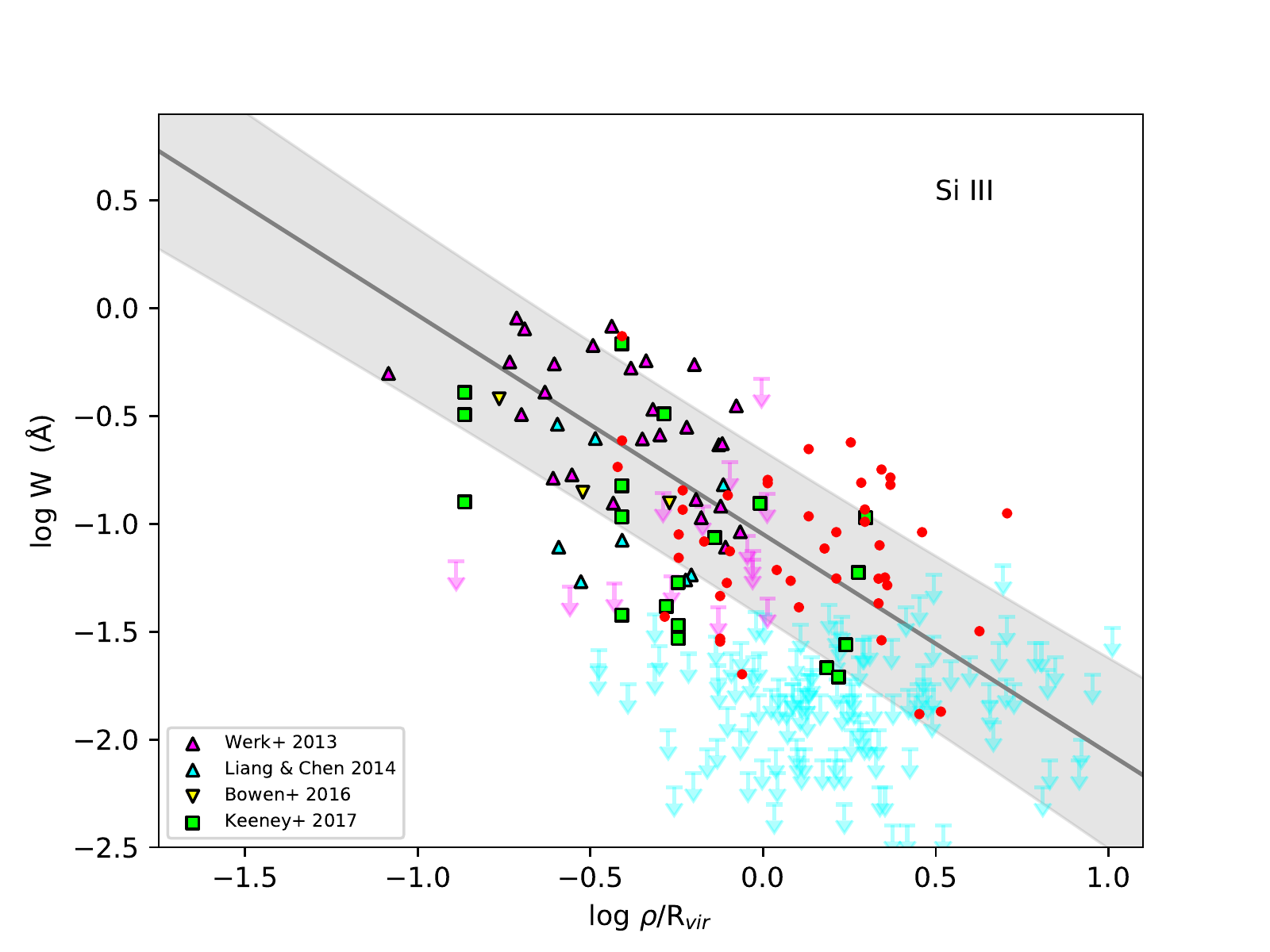}{0.4\textwidth}{(e)}}
\gridline{\fig{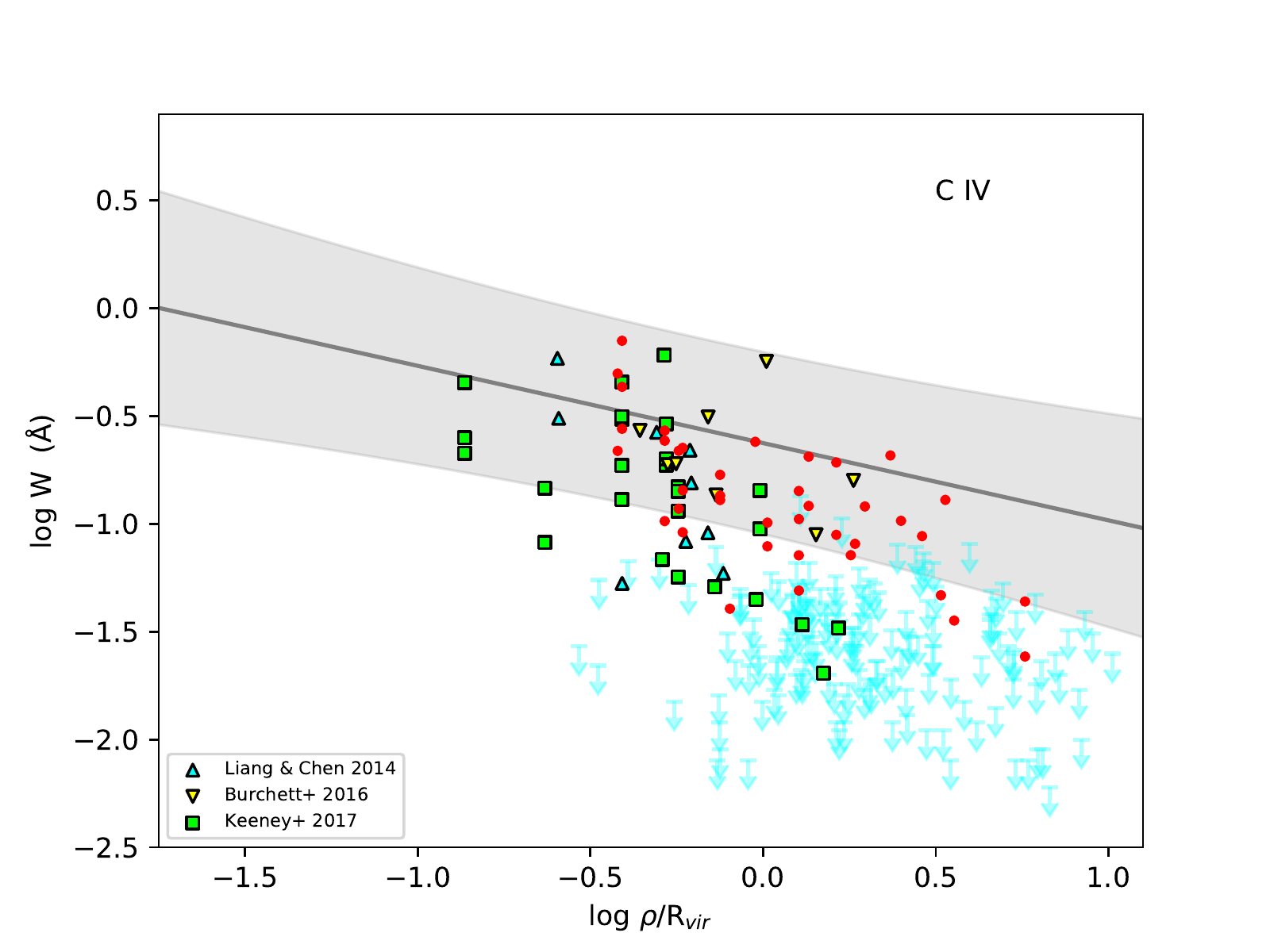}{0.4\textwidth}{(f)}
          \fig{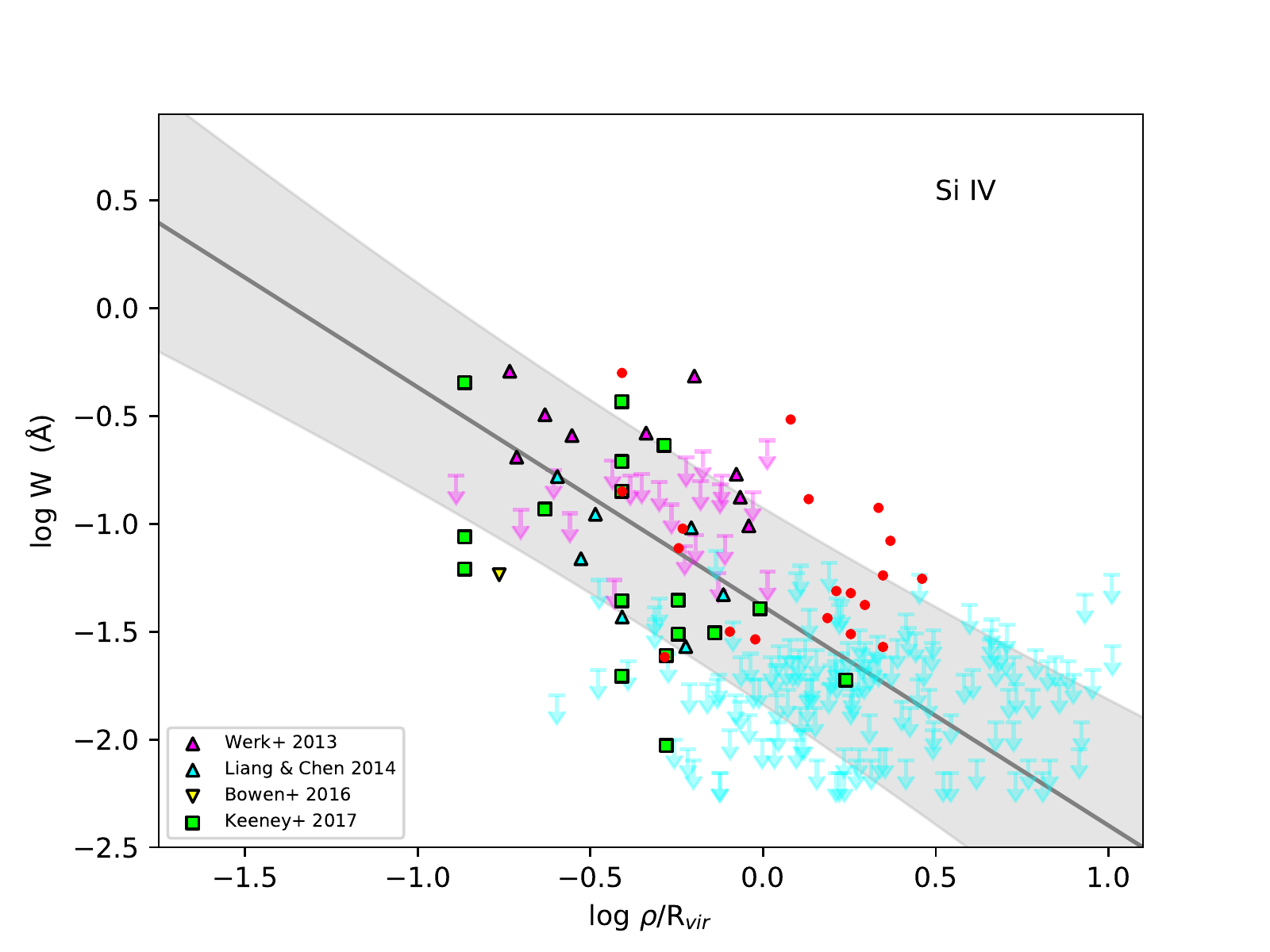}{0.4\textwidth}{(g)}}
\caption{Absorber equivalent width versus ratio of impact parameter
to virial radius for all unique galaxy-absorber pairs for (a) \lya\, (b) \ion{C}{2}
(c) \ion{Si}{2}, (d) \ion{C}{3}, (e) \ion{Si}{3}, (f) \ion{C}{4} and (g) \ion{Si}{4}, and (h) \ion{O}{6}.
Black lines and gray shaded
regions show the fits to the literature points and  $1\sigma$ confidence intervals.
Red points are the unique galaxy-absorber
pairs identified by the virial radius method, with shading
indicating the value of $P_{\rm ion}$ ($=P(r_{\rm vir})$ for single component \lya\ absorbers).
Upper limits from \cite{Prochaska2011b} (yellow), \cite{Werk2013} (magenta), \cite{Tumlinson2013} (green), and \cite{Liang2014} (cyan).
In panel (a), pairs with $P(r_{\rm vir})> 0.2$ are shown on the top,
and those with $P(r_{\rm vir}) < 0.2$ are on the bottom.}
\end{figure}

\renewcommand{\thefigure}{\arabic{figure} (Cont.)}
\addtocounter{figure}{-1}

\begin{figure}
\gridline{\fig{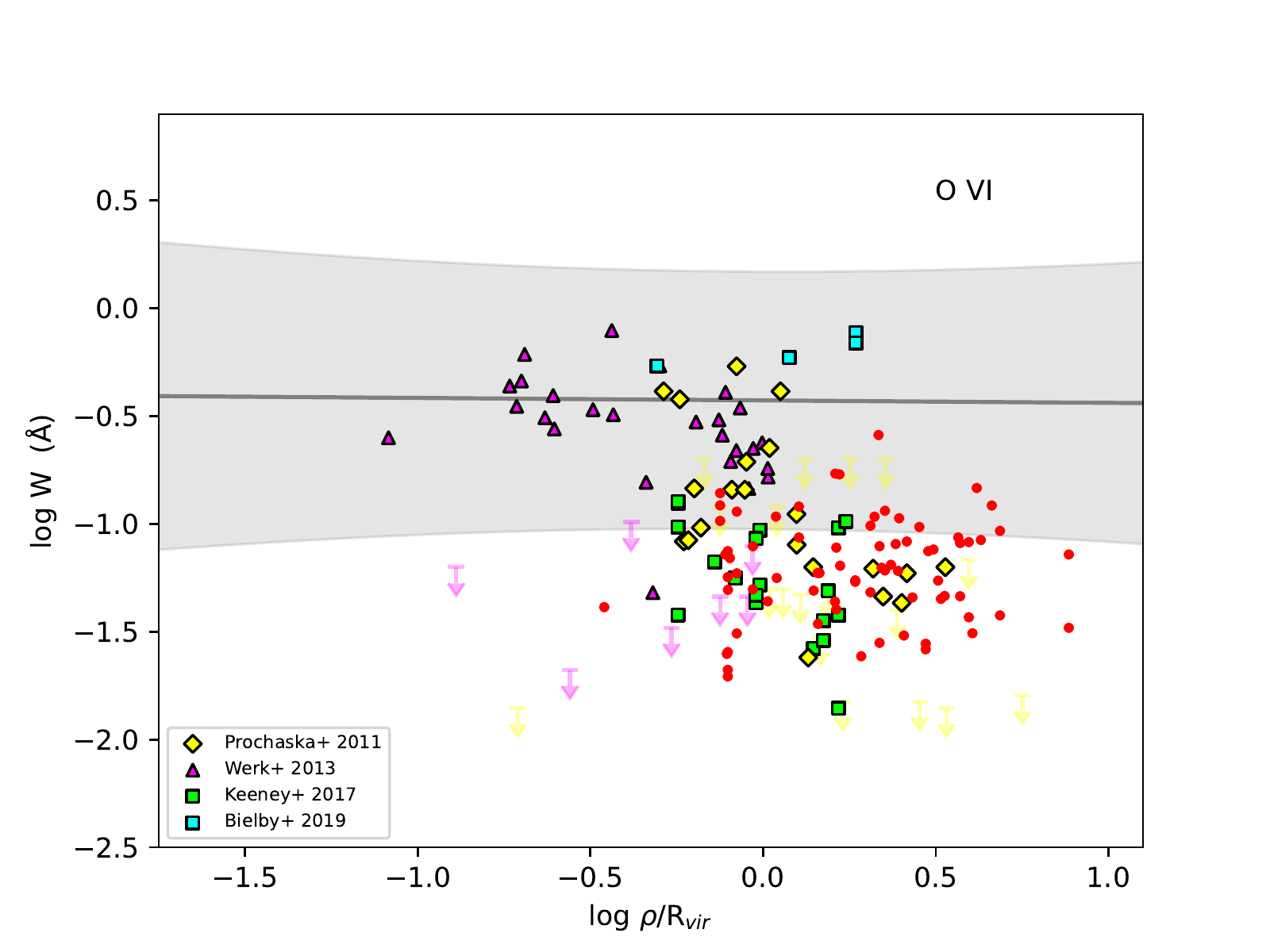}{0.4\textwidth}{(h)}}
\caption{Absorber equivalent width versus ratio of impact parameter
to virial radius for all unique galaxy-absorber pairs for (a) \lya\, (b) \ion{C}{2}
(c) \ion{Si}{2}, (d) \ion{C}{3}, (e) \ion{Si}{3}, (f) \ion{C}{4}, (g) \ion{Si}{4}, and (h) \ion{O}{6}.
Black lines and gray shaded
regions show the fits to the literature points and  $1\sigma$ confidence intervals.
Red points are the unique galaxy-absorber
pairs identified by the virial radius method, with shading
indicating the value of $P_{\rm ion}$ ($=P(r_{\rm vir})$ for single component \lya\ absorbers).
Upper limits from \cite{Prochaska2011b} (yellow), \cite{Werk2013} (magenta), \cite{Tumlinson2013} (green), and \cite{Liang2014} (cyan).
In panel (a), pairs with $P(r_{\rm vir})> 0.2$ are shown on the top,
and those with $P(r_{\rm vir}) < 0.2$ are on the bottom.}
\end{figure}

\renewcommand{\thefigure}{\arabic{figure}}

\subsection{Stellar Mass Method}
\begin{figure}
\gridline{\fig{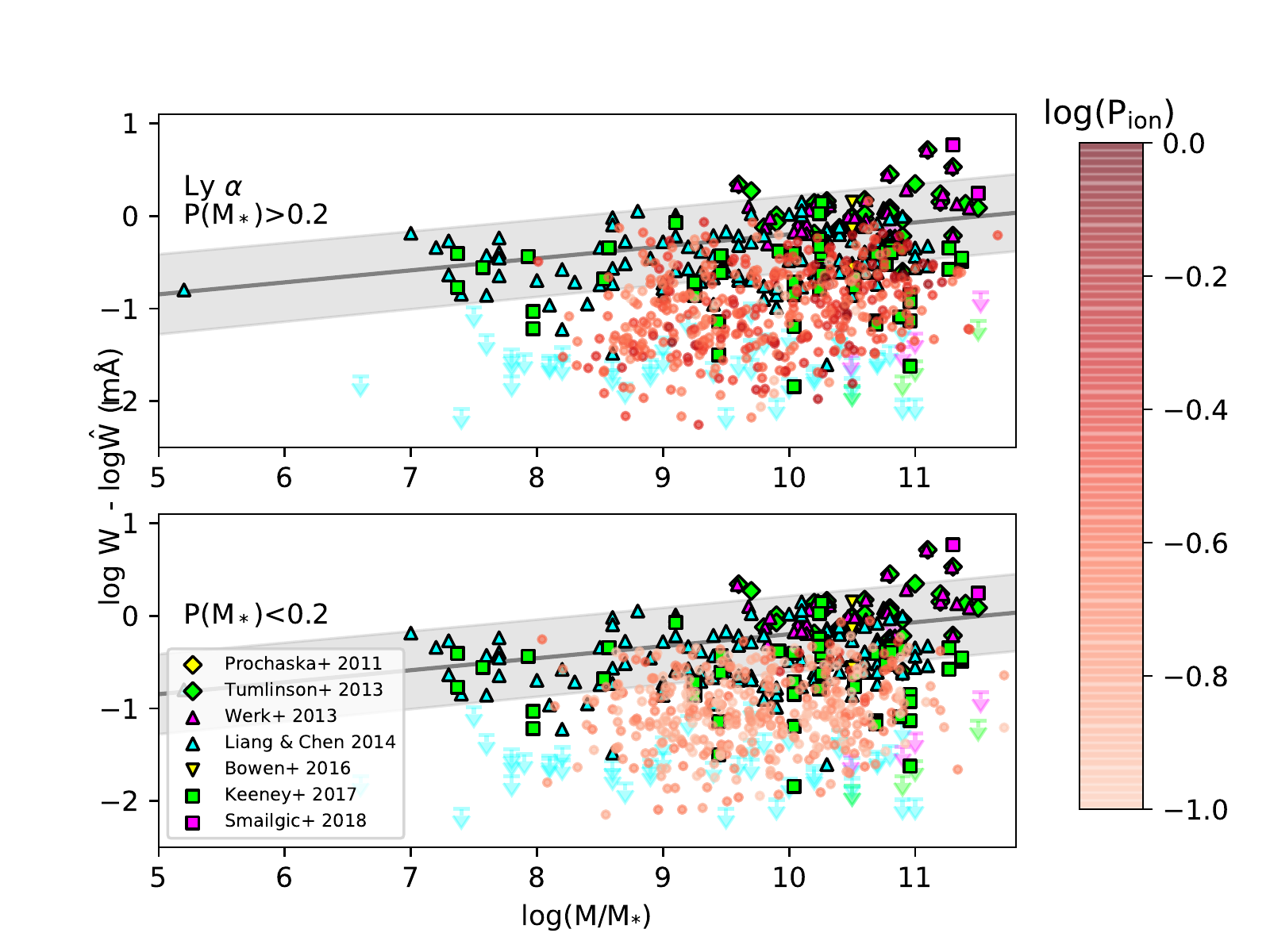}{1.0\textwidth}{(a)}}
\caption{Difference of absorber equivalent width and
$\hat{W}$ versus galaxy stellar mass in solar masses
for all unique galaxy-absorber pairs for (a) \lya\, (b) \ion{C}{2}
(c) \ion{Si}{2}, (d) \ion{C}{3}, (e) \ion{Si}{3}, (f) \ion{C}{4}, (g) \ion{Si}{4}, and (h) \ion{O}{6}. Black lines and gray shaded
regions show the fits to the literature points and  $1\sigma$ confidence intervals.
Red points are the unique galaxy-absorber
pairs identified by the virial radius method, with shading
indicating the value of $P_{\rm ion}$ ($=P(M_*)$ for single component \lya\ absorbers).
Upper limits from \cite{Prochaska2011b} (yellow), \cite{Werk2013} (magenta), \cite{Tumlinson2013} (green), and \cite{Liang2014} (cyan).
In panel (a), pairs with $P(M_*)> 0.2$ are shown on the top,
and those with $P(M_*) < 0.2$ are on the bottom.}~\label{fig:mstar}
\end{figure}

The second statistic we define, $P(M_*)$, uses the CGM fundamental plane, a relationship between absorber 
equivalent width, galaxy-absorber impact parameter and galaxy stellar mass proposed by
\cite{Bordoloi2018}. Similar to the virial radius method, the
$P({\bf p}| m_z)$ term in Equation~\ref{equ:bayes} is calculated from a 
Gaussian probability function with mean deteremined from a linear fit to $\log{W} - \log{\hat{W}}$ versus $\log{M_*/M_{\sun}}$ 
where $\hat{W}$ is itself found from a linear fit to $W$ versus impact parameter $\rho$ and galaxy stellar masses are estimated from the 
SDSS photometry using the prescription of \cite{Taylor2011}. The 1$\sigma$ confidence limit on the linear fit sets the width of the probability distribution,
and the value of $P({\bf p}| m_z)$ is calculated from the integral about the observed absorber equivalent width.
\cite{Bordoloi2018} developed the fundamental plane for the ubiquitous \lya\ CGM absorption and its
clear dependence on galaxy stellar mass. We extend that formalism to all the ions considered.
Fit parameters are listed in Table~\ref{tab:fits} and the results are shown in Figure~\ref{fig:mstar}, with top and bottom sections for
$P(M_*)> 0.2$ and $P(M_*)> 0.2$ as defined
in Figure~\ref{fig:rvir}(a). 

\renewcommand{\thefigure}{\arabic{figure} (Cont.)}
\addtocounter{figure}{-1}

\begin{figure}
\gridline{\fig{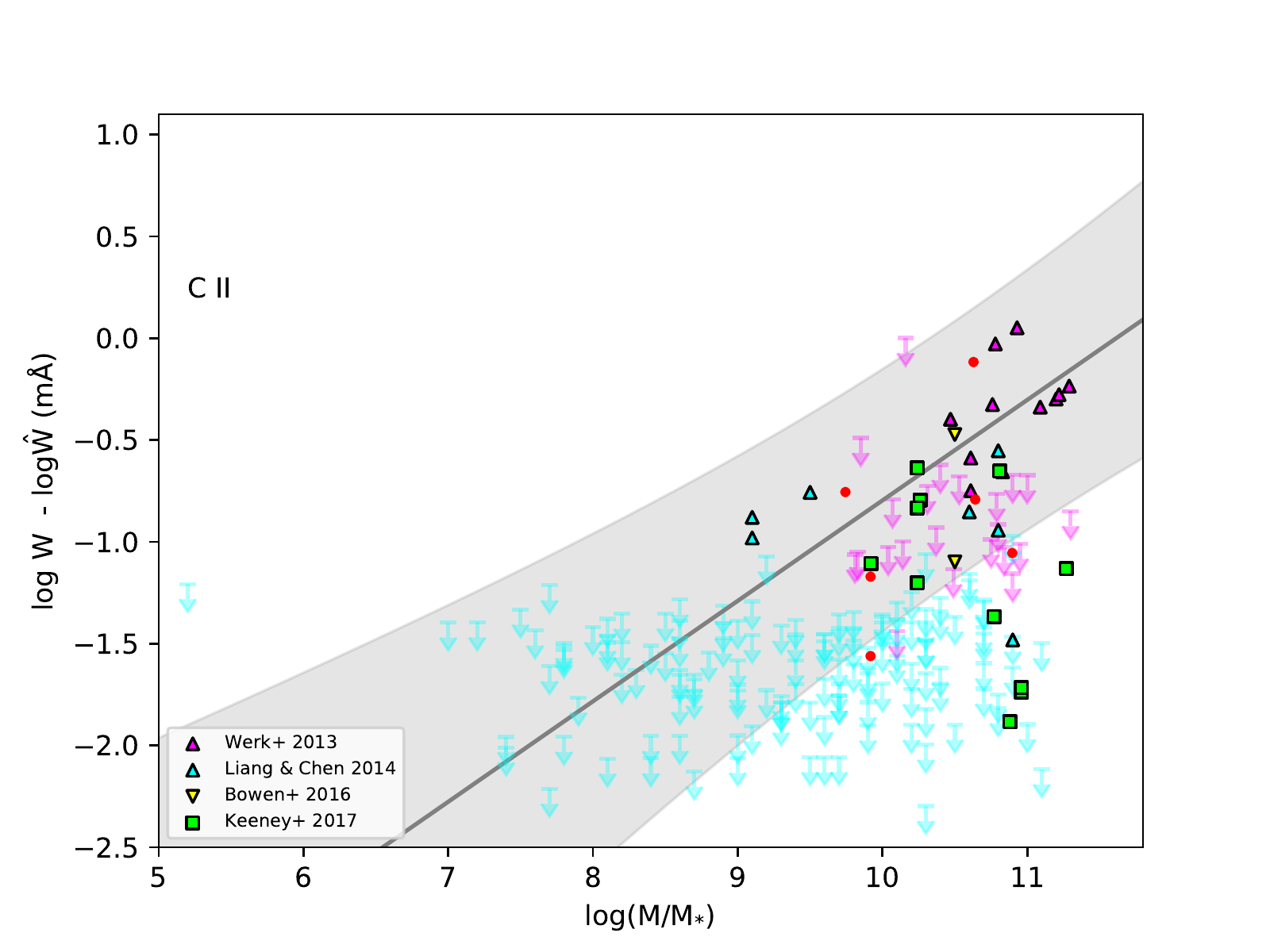}{0.4\textwidth}{(b)}
          \fig{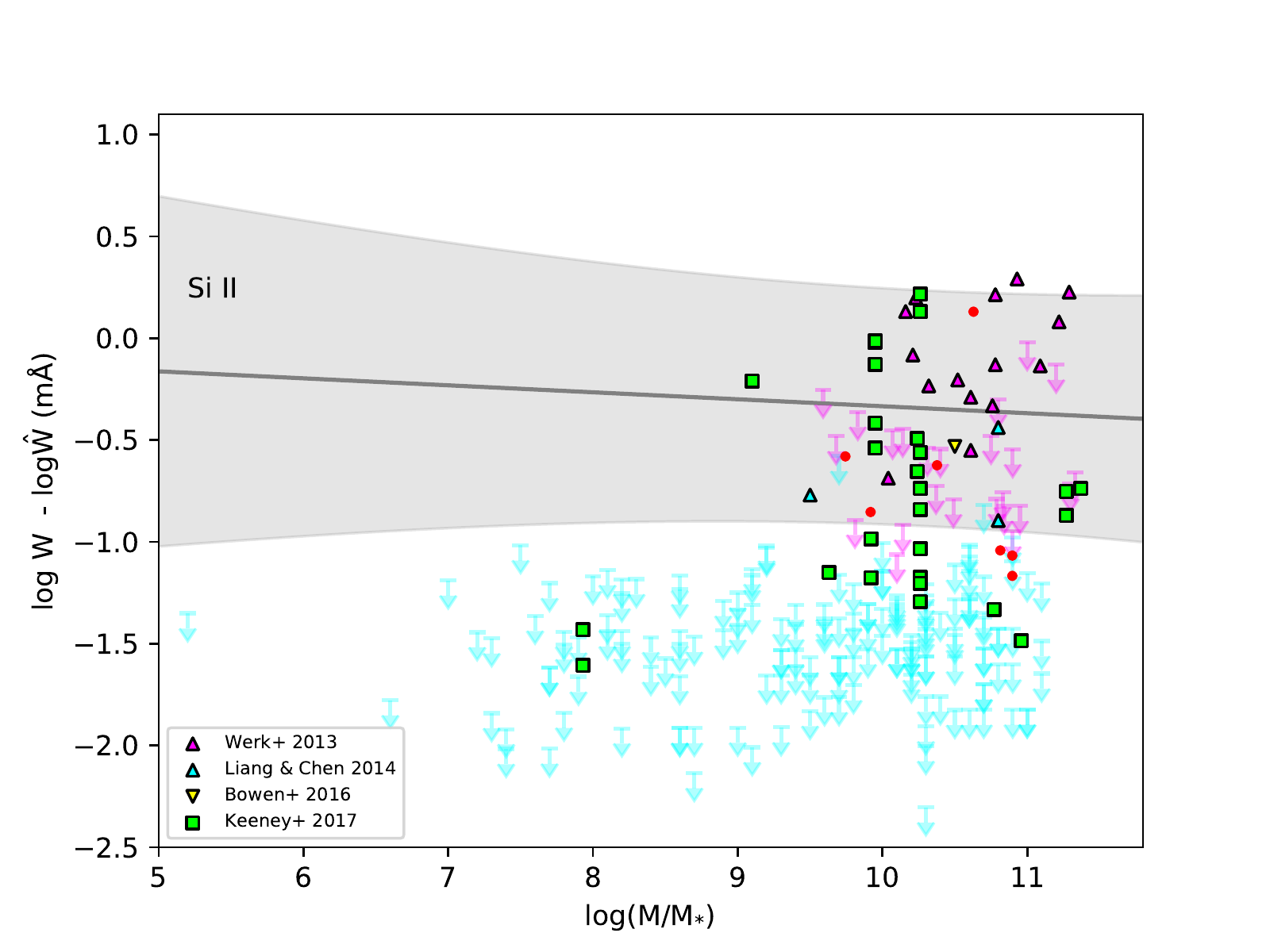}{0.4\textwidth}{(c)}}
\gridline{\fig{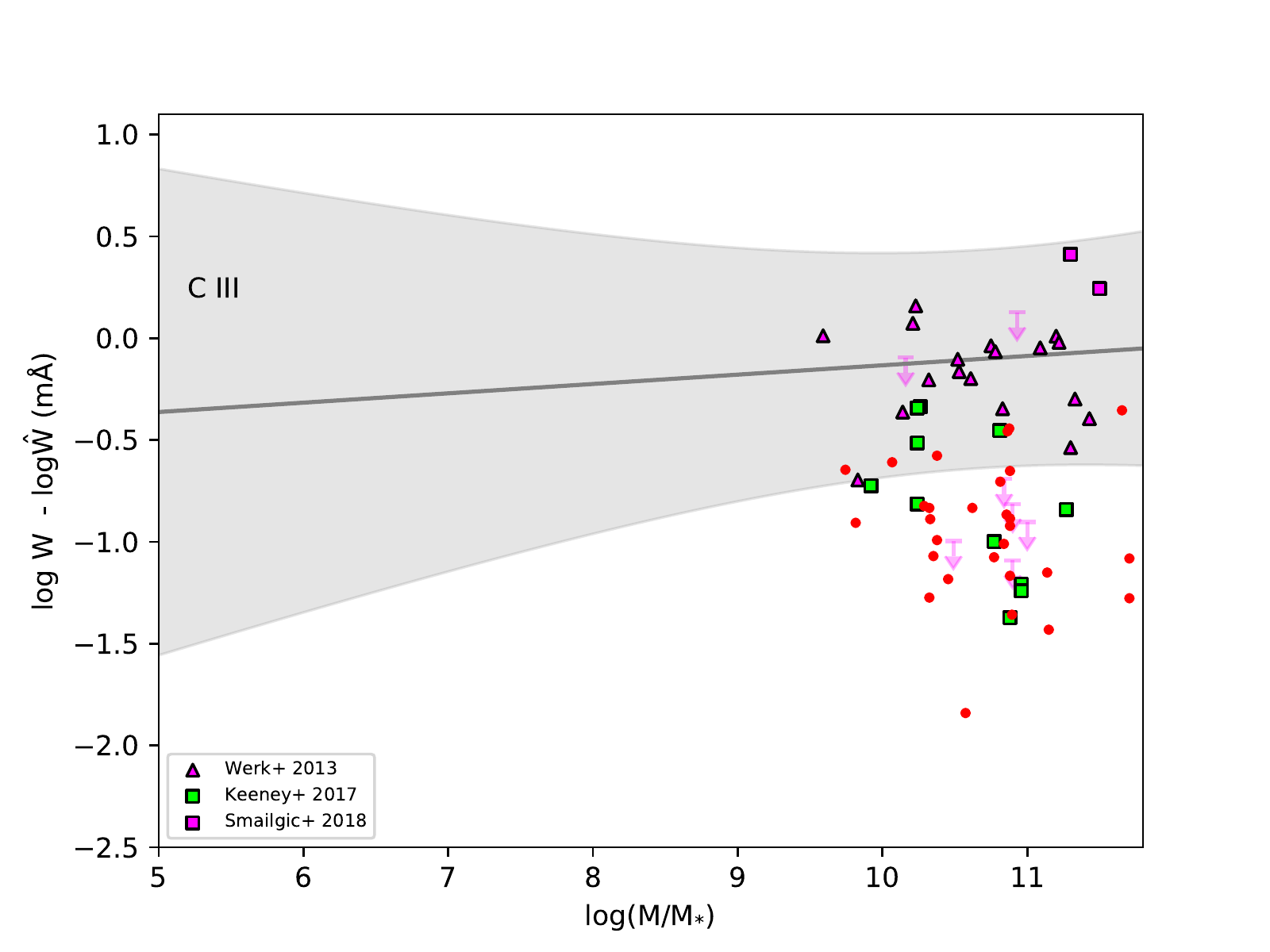}{0.4\textwidth}{(d)}
          \fig{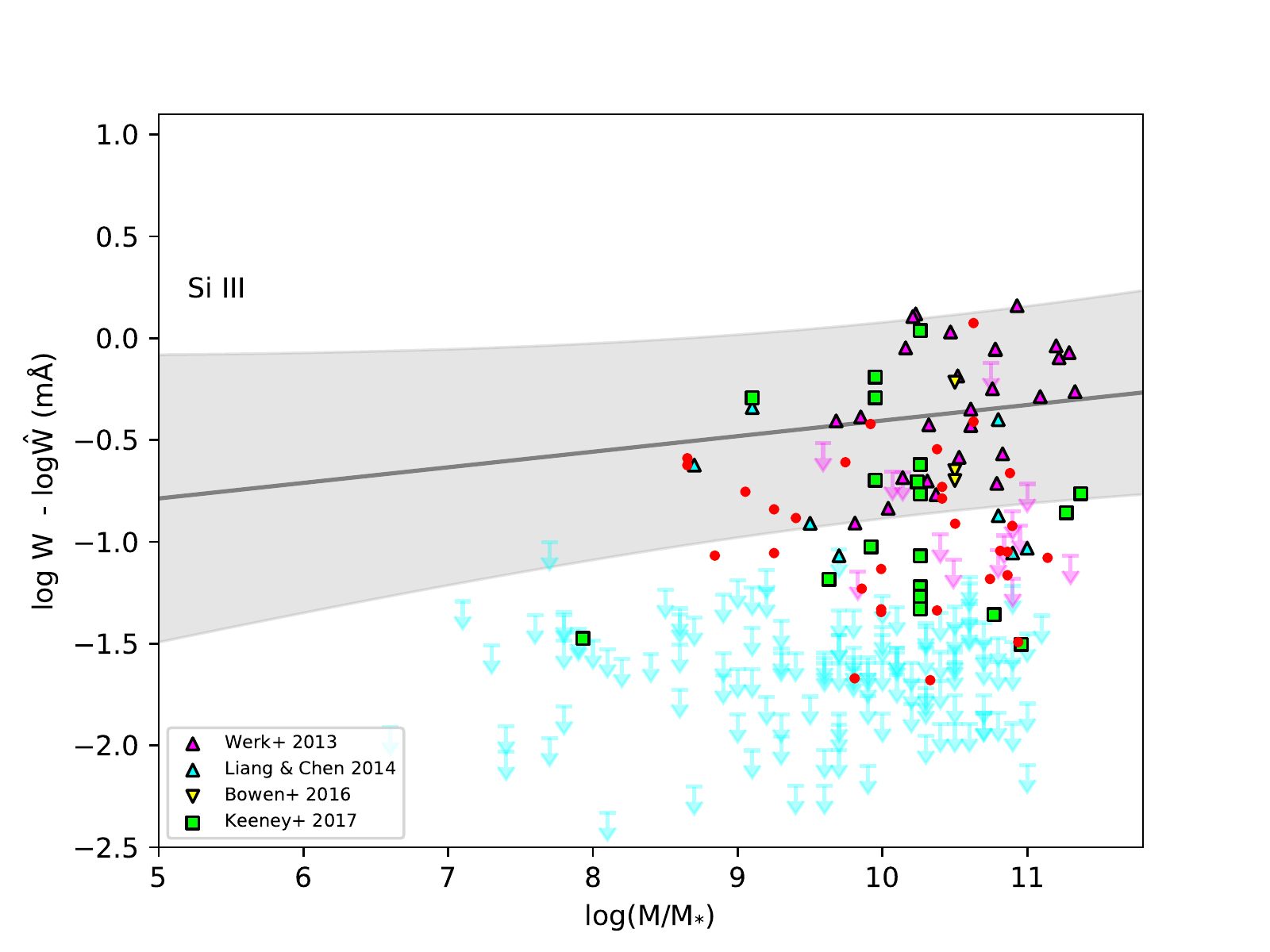}{0.4\textwidth}{(e)}}
\gridline{\fig{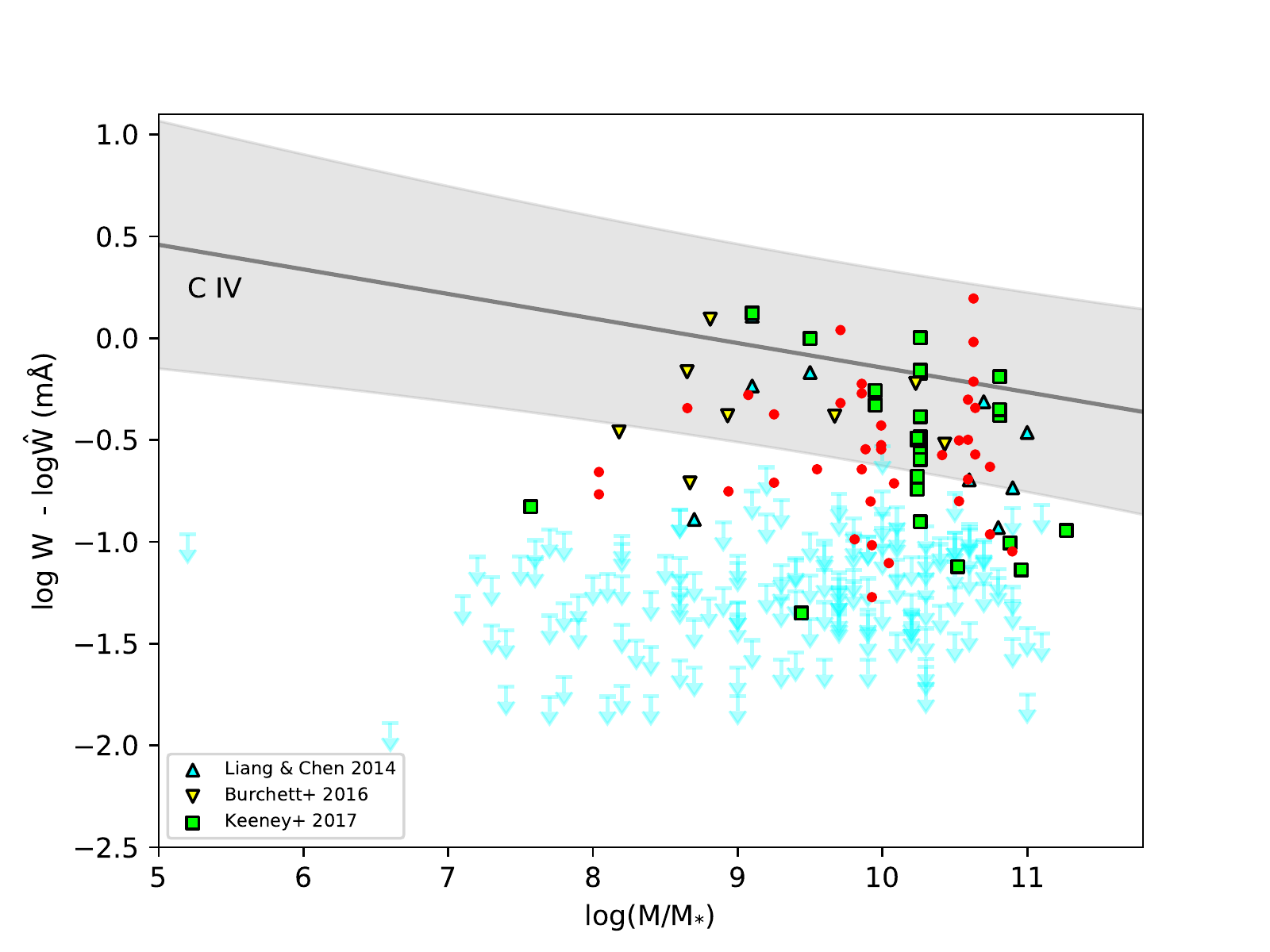}{0.4\textwidth}{(f)}
          \fig{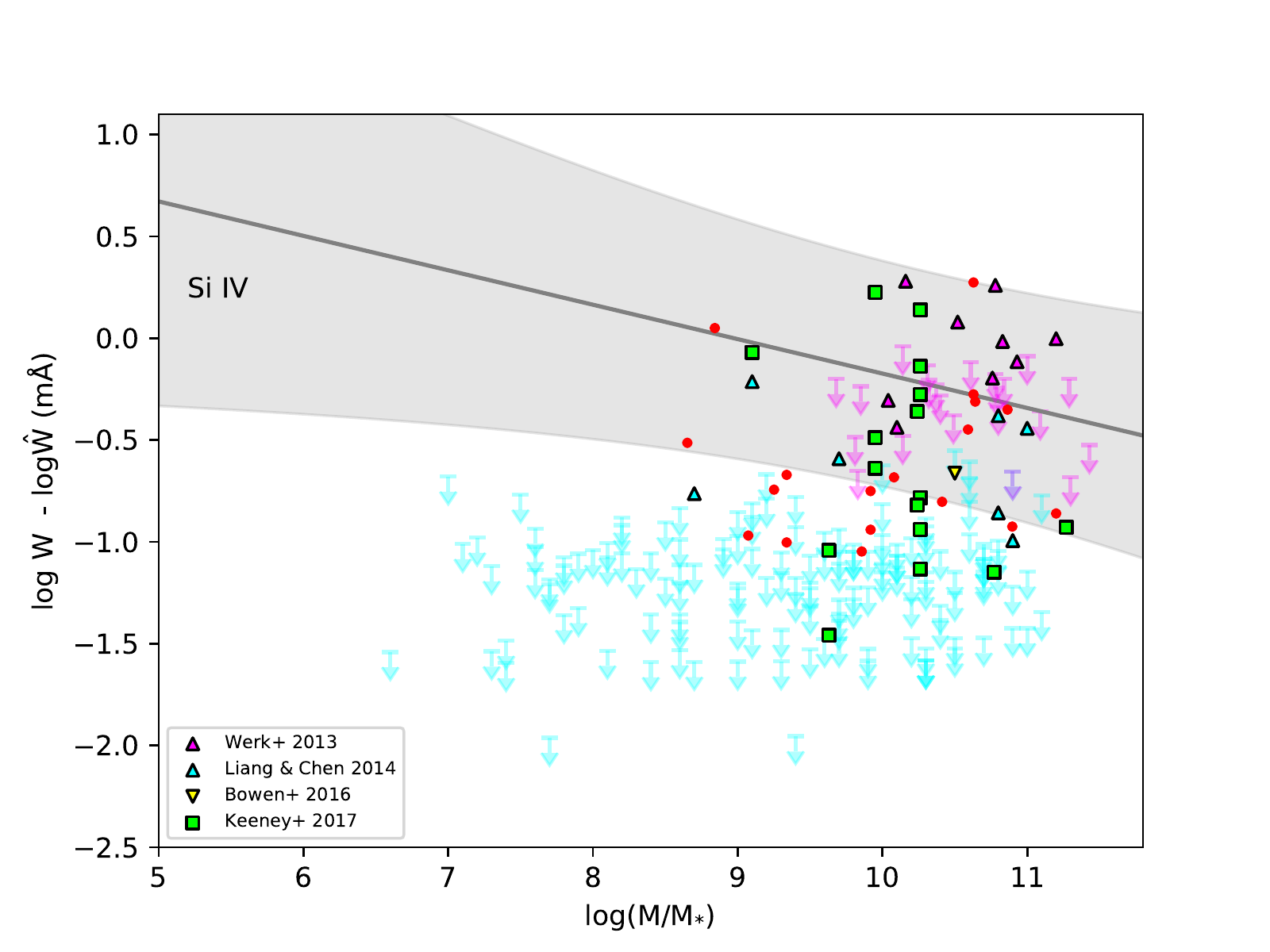}{0.4\textwidth}{(g)}}
\caption{Difference of absorber equivalent width and
$\hat{W}$ versus galaxy stellar mass in solar masses
for all unique galaxy-absorber pairs for (a) \lya\, (b) \ion{C}{2}
(c) \ion{Si}{2}, (d) \ion{C}{3}, (e) \ion{Si}{3}, (f) \ion{C}{4}, (g) \ion{Si}{4}, and (h) \ion{O}{6}. Black lines and gray shaded
regions show the fits to the literature points and  $1\sigma$ confidence intervals.
Red points are the unique galaxy-absorber
pairs identified by the virial radius method, with shading
indicating the value of $P_{\rm ion}$ ($=P(M_*)$ for single component \lya\ absorbers).
Upper limits from \cite{Prochaska2011b} (yellow), \cite{Werk2013} (magenta), \cite{Tumlinson2013} (green), and \cite{Liang2014} (cyan).
In panel (a), pairs with $P(M_*)> 0.2$ are shown on the top,
and those with $P(M_*) < 0.2$ are on the bottom.}
\end{figure}

\renewcommand{\thefigure}{\arabic{figure} (Cont.)}
\addtocounter{figure}{-1}

\begin{figure}
\gridline{\fig{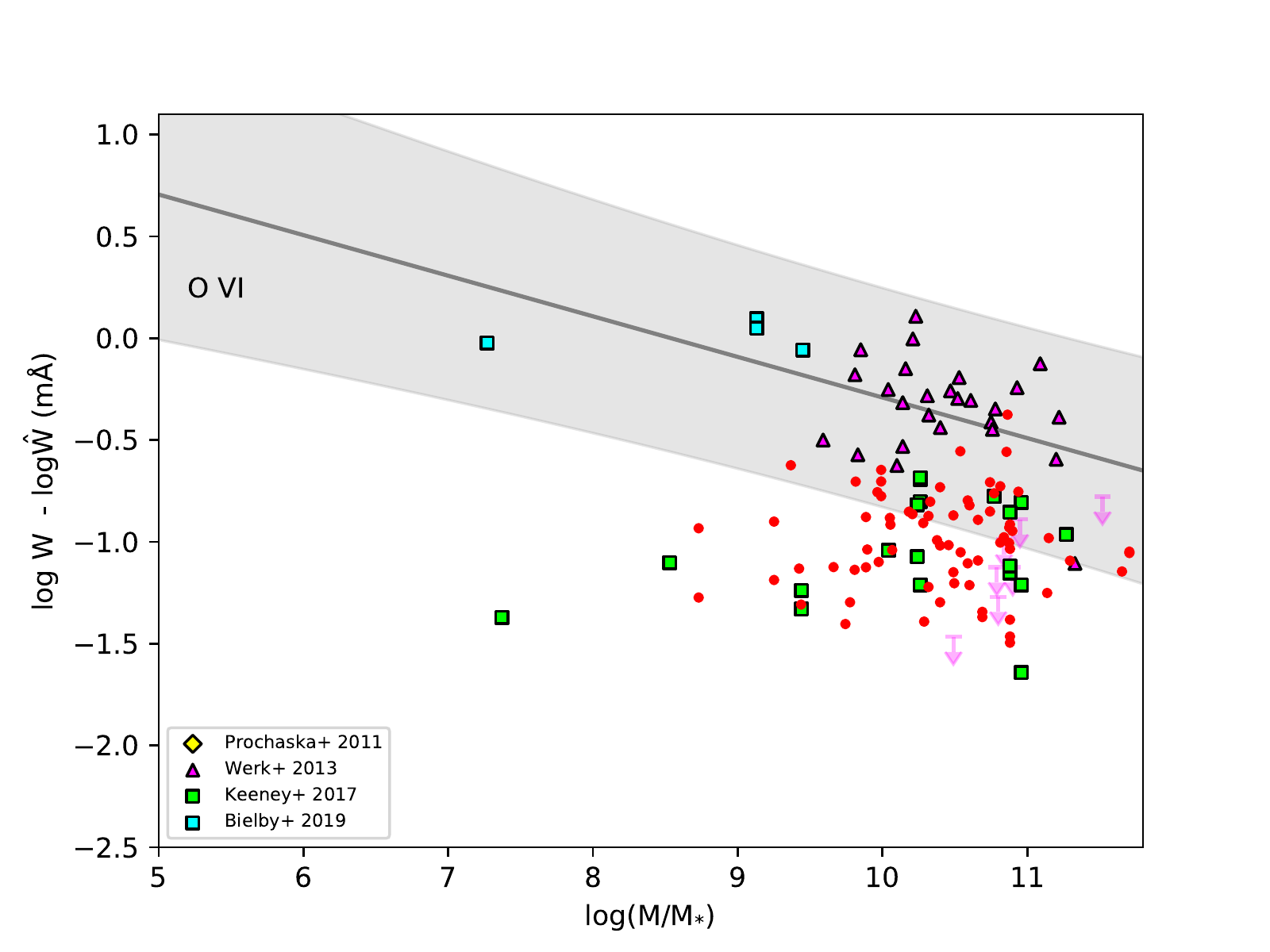}{0.4\textwidth}{(h)}}
\caption{Difference of absorber equivalent width and
$\hat{W}$ versus galaxy stellar mass in solar masses
for all unique galaxy-absorber pairs for (a) \lya\, (b) \ion{C}{2}
(c) \ion{Si}{2}, (d) \ion{C}{3}, (e) \ion{Si}{3}, (f) \ion{C}{4}, (g) \ion{Si}{4}, and (h) \ion{O}{6}. Black lines and gray shaded
regions show the fits to the literature points and  $1\sigma$ confidence intervals.
Red points are the unique galaxy-absorber
pairs identified by the virial radius method, with shading
indicating the value of $P_{\rm ion}$ ($=P(M_*)$ for single component \lya\ absorbers).
Upper limits from \cite{Prochaska2011b} (yellow), \cite{Werk2013} (magenta), \cite{Tumlinson2013} (green), and \cite{Liang2014} (cyan).
In panel (a), pairs with $P(M_*)> 0.2$ are shown on the top,
and those with $P(M_*) < 0.2$ are on the bottom.}
\end{figure}

\renewcommand{\thefigure}{\arabic{figure}}

\section{Results}
Using each of the two methods described above we identify the ten most probable galaxy matches to each absorber or absorption system, for all
$P(r_{\rm vir})> 0$ or $P(M_*) > 0$. 
Both methods method resulted in 1026 absorbers in the COS sample, including 136 metal line systems,
matched to at least one SDSS galaxy.
From among these top
ten matches, we identify unique candidate galaxy-absorber pairs.
Table~\ref{tab:stats} lists the characteristics of the galaxies, absorbers, and pairs in the virial radius method candidate list. Values are the same for the 
pairs found using the stellar mass method unless otherwise noted.
\begin{deluxetable}{lrrr}
\tablecaption{Distributions of Galaxy-Absorber Pair Characteristics\tablenotemark{1}
\label{tab:stats}
}
\tablewidth{0pt}
\tablehead{
\colhead{} &\colhead{min.} &\colhead{med.} &\colhead{max.}
}
\startdata
\\
Galaxies & & & \\
\tableline
$M_r$                &-23.0 &-19.9(-20.0) &-15.9\\
$u-r$                &0.57  &1.7  &6.9\\
$\log{M_*/M_{\sun}}$ &8.0   &9.7   &11.7\\
\tableline
\\
Absorbers & & & \\
\tableline
$z_{\rm abs}$                 &0.0016 &0.11 &0.47\tablenotemark{2}\\
$W($\lya$)$ (m${\rm \AA}$)    &5.3    &92   &2800(1400)\\
\tableline
\\
Pairs & & & \\
\tableline
$\rho$ (kpc)        &29(19) &430 &500\tablenotemark{3} \\
\enddata
\tablenotetext{1}{Values are the same for the pairs found by both methods, otherwise listed separately for pairs found by the virial radius(stellar mass) method.}
\tablenotetext{2}{$z=0.5$ is imposed as the maximum redshift for galaxy-absorber pairs.}
\tablenotetext{3}{A maximum impact parameter $\rho=500$~kpc is imposed by both the virial radius and
stellar mass algorithms}
\end{deluxetable}

As discussed above, the product of $P_{\rm ion}$, or
$P_j$ as expressed in 
Equation~\ref{equ:bayes}, over all ionic species
results in a smaller value for the overall probability, $P(r_{\rm vir})$ or $P(M_*)$, for multicomponent absorption systems.
This is consistent with the
additional constraints provided by the combination of several ions. However, this does mean that only 
relative comparisons of the total probability, $P(r_{\rm vir})$ or $P(M_*)$, 
for different galaxy matches to the same
absorber, are possible. In our assignment of unique individual galaxy-absorber pairs, we must assess the relative 
strength of different galaxy-absorber matches, and 
so for this we use the value of $P(r_{\rm vir})$  or $P(M_*)$ of \lya\ only in the case of a multicomponent system so that the presence of
metals does not penalize a particular absorber in the ranking, when in fact the presence of metals should
improve the chances that it is associated with a galaxy.

The 849 \lya\ lines in 636 unique galaxy-absorber pairs satisfying the 10\% probability criterion identified with virial radius method are shown as red points,
in Figure~\ref{fig:rvir} and tabulated in 
Table~\ref{tab:rvir}. 
In Figure~\ref{fig:rvir}(a) we show the results for \lya, shaded by the value of $P(r_{\rm vir})$.
Of these candidate pairs, 87 of the absorbers are metal line systems.
Table~\ref{tab:mstar} lists the 632 galaxy-absorber pairs comprising 831 \lya\ lines identified using the CGM fundamental plane, the stellar mass method. 
These pairs are shown in Figure~\ref{fig:mstar}
Of these candidate pairs, 83 of the absorbers are metal line systems. 
The red points in Figures~\ref{fig:rvir} and \ref{fig:mstar} do generally lie within the confidence limits on the fits to the 
literature pairs listed in Table~\ref{tab:fits}, but
linear fits to the results for each ion
do yield different slopes and intercepts. 
However, there is no systematic difference, and
this is likely due to the relative lack of
results at low impact parameter.
\begin{longrotatetable}
\begin{deluxetable*}{lllllrrllllllrrlll}
\tabletypesize{\tiny}
\tablecaption{Galaxy-Absorber Pairs: P($r_{\rm vir}$)  \label{tab:rvir}}
\tablewidth{0pt}
\tablehead{
\colhead{QSO} &\colhead{P$_{\rm tot}$} &\colhead{$\rho$} &\colhead{Ion} &\colhead{$z_{\rm abs}$} &\colhead{W} &\colhead{$\sigma_{\rm W}$}
&\colhead{P$_{\rm ion}$} &\colhead{$z_{\rm phot}$}  &\colhead{RA} &\colhead{Dec} &\colhead{$r$} &\colhead{M$_r$} &\colhead{$r_{\rm vir}$}
&\colhead{$\log{M_*/M_\odot}$} &\colhead{SDSS ID} &\colhead{K18 flag\tablenotemark{1}} \\
\colhead{} &\colhead{} &\colhead{(kpc)} &\colhead{} &\colhead{} &\colhead{(m\AA)} &\colhead{(m\AA)}
&\colhead{} &\colhead{} &\colhead{(deg)} &\colhead{(deg)} &\colhead{} &\colhead{} &\colhead{(kpc)} &\colhead{} &\colhead{} &\colhead{}
}
\startdata
1es1028  &2.5207E-01 &254.687 &\lya &2.427E-03 &89.782 &12.969 &2.671E-01 &2.154E-02 &158.921 &50.100 &16.465 &-18.431 &70.329 &9.032 &1237657630058938491  &\dots \\
1es1028  &2.5207E-01 &254.687 &\lya &3.224E-03 &270.129 &28.907 &8.360E-02 &2.154E-02 &158.921 &50.100 &16.465 &-18.431 &70.329 &9.032 &1237657630058938491  &\dots \\
1es1028  &1.1958E-01 &343.842 &\lya &2.188E-02 &79.266 &8.807 &1.731E-01 &2.114E-02 &157.506 &50.847 &15.275 &-19.579 &91.705 &9.536 &1237658800956833843  &\dots \\
1es1028  &1.1958E-01 &343.842 &\lya &2.258E-02 &86.057 &9.779 &1.785E-01 &2.114E-02 &157.506 &50.847 &15.275 &-19.579 &91.705 &9.536 &1237658800956833843  &\dots \\
1es1028  &1.0094E-01 &371.030 &\lya &4.484E-02 &135.906 &21.056 &1.009E-01 &1.033E-01 &157.917 &50.995 &19.433 &-18.974 &79.386 &9.090 &1237658800956965095 &1 \\
1es1028  &6.0033E-01 &447.070 &\lya &5.098E-02 &194.105 &9.515 &6.003E-01 &4.761E-02 &157.988 &50.965 &17.823 &-18.847 &77.109 &9.053 &1237658800956965110 &2 \\
1es1028  &1.8408E-01 &335.835 &\lya &7.604E-02 &35.315 &13.011 &1.841E-01 &1.739E-01 &157.729 &50.911 &19.989 &-19.632 &92.899 &9.806 &1237658800956899418  &\dots \\
1es1028  &1.1389E-01 &477.845 &\lya &8.836E-02 &78.099 &12.863 &1.139E-01 &1.649E-01 &157.905 &50.831 &18.799 &-20.787 &125.474 &10.287 &1237657589242331520 &3 \\
1es1028  &1.2180E-01 &479.116 &\lya &1.071E-01 &86.713 &7.226 &1.218E-01 &1.739E-01 &157.779 &50.954 &19.406 &-20.316 &110.620 &10.152 &1237658800956899891  &\dots \\
1es1028  &2.1460E-01 &469.197 &\lya &1.125E-01 &115.956 &12.584 &2.146E-01 &1.552E-01 &157.732 &50.913 &19.508 &-19.804 &96.968 &9.525 &1237658800956899417 &3 \\
\enddata
\tablenotetext{1}{Absorption flag as in \cite{Keeney2018}:
   -1 = object has $z < 0.001$ and is likely a star;
    0 = galaxy is not within 1000 \kms\ of an absorber;
    1 = galaxy is within 1000 \kms\ of an absorber but is not the closest galaxy;
    2 = galaxy is closest in this Table to an absorber, but a closer galaxy
        is known from SDSS or other sources;
    3 = closest known galaxy to an absorber.}
\end{deluxetable*}
\end{longrotatetable}

\begin{longrotatetable}
\begin{deluxetable*}{lllllrrllllllrrlll}
\tabletypesize{\tiny}
\tablecaption{Galaxy-Absorber Pairs: P($M_*$) \label{tab:mstar}}
\tablewidth{0pt}
\tablehead{
\colhead{QSO} &\colhead{P$_{\rm tot}$} &\colhead{$\rho$} &\colhead{Ion} &\colhead{$z_{\rm abs}$} &\colhead{W} &\colhead{$\sigma_{\rm W}$}
&\colhead{P$_{\rm ion}$} &\colhead{$z_{\rm phot}$}  &\colhead{RA} &\colhead{Dec} &\colhead{$r$} &\colhead{M$_r$} &\colhead{$r_{\rm vir}$}
&\colhead{$\log{M_*/M_\odot}$} &\colhead{SDSS ID} &\colhead{K18 flag\tablenotemark{1}}\\
\colhead{} &\colhead{} &\colhead{(kpc)} &\colhead{} &\colhead{} &\colhead{(m\AA)} &\colhead{(m\AA)}
&\colhead{} &\colhead{} &\colhead{(deg)} &\colhead{(deg)} &\colhead{} &\colhead{} &\colhead{(kpc)} &\colhead{} &\colhead{} &\colhead{}
}
\startdata
1es1028  &1.4423E-01 &50.251 &\lya &2.427E-03 &89.782 &12.969 &1.322E-01 &2.114E-02 &157.506 &50.847 &15.275 &-19.579 &91.705 &9.536 &1237658800956833843  &\dots \\
1es1028  &1.4423E-01 &50.251 &\lya &3.224E-03 &270.129 &28.907 &2.023E-01 &2.114E-02 &157.506 &50.847 &15.275 &-19.579 &91.705 &9.536 &1237658800956833843  &\dots \\
1es1028  &1.7174E-01 &434.062 &\lya &2.188E-02 &79.266 &8.807 &1.803E-01 &3.214E-02 &157.440 &50.797 &15.772 &-20.018 &102.384 &9.779 &1237658800956833808  &\dots \\
1es1028  &1.7174E-01 &434.062 &\lya &2.258E-02 &86.057 &9.779 &1.830E-01 &3.214E-02 &157.440 &50.797 &15.772 &-20.018 &102.384 &9.779 &1237658800956833808  &\dots \\
1es1028  &6.0011E-01 &447.070 &\lya &5.098E-02 &194.105 &9.515 &6.001E-01 &4.761E-02 &157.988 &50.965 &17.823 &-18.847 &77.109 &9.053 &1237658800956965110 &2 \\
1es1028  &1.1362E-01 &444.346 &\lya &7.604E-02 &35.315 &13.011 &1.136E-01 &2.860E-01 &157.903 &50.964 &20.549 &-20.139 &105.635 &9.487 &1237658800956965086  &\dots \\
1es1028  &1.1236E-01 &477.845 &\lya &8.836E-02 &78.099 &12.863 &1.124E-01 &1.649E-01 &157.905 &50.831 &18.799 &-20.787 &125.474 &10.287 &1237657589242331520 &3 \\
1es1028  &1.6630E-01 &408.069 &\lya &9.436E-02 &101.429 &9.138 &1.663E-01 &1.739E-01 &157.729 &50.911 &19.989 &-19.632 &92.899 &9.806 &1237658800956899418  &\dots \\
1es1028  &1.1417E-01 &479.116 &\lya &1.071E-01 &86.713 &7.226 &1.142E-01 &1.739E-01 &157.779 &50.954 &19.406 &-20.316 &110.620 &10.152 &1237658800956899891  &\dots \\
1es1028  &2.0735E-01 &469.197 &\lya &1.125E-01 &115.956 &12.584 &2.074E-01 &1.552E-01 &157.732 &50.913 &19.508 &-19.804 &96.968 &9.525 &1237658800956899417 &3 \\
\enddata
\tablenotetext{1}{Absorption flag as in \cite{Keeney2018}:
   -1 = object has $z < 0.001$ and is likely a star;
    0 = galaxy is not within 1000 \kms\ of an absorber;
    1 = galaxy is within 1000 \kms\ of an absorber but is not the closest galaxy;
    2 = galaxy is closest in this Table to an absorber, but a closer galaxy
        is known from SDSS or other sources;
    3 = closest known galaxy to an absorber. }
\end{deluxetable*}
\end{longrotatetable}

The galaxies for which there is a spectroscopic redshift measured by SDSS provide a useful check on our methods. For those
222 galaxies, in pairs identified by the two methods, the galaxy spectroscopic redshift lies 
within 400 \kms\ of the absorber redshift in  55(57) of the cases for the virial radius(stellar mass) method, 
indicating that our candidate galaxy-absorber pair lists may contain a fairly high 
rate of spurious matches.  
The rate of velocity misalignment is $\sim$48-57\% if we consider only pairs with
$\rho< 200$ kpc or $\rho/r_{\rm vir} < 2$ or absorbers with $W> 200$ m\AA, which constitute 10\%, 12\% and 23\%, respectively, of the 
pairs in Table~\ref{tab:rvir} and 11\%, 14\%, and 23\% of those listed in Table~\ref{tab:mstar}. 
For metal line systems, the overall rate of velocity misalignments is 35\% and 29\% for the two methods.
The SDSS spectroscopic redshift measurements are primarily                                         
concentrated in the lower redshift half of our sample $z<0.1$, where the offsets with photometric redshifts are large compared to the photometric redshift error 
\citep{Beck2016}.
Thus, the false positive rate may be lower than estimated here in the overall candidate list.

Individual sightlines are discussed in Appendix~\ref{sec:sightlines}. 
These notes highlight the importance of investigating both the ranked galaxy-absorber outcomes of the methods and the final
galaxy-absorber pairs after the uniqueness criterion has been imposed.
Galaxies particularly worthy of follow-up investigations are noted.

\subsection{Comparison with the Literature}
Our virial radius and stellar mass Bayesian methods identified 402 and 411
candidate galaxy-absorber pairs respectively along the QSO lines of sight in common with previous studies.
Specific comparisons with 81 pairs found by previous studies are tabulated in
Table~\ref{tab:lit}. In 34 cases, the galaxy is excluded from our catalog of SDSS galaxies, usually on the basis of the photometric redshift quality cut we employed.
Of the 47 pairs for which the galaxy is part of our SDSS catalog, we recover the galaxy-absorber pair as the top ranked match with one or both methods in 19 cases, and in another
15 cases, the pair was recovered but with a lower ranking for both methods.
The 13 unrecovered pairs are due to either an impact parameter larger than our threshold
of 500~kpc or to a large  mismatch between the photometric and spectroscopic redshifts, which, by construction has been measured for all of these galaxies.
Our methods do not use the spectroscopic redshifts even when available, but they do provide insight into why we do not recover all the 
pairings found in the targeted studies.
The cases in which the poor photometric redshift estimate is the reason we do not recover a pair correspond to galaxies with $z< 0.01$ where the photometric redshift techniques are expected
to have trouble given their overall accuracy. Nevertheless, 
there are several cases listed in Table~\ref{tab:lit} of $z< 0.01$ galaxy-absorber pairs that are recovered by our method, even as the top-ranked match.
These QSO fields are discussed in detail Appendix~\ref{sec:sightlines} along with all the others in our sample.
\begin{longrotatetable}
\begin{deluxetable*}{lllllllllllll}
\tabletypesize{\scriptsize}
\tablecaption{Galaxy-Absorber Pairs in the Literature \label{tab:lit}}
\tablewidth{0pt}
\tablehead{
\colhead{QSO} &\colhead{$\rho$} &\colhead{Galaxy} &\colhead{Mag.} &\colhead{$z_{\rm gal}$} &\colhead{$z_{\rm abs}$} &\colhead{W\tablenotemark{1}} &\colhead{$\sigma_{\rm W}$\tablenotemark{1}} 
&\colhead{Ref.\tablenotemark{2}} &\colhead{SDSS} 
&\colhead{Rank\tablenotemark{4}} 
&\colhead{Rank\tablenotemark{4}} &\colhead{$z_{\rm phot}$}  \\
\colhead{} &\colhead{(kpc)} &\colhead{} &\colhead{Filter} &\colhead{} &\colhead{} &\colhead{(m\AA)} &\colhead{(m\AA)} 
&\colhead{} &\colhead{Note\tablenotemark{3}} 
&\colhead{($r_{\rm vir}$)} 
&\colhead{($M_*$)} &\colhead{}
}
\startdata
1ES 1028+511 &90  &UGC 5740                  &15.1B  &0.00216 &0.00243 &13.5  &0.19 &S13 &3 & & & \\
1ES 1028+511 &25  &SDSS J103108.88+504708.7  &16.3g  &0.00311 &0.00320 &17.2  &22   &S13 &3 & & & \\
1ES 1028+511 &279 &SDSS J103110.35+505211.0  &18.4g  &0.137   &0.137   &238   &10   &LC14 &1 &1  &1 &0.136  \\
1SAX J1032.3+5051 &65 &UGC 5740           &15.1B &0.00216 &0.00239 &13.1  &3.5  &S13  &3  &  & & \\
3C 273  &69   &SDSS J122815.96+014944.1  &16.5g &0.00303 &0.00340 &393   &8    &LC14 &1  &1  &1 &0.0136 \\
3C 273  &118  &SDSS J123103.89+014034.4  &14.8B &0.00368 &0.00337  &350 &36 &B96,P02,S13 &3 & & & \\
3C 273  &126   &SDSS J122745.86+013601.8 &17.0B &0.00431 &0.00528  &410   &\dots   &B96,P02 &3 & & & \\
3C 273  &288   &NGC 4420                 &12.7g &0.00565 &0.00529  &15.38 &0.34    &S13 &2 & & & \\ 
3C 273  &80   &SDSS J122950.57+020153.7  &17.3g &0.00592 &0.00529  &15.38 &0.34 &S13   &1 &\dots &\dots  &0.0362 \\
3C 273  &771  &UGC 7625                  &16.7g &0.00745 &0.00720  &20 &5      &P02,WS09  &3 & & & \\
3C 273  &429  &SDSS J122815.88+024202.9  &15.4g &0.00762 &0.00758  &31 &7      &WS09   &1 &1 &1 &0.019  \\
FBQS J1010+3003  &48 &UGC 5478                  &14.3B  &0.00459 &0.00462 &17.8  &3.5  &S13  &3 & & & \\
FBQS J1010+3003  &181 &SDSS J101008.85+300252.5 &18.6g  &0.0874  &0.0875  &329   &12   &LC14 &1 &7 &7 &0.111  \\
FBQS J1010+3003  &252 &SDSS J100953.51+300202.2 &17.5g  &0.113   &0.114   &254   &15   &LC14 &1 &\dots &6 &0.0957  \\
MRK 106 &1030 &UGC 4800                  &14.6B &0.00811 &0.00803 &77$\tablenotemark{5}$  &12  &WS09 &3 & & &  \\
MRK 106 &418  &SDSS J091923.29+553137.2  &16.6g &0.0318  &0.0317  &217                    &6   &LC14 &3 & & &  \\
MRK 478 &645  &NGC 5727                  &14.2B &0.00497 &0.00525 &254                    &14  &WS09 &1 &\dots &\dots &0.198 \\
PG 0003+158 &193  &SDSS J000556.15+160804.1  &17.4g &0.0909 &0.0909   &805 &8	&LC14  &1 &{\bf 6} &1 &0.087 \\
PG 0026+129 &111  &WISEA J002915.37+132056.5 &15.8B &0.0394 &0.0391   &444 &16	&B97,D16 &1 &2 &2 &0.028 \\
PG 0832+251  &263  &SDSS J083335.65+250847.1  &18.1g  &0.00743 &0.00730 &248   &11   &LC14 &3 & & &  \\
PG 0832+251  &53  &NGC 2611  &14.9g  &0.0174 &0.0174   &18.4  &0.2  &S13  &1 &2 &1 &0.027 \\
PG 0832+251  &283  &SDSS J083607.41+250645.7  &15.5g  &0.0232  &0.0233  &160   &13   &LC14 &1 &2 &2  &0.0230 \\
PG 0844+349 &250 &NGC 2683  &10.2B &0.00137 &0.00137  &25$\tablenotemark{5}$ &8 &WS09 &2 & & &  \\
PG 0844+349 &372 &UGC 4621  &13.9B &0.00769 &0.00754  &6$\tablenotemark{5}$  &5 &WS09 &2 & & &  \\
PG 0953+414 &296 &NGC 3104  &13.6B &0.00204 &0.00204  &70                    &9 &WS09 &2 & & &  \\
PG 0953+414 &435 &SDSS J095638.90+411646.0  &17.6g &0.143   &0.142    &268   &5    &S13,LC14 &1 &7 &3 &0.132 \\
PG 1001+291 &84   &UGC 5427                  &14.6B &0.00165 &0.00165 &308 &112 &WS09 &3 & & &  \\
PG 1001+291 &337  &UGC 5464                  &15.8g &0.00337 &0.00357 &267 &14  &WS09 &1 &7 &7 &0.0236 \\
PG 1001+291 &167  &SDSS J100618.16+285641.9  &14.4g &0.00454 &0.0036  &297 &23  &B97,LC14  &3 & & &  \\
PG 1001+291 &1249  &UGC 5461                 &15.5g &0.0160  &0.0153  &242 &11  &WS09  &1 &\dots &\dots &0.0350 \\
PG 1001+291 &179   &SDSS J100403.24+285650.2 &18.6g &0.133   &0.134   &177 &9   &LC14  &1 &1 &1 &0.143 \\
PG 1001+291 &57   &SDSS J100402.36+285512.5  &22.3g &0.138   &0.137   &776 &18  &M14   &3 & & &  \\ 
PG 1001+291 &222  &TON0028:[KSS94] 39        &22.4g &0.214   &0.214   &736 &55  &M14   &4 & & &  \\ 
PG 1048+342 &465  &SDSS J105111.41+335935.6  &15.8g &0.0596  &0.0593 &270  &8   &LC14  &1 &1 &2  &0.0601 \\
PG 1116+215 &543  &UGC 6258                  &14.9B &0.00485 &0.00493 &91  &10  &WS09  &1 &\dots &\dots  &0.0140 \\
PG 1116+215 &1742 &NGC 3649                  &14.6B &0.0166  &0.0163 &111  &12  &WS09 &3 & &  & \\
PG 1116+215 &557  &SDSS J112045.94+211115.3  &18.1g &0.0210  &0.0195  &167  &24	&T98  &1 &\dots &\dots  &0.0217 \\
PG 1116+215  &388   &SDSS J111843.28+212723.0  &17.1B &0.0323  &0.0322 &108   &6     &LC14 &1 &1 &1 &0.0343 \\
PG 1116+215 &746  &SDSS J111909.56+210243.4  &18.1g &0.0410  &0.0412  &164  &37	&T98  &1 &\dots  &\dots &0.0505 \\
PG 1116+215 &601  &SDSS J111924.26+211029.9  &16.6g &0.0590  &0.0590  &157  &24	&T98  &1 &\dots &\dots &0.0687 \\   
PG 1116+215  &256   &SDSS J111905.34+211537.7 &17.2B &0.0590  &0.0590 &13.5 &0.1 &S13 &1 &7 &6 &0.0411 \\
PG 1116+215 &131  &SDSS J111905.51+211733.0  &19.3g &0.0600  &0.0590 &13.5 &0.1 &S13  &1 &\dots &10 &0.0983 \\
PG 1116+215 &677  &SDSS J111942.04+212610.3  &19.3g &0.0613 &0.0608  &62   &22	&T98  &1 &\dots &\dots &0.0904 \\
PG 1116+215  &138   &SDSS J111906.68+211828.7  &18.1B &0.138   &0.138  &471   &5     &LC14 &1 &{\bf 4} &1  &0.134 \\
PG 1121+422 &123 &SDSS J112418.74+420323.1  &17.8g &0.0245 &0.0245   &512   &9    &LC14 &1 &1 &1 &0.0325  \\
PG 1121+422 &213 &SDSS J112457.15+420550.8  &17.1g &0.0337 &0.0338   &198   &6    &LC14 &1 &2 &2  &0.0421  \\
PG 1216+069 &12	 &VCC 381  &16.6HI &0.00160 &0.00550 &1630  &160 &B96 &3 & & &  \\
PG 1216+069 &77 &VCC 538  &15.4B &0.00167 &0.00550  &1630  &160 &B96 &1 &\dots &\dots &0.384 \\
PG 1216+069 &42	 &IC 3115 &13.7B  &0.00244 &0.00550 &1630  &160 &B96 &2 & & &  \\ 
PG 1216+069 &72  &VCC 446  &15.5B &0.00283 &0.00550  &1630  &160 &B96 &1 &9 &\dots &0.0128 \\
PG 1216+069 &146 &UGC 7423 &15.6B &0.00419 &0.00550  &1630  &160 &B96 &1 &\dots &\dots &0.0243  \\
PG 1216+069 &192 &VCC 340  &15.4B &0.00504 &0.00550  &1630  &160 &B96 &1 &10 &\dots &0.0116  \\
PG 1216+069 &184 &VCC 329  &16.8B &0.00541 &0.00550  &1630  &160 &B96 &1 &\dots &\dots &0.0913 \\
PG 1216+069 &184 &NGC 4260 &12.7B &0.00653 &0.00550  &1630  &160 &B96 &2 & & & \\
PG 1216+069 &207 &VCC 223  &16.5B &0.00690 &0.00550  &1630  &160 &B96 &1 &\dots &\dots &0.0159 \\
PG 1216+069 &185 &NGC 4241 &13.0B &0.00745  &0.00550 &1630  &160 &B96 &2 & & &  \\
PG 1216+069 &16  &VCC 415  &15.1B  &0.00854 &0.00550 &1630  &160 &B96 &3 & & &  \\ 
PG 1216+069 &344 &SDSS J121930.87+064334.4  &16.3g &0.0799 &0.0794   & 520  &110 &B96 &3 & & &  \\
PG 1216+069 &344 &SDSS J121930.87+064334.4  &16.3g &0.0799 &0.0784   & 450  &110 &B96 &3 & & &  \\
PG 1216+069 &500 &SDSS J121930.86+064334.4  &16.2g &0.0804  &0.0805  &13.87 &0.28 &S13   &3 & &  & \\
PG 1216+069 &95  &SDSS J121923.43+063819.7  &18.0R &0.1242 &0.124    &1376  &173  &M14 &1 &6 &1  &0.132 \\
PG 1229+204 &112 &UGC 7697                  &15.0B  &0.00846 &0.00859 &290   &70    &C05  &3 & & &  \\
PG 1259+593 &58  &SDSS J130207.44+584153.8  &14.4g  &0.00221 &0.00229 &13.83 &0.24  &S13  &3 & & &  \\
PG 1259+593 &55  &UGC 8146  &14.4B &0.00223 &0.00226   &330  &80 &C05 &3 & & & \\
PG 1259+593 &595 &UGC 8040  &14.7B &0.00841 &0.00759   &291  &11 &WS09   &1 &5 &5  &0.0194 \\
PG 1259+593 &136  &SDSS J130101.05+590007.1  &17.1g &0.0462  &0.0460    &15.51 &0.28  &S13  &1 &1 &1  &0.0433 \\
PG 1259+593 &135  &SDSS J130116.43+590135.7  &21.6g &0.197   &0.196     &44   &24   &M14 &1 &7 &8  &0.352 \\
PG 1259+593 &280  &SDSS J130109.88+590315.3  &20.6g &0.241   &0.241     &49   &15   &M14 &1 &5 &5  &0.292 \\
PG 2349-014 &198 &SDSS J235142.21-010100.9   &17.1g &0.0385 &0.0381   &418   &41   &B97,D16 &1 &1 &1 &0.0433 \\
Q 1230+115  &339  &UGC 7625   &16.7g &0.00745 &0.00769  &338   &21   &P02,WS09 &1 &\dots &\dots &0.114 \\
SBS 1108+560 &20  &M 108     &10.7B &0.00232 &0.00222  &14.3  &4.0  &S13  &2 & & & \\
SBS 1108+560 &20  &M 108     &10.7B &0.00232 &0.00259  &14.2  &4.0  &S13  &2 & & & \\
SBS 1108+560 &437 &WISEA J111125.60+554435.4  &18.4g &0.137   &0.138    &542   &5    &LC14 &1 &1 &1 &0.133  \\
SBS 1122+594 &32  &IC 691     &14.2g &0.00401 &0.004    &993   &15   &S13,LC14 &2 & & & \\
SBS 1122+594 &336 &SDSS J112517.67+590828.8  &17.2g &0.0578  &0.0578   &236   &11   &LC14 &1 &1 &1  &0.0645 \\
SDSS J080908.13+461925.6 &63 &SDSS J080913.17+461842.7  &17.1g  &0.0466  &0.0464  &1026 &10   &LC14 &1 &1 &2 &0.0636 \\
SDSS J092554.43+453544  &342 &SDSS J092721.06+454158.8  &17.1g &0.0171  &0.0170   &183   &10   &LC14 &1 &2 &2  &0.0338 \\
SDSS J092554.43+453544	&244 &SDSS J092617.38+452924.9  &17.1g &0.0270  &0.0261   &453   &10   &LC14 &1 &1 &2 &0.0326 \\
SDSS J094952.91+390203  &166 &SDSS J095002.76+390308.7 &17.8g &0.0658  &0.0669   &165   &12   &LC14 &1 &1 &1 &0.0584\\
TON 236 &193 &SDSS J152827.39+282738.6  &18.4g &0.0451  &0.0451   &160   &8    &LC14 &3 & & &  \\
TON 580	&249 &SDSS J113056.11+311445.6  &18.5g &0.0745  &0.0744   &324   &11   &LC14 &1 &1 &1  &0.0872 
\enddata
\tablenotetext{1}{Lyman $\alpha$ rest equivalent width. For \cite{Stocke2013} entries, these columns list $\log{{\rm N_{\rm HI}}}$ and its error}
\tablenotetext{2}{B97: \cite{Bowen1997}; T98: \cite{Tripp1998}; P02 \cite{Penton2002}; C05 \cite{Cote2005}; WS09: \cite{WakkerSavage2009}; S13: \cite{Stocke2013}; LC14: \cite{Liang2014}; M14: \cite{Mathes2014}; D16: \cite{Danforth2016}}
\tablenotetext{3}{Galaxy is present in SDSS catalog (1);Galaxy is not in SDSS catalog, $r < 14$ (2), photometric redshift cut (3), no photometric redshift (4)}
\tablenotetext{4}{Rank of galaxy-absorber pair with $P(r_{\rm vir})$ or $P(M_*)$. No entry indicates that this galaxy is not recovered in top 10 highest probability matches to the absorber for that method.}
\tablenotetext{5}{Lyman $\beta$}
\end{deluxetable*}
\end{longrotatetable}

In a large study of galaxy redshifts in QSO fields, 
\cite{Keeney2018} report spectroscopic redshifts for galaxies in 25 of our fields and flag the galaxies that are closest to an absorber in their Table 6. For the galaxy-absorber pairs
in our Table~\ref{tab:rvir} and Table~\ref{tab:mstar} that are in common with their sample, we reproduce their flag in Column 17 and we discuss the cases in which 
these new spectroscopic measurements confirm or refute our candidate galaxy-absorber pairs. The comparison with the \cite{Keeney2018} results is a useful exercise
for both demonstrating the reliability of these methods and uncovering cases in which our methods would recommend spectroscopic follow-up that was not included in their sample.
Our methods recover 86\% of the galaxies in their Table 6 with impact parameter less than 500~kpc and which are flagged as the closest galaxy to an absorber (their flag 3) to be a member of
a candidate galaxy-absorber pair. Including galaxies within 500~kpc that these authors mark as lying within 1000 \kms\ of an absorber but not necessarily the closest galaxy to an
absorber, their flags 1 or 2, we recover 73\%.

\subsection{Galaxy Properties}
\begin{figure}
\gridline{\fig{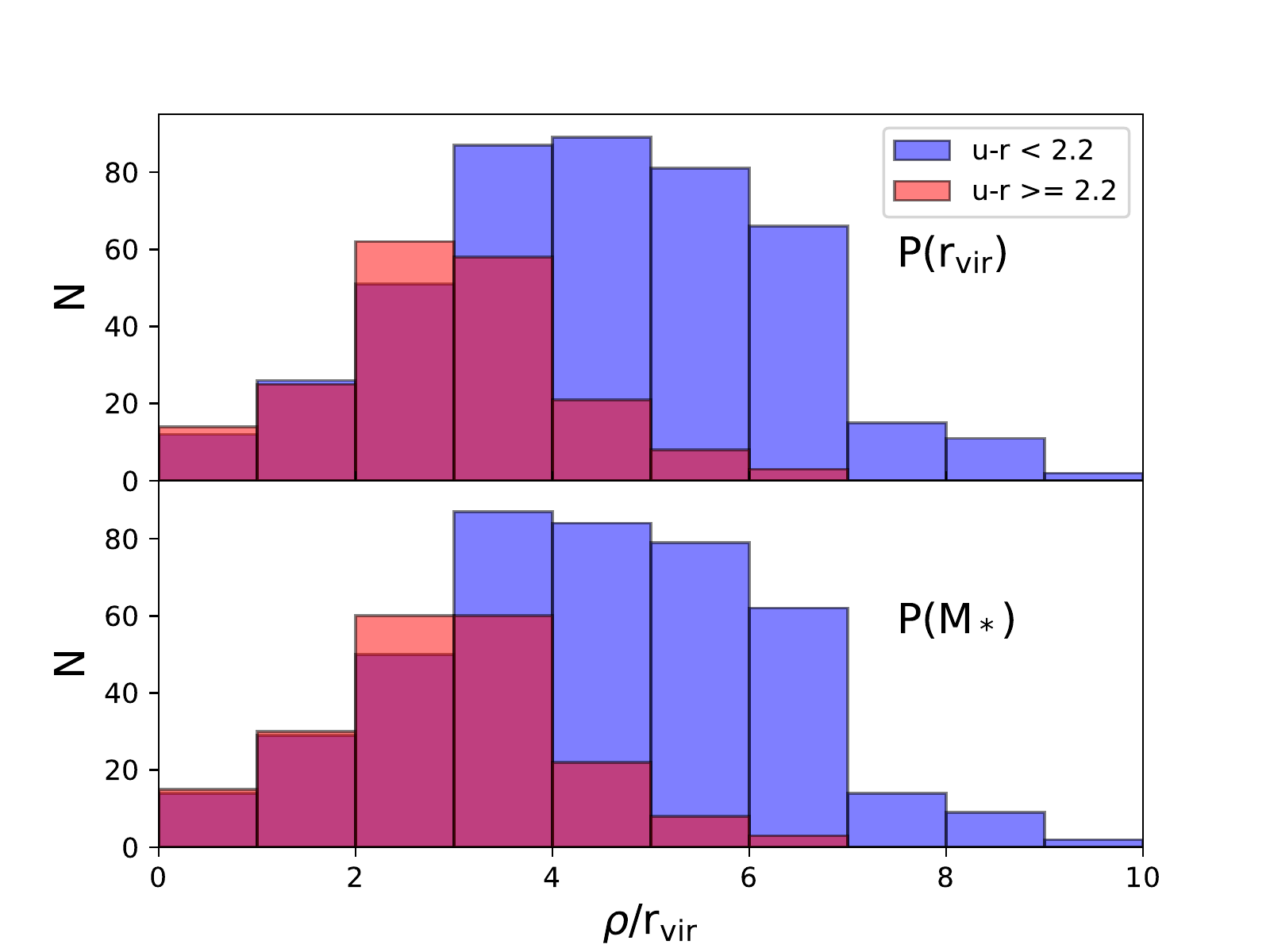}{0.5\textwidth}{(a)}
          \fig{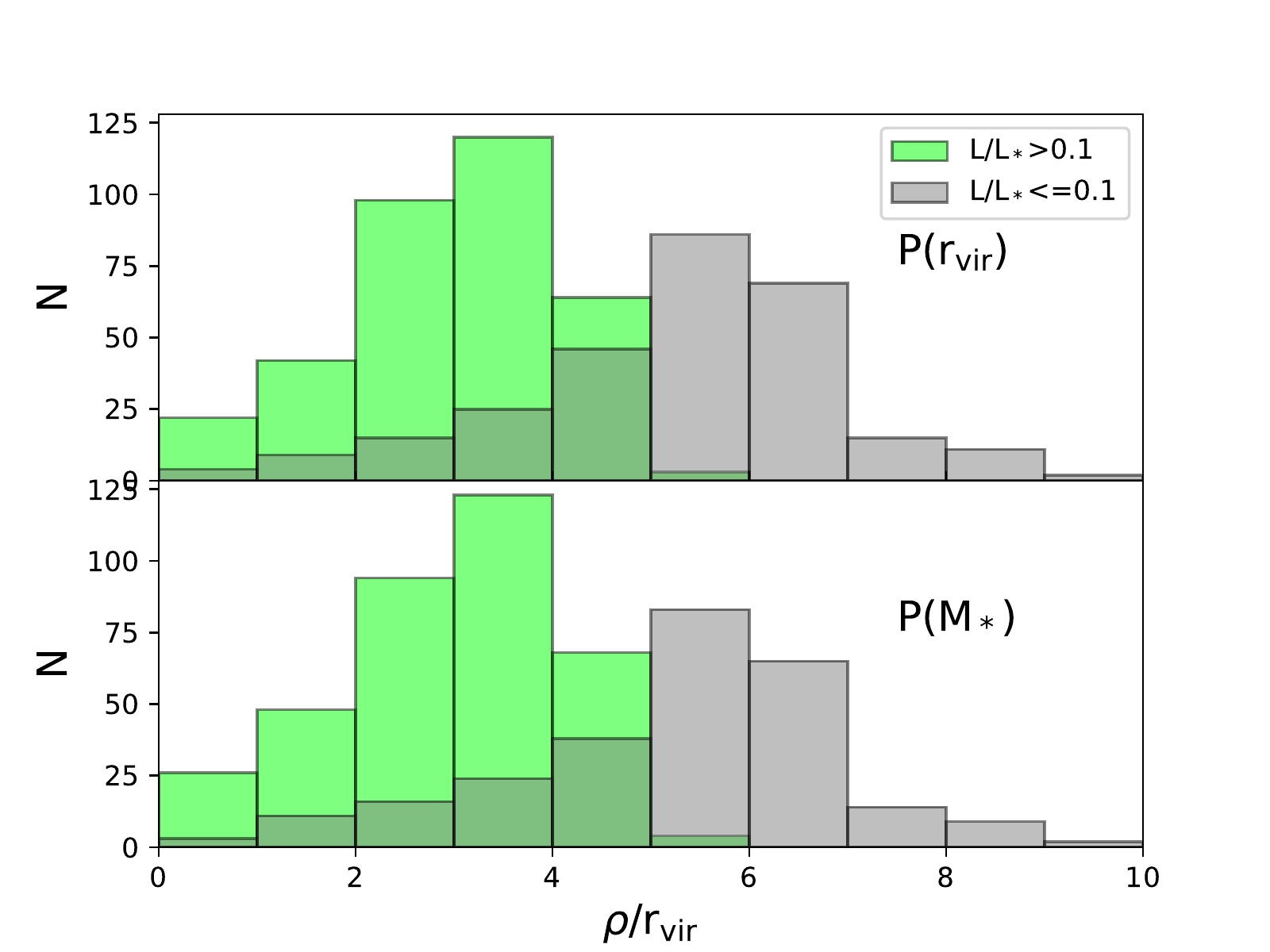}{0.5\textwidth}{(b)}}
\gridline{\fig{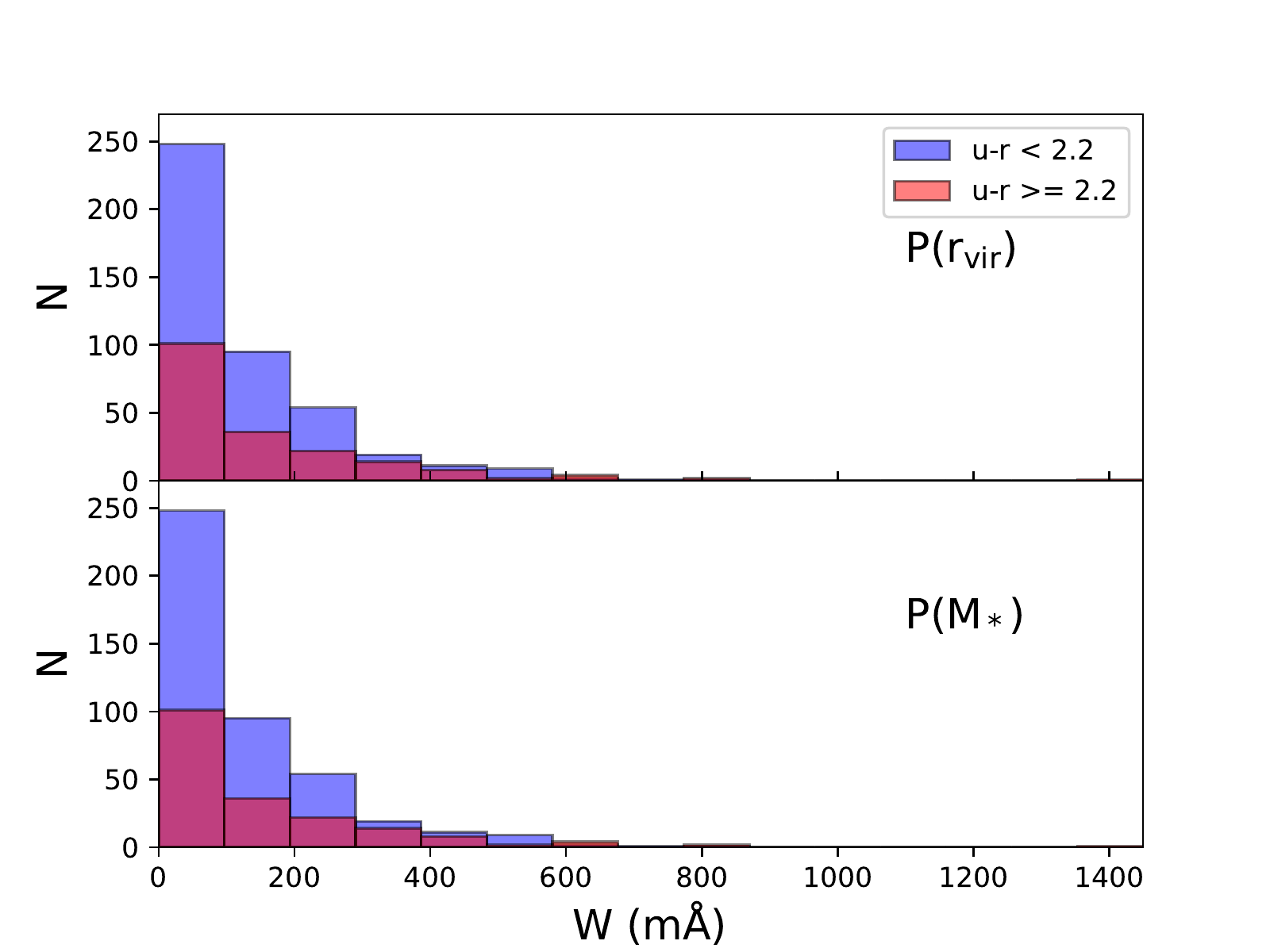}{0.5\textwidth}{(c)}
          \fig{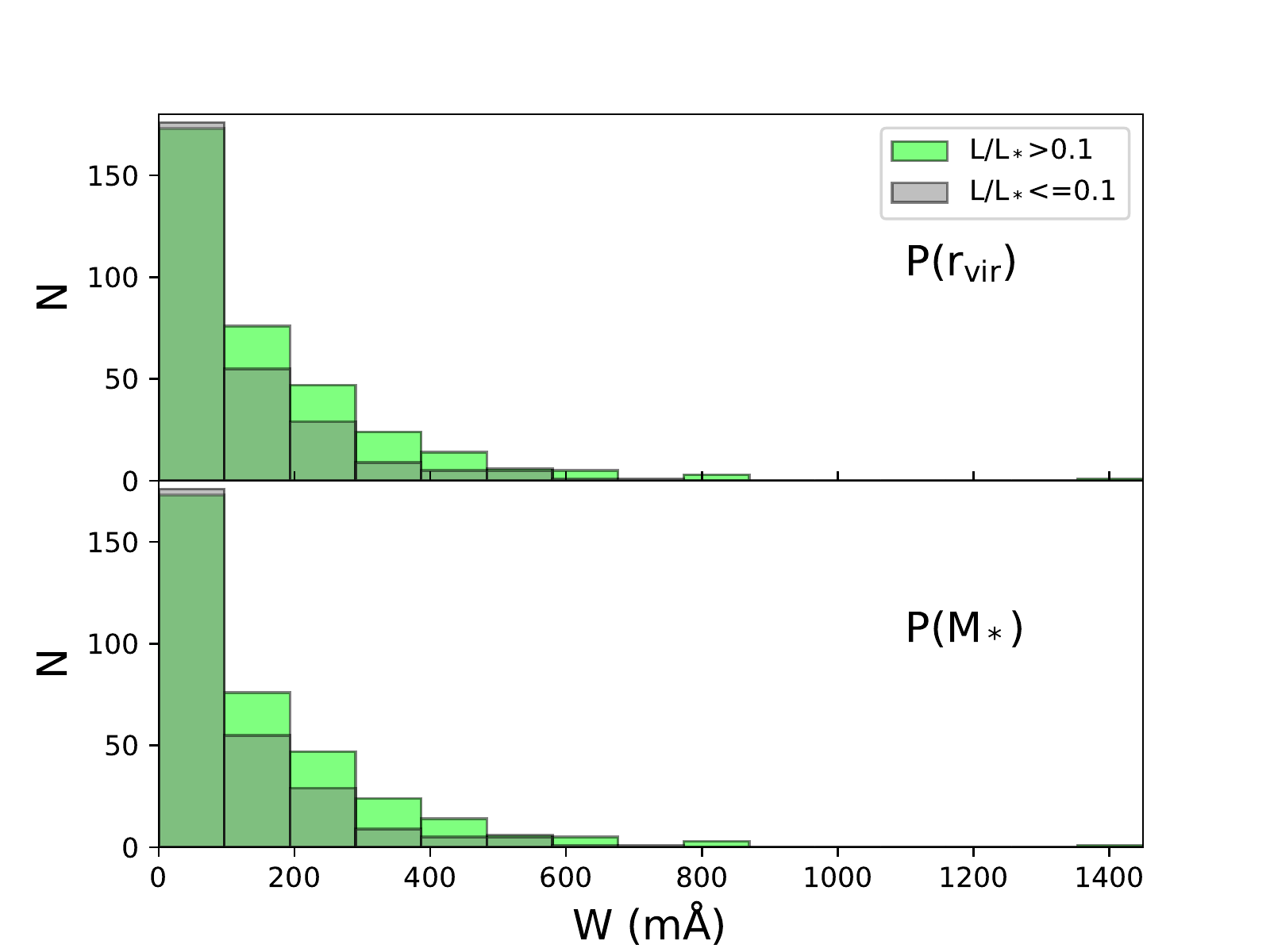}{0.5\textwidth}{(d)}}
\caption{Distributions of galaxy-absorber impact parameter for (a)
red ($u-r> 2.2$) versus blue galaxies and for (b) low- ($L/L^* < 0.1$) versus high luminosity galaxies.
Distributions of absorber rest \lya\ equivalent width for (c)
red ($u-r> 2.2$) versus blue galaxies and for (d) low- ($L/L^* < 0.1$) versus high luminosity galaxies.
}~\label{fig:hitsgalprops}
\end{figure}

In Figure~\ref{fig:hitsgalprops}, we show the distributions in galaxy-absorber impact parameter and \lya\ rest equivalent width for the unique galaxy-absorber pairs
identified by both methods. To investigate trends with galaxy properties, the sample is divided into red and blue, $u-r$ greater or less than 2.2, to approximately delineate 
early and late morphological types
\citep{Strateva2001}. We also compare low and high luminosity galaxies, defined here as less or greater than 0.1$L^*$.
There is a distinct difference in the equivalent width distributions for these two subsamples, 
in the sense that redder and more luminous galaxies show larger 
equivalent width absorption at lower
impact parameter. The median values of $\rho/r_{\rm vir}$ differ by a factor $\sim$1.5 for the red versus blue (2.9 versus 4.5) and
by a factor of $\sim$1.7 for higher versus lower luminosity
(3.1 versus 5.4) subsamples. Table~\ref{tab:ks} lists the results of
Kolmogorov-Smirnov tests, indicating 
that these samples are not drawn from the same parent distribution. 
The most plausible CGM absorbers arising from our analysis are those
pairs with impact parameter within $2r_{\rm vir}$, which are appoximately evenly split between the red and blue subsamples, but dominated by 
$ > 0.1L^*$ galaxies.
\begin{deluxetable}{lrc}
\tablecaption{Galaxy-\lya\ Absorber Pair Distribution
\label{tab:ks}
}
\tablewidth{0pt}
\tablehead{
\colhead{} &\colhead{D$_{\rm max}$}  &\colhead{$P_{\rm KS}$} 
}
\startdata
\multicolumn{3}{l}{Unique galaxy-absorber pairs}\\
\tableline
\multicolumn{3}{l}{ $u-r > 2.2$ vs. $u-r < 2.2$} \\
\tableline
\multicolumn{3}{l}{Virial Radius Method} \\
\tableline
$\rho$ &0.43  &$7.7\times10^{-24}$ \\
$W$    &0.10  &0.12 \\
\tableline
\multicolumn{3}{l}{Stellar Mass Method} \\
\tableline
$\rho$ &0.42 &0. \\
$W$    &0.11 &0.067 \\
\\
\tableline
\multicolumn{3}{l}{$L/L^* < 0.1$ vs. $L/L^* > 0.1$} \\
\tableline
\multicolumn{3}{l}{Virial Radius Method} \\
\tableline
$\rho$ &0.70  &$6.6\times10^{-16}$ \\
$W$    &0.15  &$8.7\times10^{-4}$ \\
\tableline
\multicolumn{3}{l}{Stellar Mass Method} \\
\tableline
$\rho$ &0.69  &$1.8\times10^{-74}$ \\
$W$    &0.12  &0.019 \\
\\
\tableline
\multicolumn{3}{l}{Hits vs. Misses within 500~kpc} \\
\tableline
\multicolumn{3}{l}{Virial Radius Method} \\
\tableline
$\rho$  &0.20  &$1.6\times10^{-8}$ \\
$\rho$\tablenotemark{1}  &0.26  &$8.2\times10^{-5}$ \\
$u-r$   &0.21   &$9.6\times10^{-9}$  \\
$\log{M_*/M_{\sun}}$ &0.21  &$4.4\times10^{-9}$\\
$\phi$  &0.04  &0.65 \\
\tableline
\multicolumn{3}{l}{Stellar Mass Method} \\
\tableline
$\rho$  &0.18  &$5.6\times10^{-7}$\\
$\rho$\tablenotemark{1}  &0.30  &$7.3\times10^{-6}$  \\
$u-r$   &0.20   &$1.9\times10^{-8}$ \\
$\log{M_*/M_{\sun}}$ &0.20  &$1.9\times10^{-8}$\\
$\phi$  &0.05  &0.52 \\
\enddata
\tablenotetext{1}{Metal line systems}
\end{deluxetable}

To test the robustness of these conclusions, we generated 100 realizations of random pairings between the \lya\ absorbers in our final samples of unique pairs
and SDSS galaxies which lie within 500~kpc of the QSOs. The impact parameter distributions for red versus blue and high versus low luminosity subsamples as 
defined above are shifted to smaller impact parameter and less distinctly bimodal than the distributions of hits versus misses, 
with medians of 2.0~kpc(2.4~kpc) for red(blue) and 3.1~kpc(2.0~kpc) for high(low) luminosity. 

\begin{figure}
\gridline{\fig{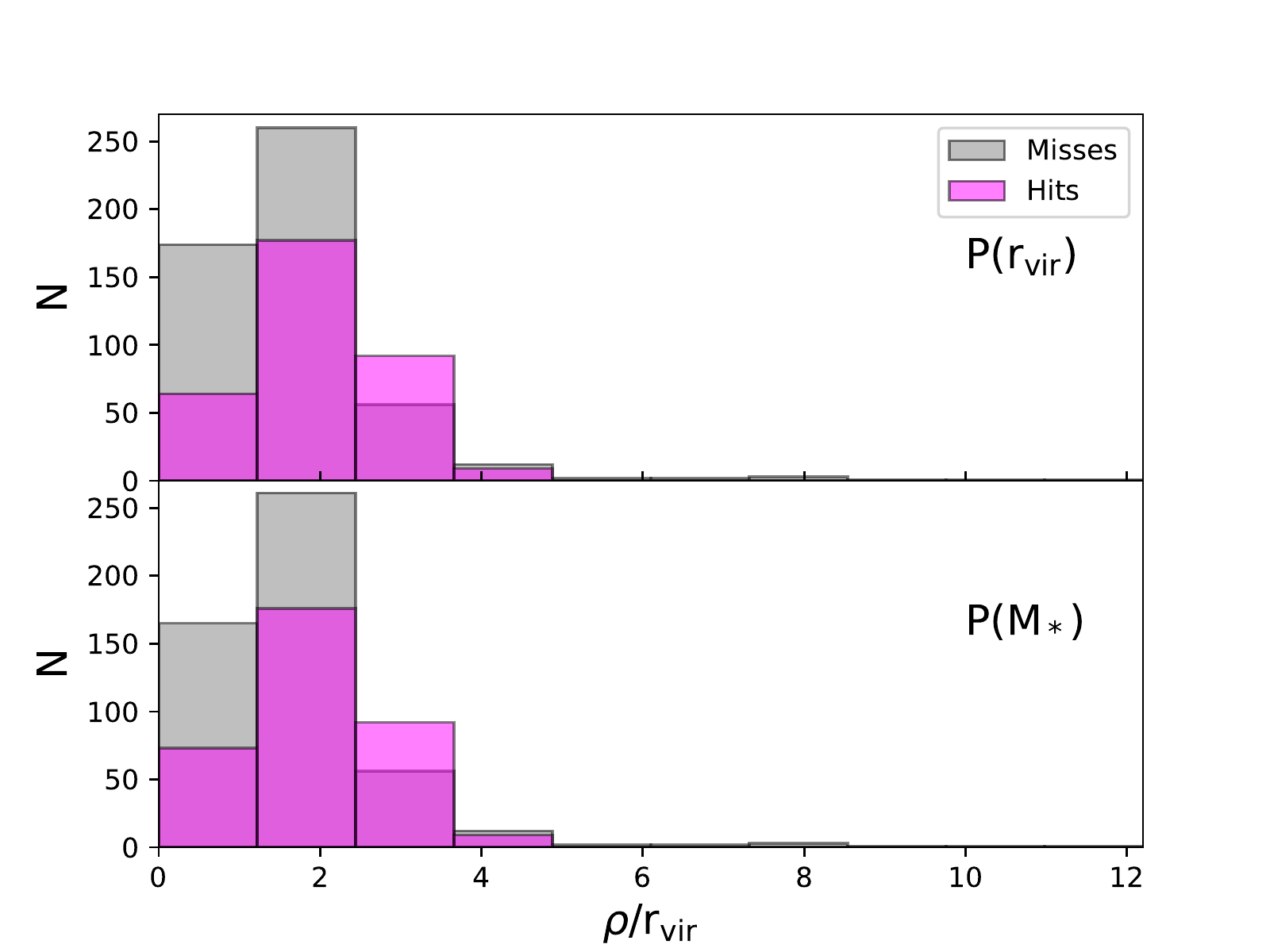}{0.5\textwidth}{(a)}
          \fig{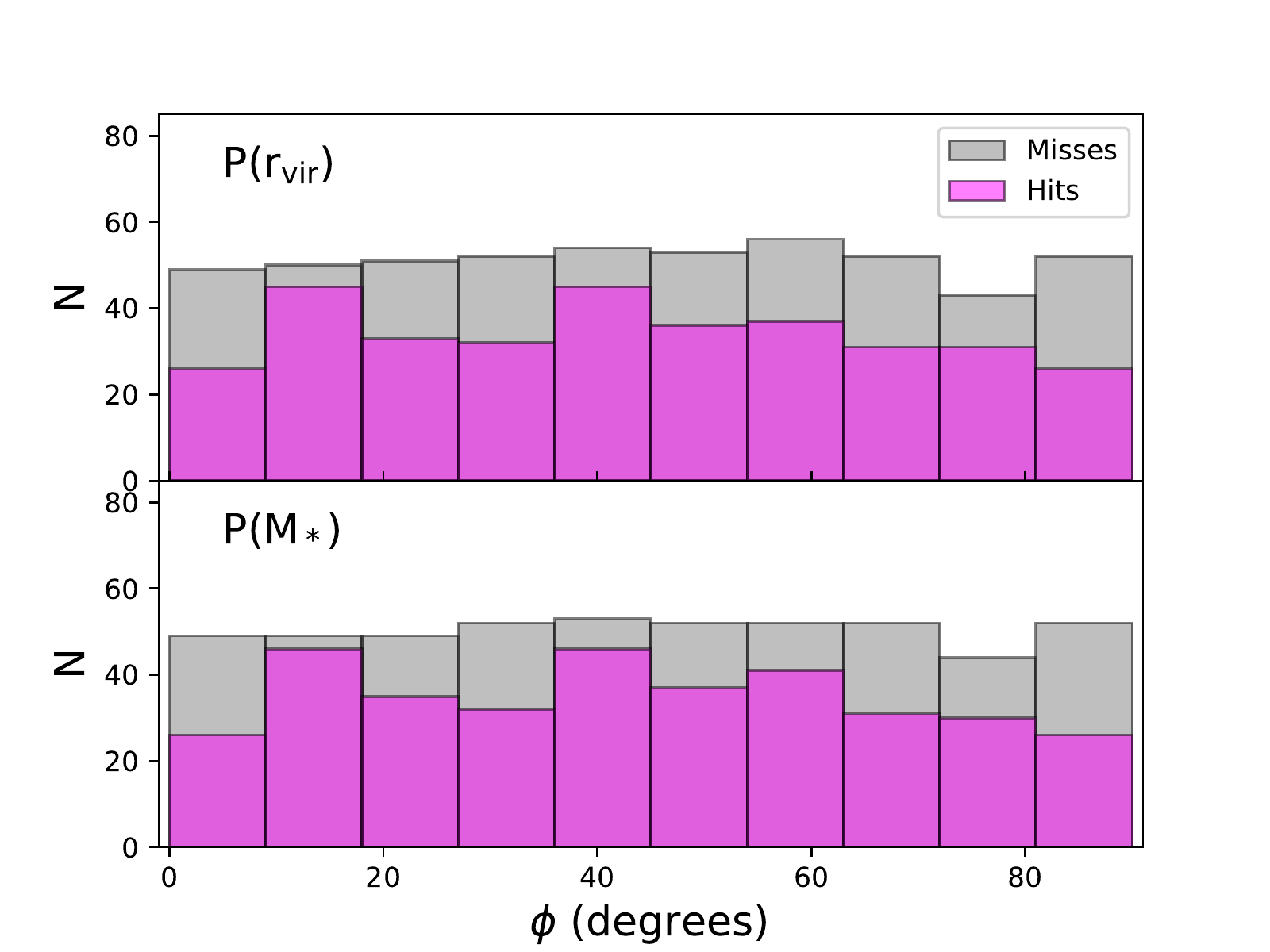}{0.5\textwidth}{(b)}}
\gridline{\fig{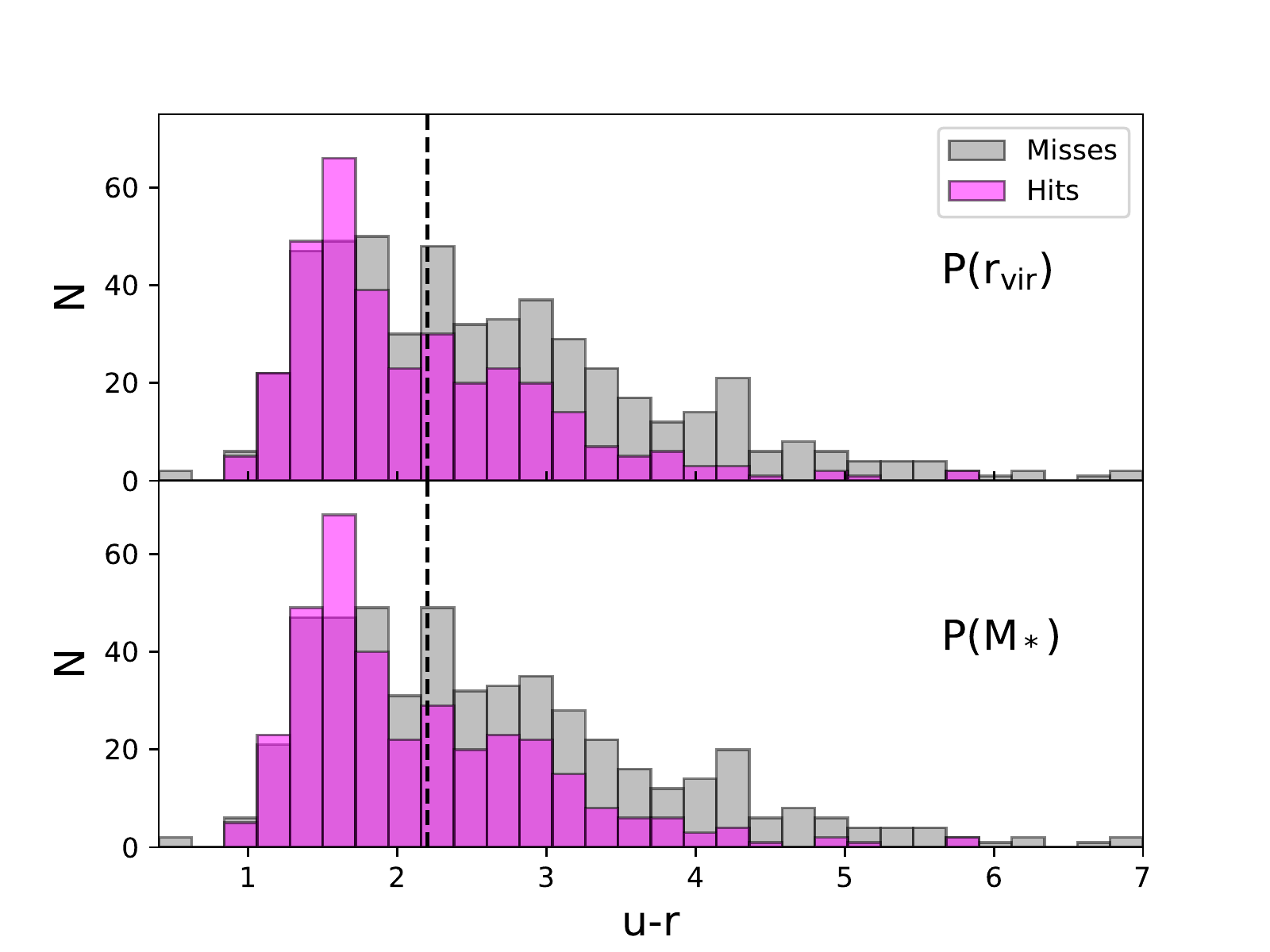}{0.5\textwidth}{(c)}
          \fig{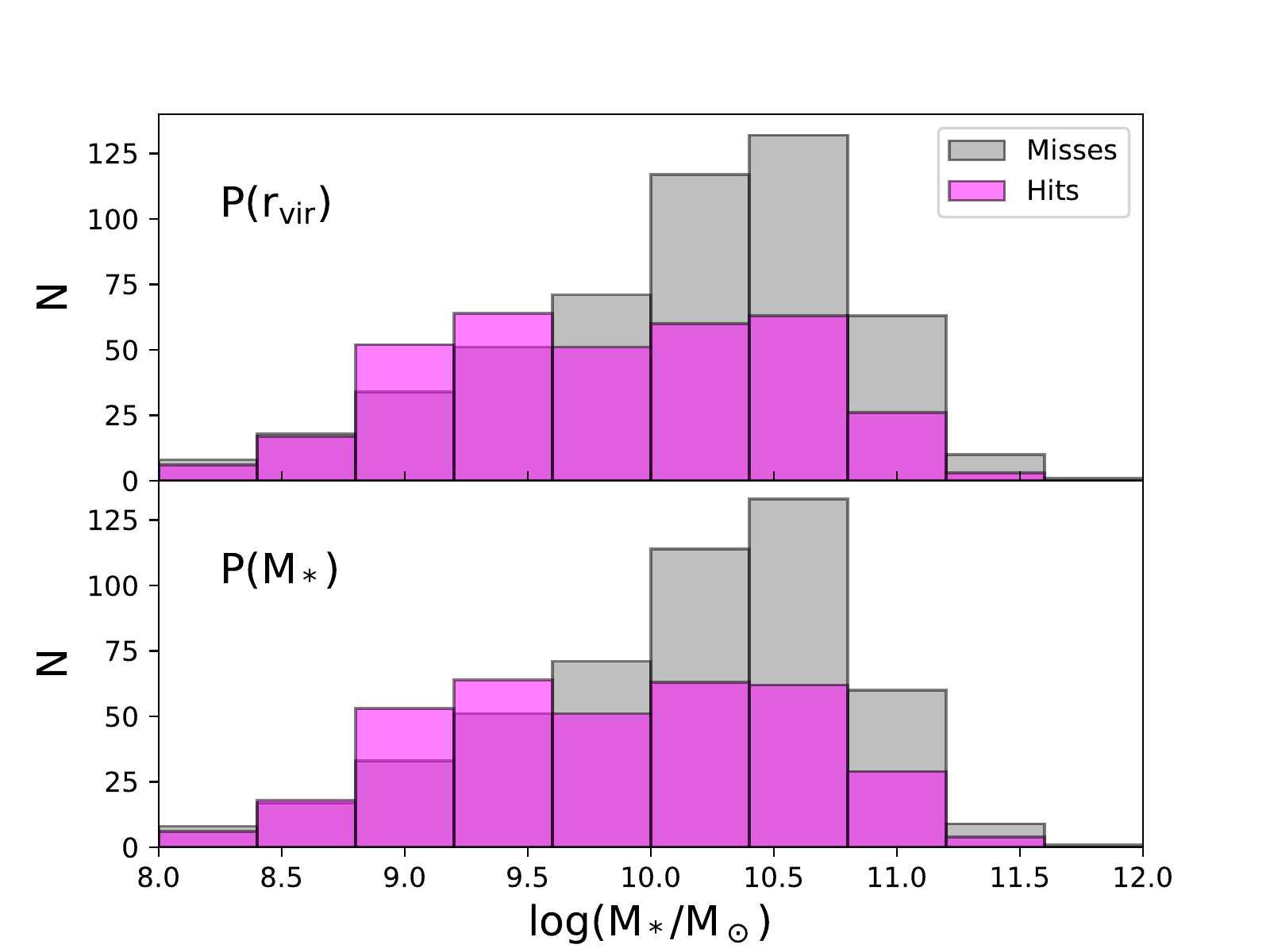}{0.5\textwidth}{(d)}}
\caption{Distributions of galaxies paired with absorbers (``hits") and galaxies not paired with absorbers (``misses")
within 500~kpc of the QSO sightlines: (a) galaxy-absorber impact parameter; (b) galaxy orientation angle; (c) galaxy $u-r$ color, vertical
dashed line at $u-r=2.2$ is the boundary between early and late morphological types in the bimodal SDSS galaxy color distribution \citep{Strateva2001};
(d) galaxy stellar mass.
}~\label{fig:hitsmisses}
\end{figure}

In Figure~\ref{fig:hitsmisses}, we compare the properties of galaxies within 500~kpc of the QSOs that are found to be paired with \lya\ absorbers using both methods, 
with those that are not paired with any absorber.
We label the former ``hits" and the latter ``misses", somewhat similarly to \cite{Stocke2013} who
define a ``miss" to be a super-$L^*$ galaxy with $\rho \le 1$~Mpc that shows no absorption in the QSO spectrum.
However, we consider the entire SDSS photometric sample, and our galaxy-absorber associations, or lack thereof, are less secure than in spectroscopic surveys of galaxies
in QSO fields.
The galaxy properties we explore here include those discussed previously, impact parameter normalized by galaxy virial radius, 
$u-r$ as a proxy for morphological type, and the stellar mass estimated from the $g-i$ color 
\citep{Taylor2011}. We also include here an estimate of the galaxy orientation to the QSO, $\phi$, defined to be between $0^\circ$ for a galaxy with 
its major axis aligned with the
galaxy-QSO direction and $90^\circ$ for one with its minor axis aligned as such.
There is evidence of a difference in the
distributions in $\rho/r_{\rm vir}$, $u-r$, and stellar mass, in the sense that the hits are bluer (median $u-r=$1.8 versus 2.3),
lower in stellar mass (median $\log(M_*/M_\sun)=9.8$ versus 10.2) and found at larger impact
parameter than misses (median $\rho/r_{\rm vir}=$1.9 versus 1.5).
For metal line systems, however, there is a preference for smaller impact parameter among the hits, median $\rho/r_{\rm vir}=$1.2 versus 1.8 for misses.
Results of the K-S tests are listed in  Table~\ref{tab:ks} .

\begin{figure}
\gridline{\fig{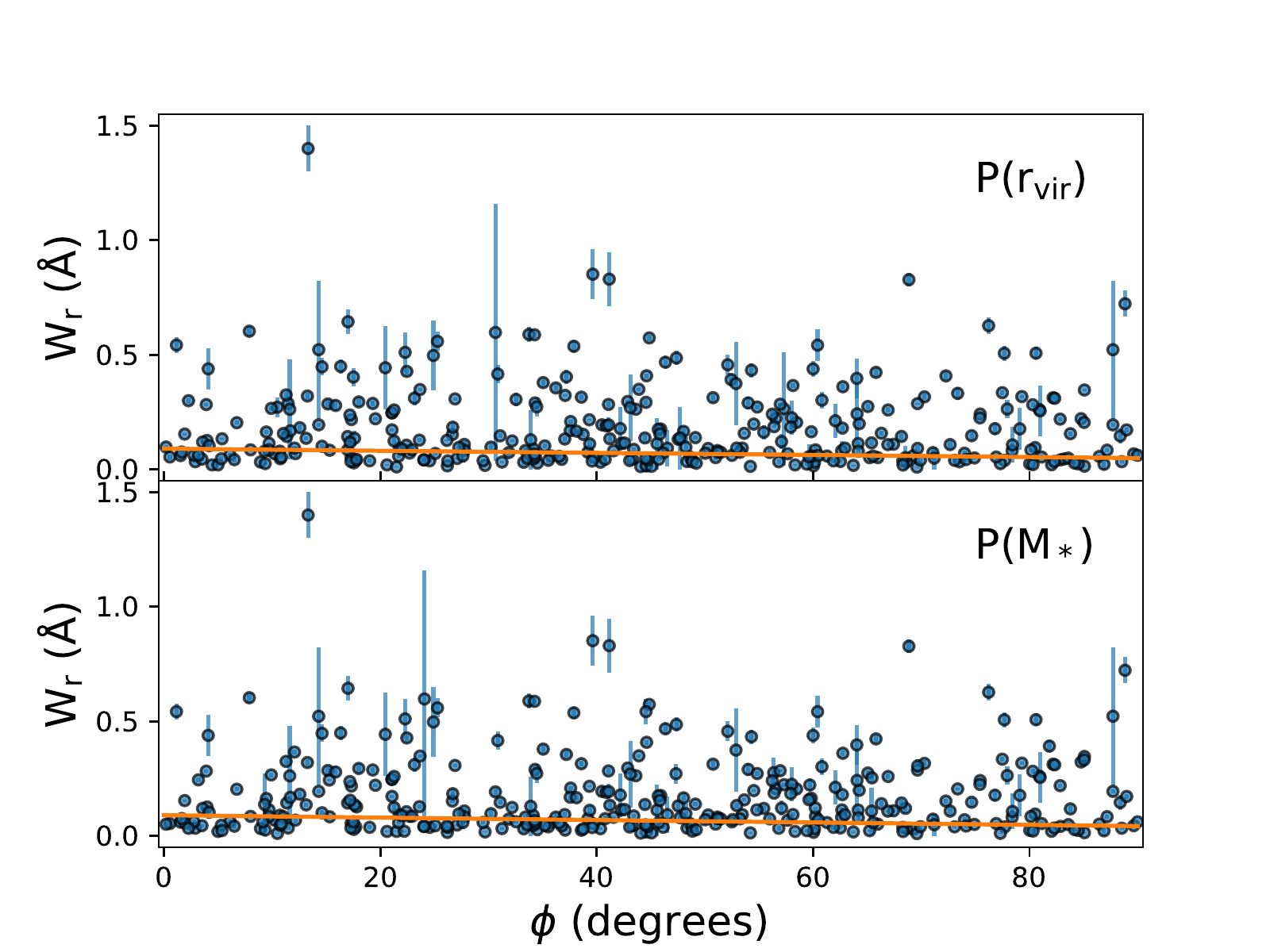}{0.5\textwidth}{(a)}
          \fig{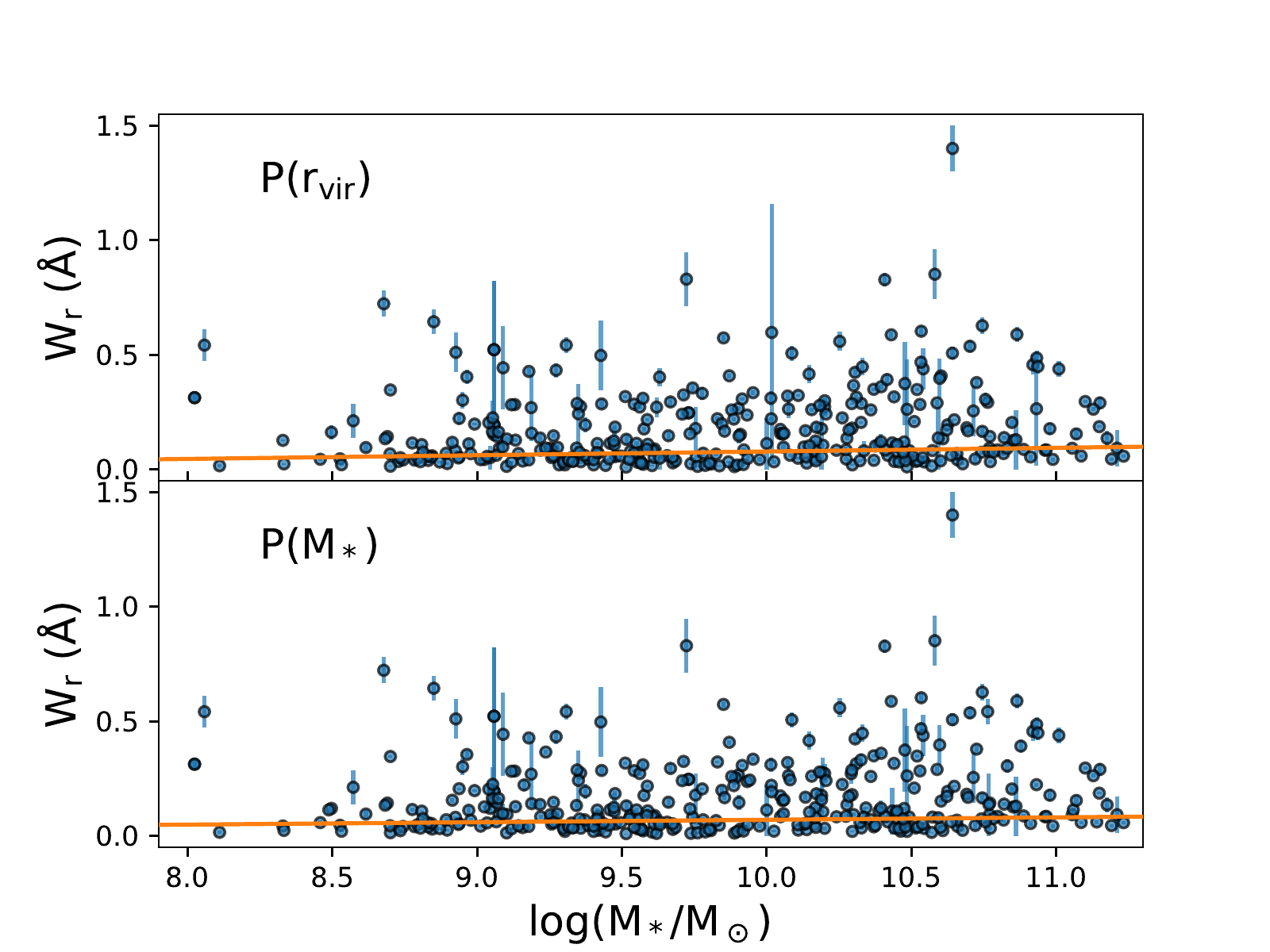}{0.5\textwidth}{(b)}}
\caption{(a) \lya\ rest equivalent width versus galaxy orientation angle ; (b) \lya\ rest equivalent width versus galaxy stellar mass
}~\label{fig:ewcover}
\end{figure}

In Figure~\ref{fig:ewcover} we show absorber \lya\ equivalent width versus galaxy orientation angle and versus galaxy stellar mass.
Pearson tests show no correlation between $W$ and either $\phi$ or $M_*$ at high significance.

\section{Discussion}
This study represents a truly blind survey of the CGM around galaxies by leveraging the extensive photometric data from SDSS in a way that 
targeted spectroscopic studies cannot. The COS-Halos sample \citep{Tumlinson2013} probes the CGM within $\sim$150~kpc of 44 galaxies. \cite{Keeney2018} present 
extensive spectroscopic data for nearly 9000 galaxies in Mpc-scale fields of 47 QSOs, 90\% complete to $L\sim0.1L^*$ for $z<0.1$.
Here, we include over ten times as many galaxies in similarly sized fields around each QSO in the sample and with the photometry alone, can reach similar completeness levels to
$z \sim 0.3$.
Of course, this depth comes at the expense of precise redshift information, which is absolutely critical for establishing
physical correspondences between galaxies and absorbers. In order to compensate, we have devised two Bayesian statistics, which build on the
results of the previous targeted and detailed studies and which in general reliably reproduce their results, within the parameters of our analysis.

The results of our Bayesian analysis show consistency in the sense that they are 
similar using either of the two statistics we developed: the virial radius method based on the empirical anticorrelation between
absorption equivalent width in \lya\ and several metal species and galaxy-absorber impact parameter relative to galaxy virial radius; and the
stellar mass method based on a scaling with galaxy stellar mass instead of virial radius. The specific SDSS 
galaxies paired with each COS absorber are usually the same, although 
specific differences can be found in Tables~\ref{tab:rvir} and \ref{tab:mstar} and
a few cases are outlined in the notes on individual sightlines in 
Appendix~\ref{sec:sightlines}. None of the relationships between the absorbers we identify as candidate CGM absorbers 
with galaxy properties that we investigated differed between the two methods so the general conclusions we draw apply to both methods.

The candidate galaxy-absorber pairs identified through the virial radius method, shown on the $W$ versus $\rho/r_{\rm vir}$ relation in Figure~\ref{fig:rvir} 
cluster at $\rho=3-5r_{\rm vir}$ and \lya\ equivalent width $\sim 100$ m\AA, weaker absorption than found in the literature for the
spectroscopically identified pairs. 
The results of \cite{Keeney2018} suggest that the covering fraction is $\sim$60\% or higher out to $4r_{\rm vir}$ and
\cite{Liang2014} find the \lya\ covering fraction is as high as 60\% out to 500~kpc from the SDSS spectroscopic redshift sample,
although \cite{Stocke2013} associate these ``hits" falling outside galaxy virial radii as IGM rather than CGM.

There is a less pronounced tendency for $\rho > r_{\rm vir}$ among metal line systems, 
consistent with the findings of
\cite{Liang2014} who report a
sharp cutoff in metal line absorption for $\rho>200$~kpc, with
lower ionization species declining more rapidly than higher ions like \ion{C}{4}, consistent with our fits to the literature data tabulated in Table~\ref{tab:fits}.
The candidate galaxy-absorber pairs found using the stellar mass method shown in Figure~\ref{fig:mstar} 
are spread over similar range of $M_*$ found by the studies using spectroscopic galaxy redshifts.

\cite{Lanzetta1995} suggested that 30-60\% of the \lya\ features in QSO spectra could arise in the CGM of luminous galaxies.
By contrast, \cite{Stocke2013} find that only 5\% of QSO sightlines pass within $r_{\rm vir}$ of a galaxy and likely to be
associated with the CGM, if the strict definition is gas within the virial radius.
Of the $\sim$1450 \lya\ absorption lines in the COS sample with significance level greater than four, 
we find that 57-59\% are uniquely paired with galaxies within our probability threshold, although only $\sim$6\%
lie along sightlines
within $r_{\rm vir}$, as shown in Figure~\ref{fig:rvir}. 
The $\sim$30-40\% false positive rate for pairs with $\rho < r_{\rm vir}$ places our
estimate of the percentage of QSO absorbers arising in the CGM even lower for this 
SDSS sample, with a median $L \sim 0.1L^*$.

Using all SDSS galaxies within 500~kpc of the sample QSOs and all the galaxy-absorber pairs in our ranked lists to estimate 
the \lya\ covering fraction from the galaxy ``hits" and ``misses",
we find the $f_c \sim 0.61$ for all SDSS galaxies within $\rho/r_{\rm vir}=3$,
with no distinction between
the lower luminosity and higher luminosity subsamples. This is a crude estimate, however, given the imprecision with which 
photometric redshifts allow us 
to establish velocity alignment between specific absorbers and galaxies.
The estimate at higher impact parameters, $\rho/r_{\rm vir}=3-6$, 
rises to $f_c \sim 0.84$, reflecting the overall tendency of our method to identify
pairs at larger impact parameter.
Among the candidate galaxy-absorber pairs, we 
find that the impact parameter distribution is significantly different for galaxies of  different color and luminosity
in the sense that there is a preference for
absorption at larger impact parameter among bluer, lower luminosity galaxies.
This is consistent with \cite{Keeney2018} who find a shallower drop in the covering fraction for $L < L^*$, 
and higher covering fractions at $\rho=3-4r_{\rm vir}$ for lower luminosity, emission line galaxies.
We also find a difference in the color distribution between galaxies within 500~kpc of QSO sightlines that show absorption (``hits")
and those that do not (``misses"). There is a distinct preference for bluer galaxies
among the hits, again, consistent with a tendency for star forming galaxies to show higher covering fraction.

The similarity of the galaxy orientation angle distributions between hits and misses  and the
lack of any significant correlation between $\phi$ and $W$ are in agreement with previous results, e.g.
\cite{Borthakur2015}, who find that \lya\  equivalent width does not correlate with
orientation of the galaxy disk relative to the quasar sightline indicating a spherical
\ion{H}{1} distribution in the CGM, and \cite{Pointon2019}, who find no dependence of CGM metallicity on orientation angle.

Galaxy hits show a slight preference
for larger impact parameter, which may indicate that we are 
probing group environments or large scale structure \citep{Tejos2014,Burchett2020,Wilde2020}.
However, metal line systems show the opposite preference, 
with hits tending to be found $\sim0.7r_{\rm vir}$ closer to QSO sightlines than misses.
We also find a preference for absorption by smaller $M_*$ galaxies,
although we find no dependence of absorption equivalent width on stellar mass.

\section{Summary}
We present a method for using galaxy photometric redshifts combined with other CGM information in the literature to identify candidate galaxy-absorber pairs
for follow up spectroscopic confirmation. 
To develop this method, we use a sample of 43 HST/COS QSO spectra and the SDSS galaxy photometry in the QSO fields within 3$^\circ$ of the sightline.
The photometric data are $>90$\% complete to $L=0.5L^*$ for $z=0.5$, the maximum redshift we consider.
Our basic results can be summarized as:
\begin{itemize}
\item{We find over  600 unique galaxy-absorber pairs using two different methods with median $\rho \sim$430~kpc, including $\sim$85
pairs with metal line systems with median $\rho \sim$250~kpc.
Up to 75\% of these pairs may be spurious due to misalignments between the photometric redshift estimates of the galaxies and their true redshifts, 
but results may be culled for higher probability associations to establish samples for spectroscopic follow-up in some heretofore unsurveyed QSO fields.}
\item{Of the 47 galaxy-absorber pairs in the literature which are part of our SDSS galaxy sample, we recover $\sim$34 with our two statistics.}
\item{We discuss
over 30 interesting individual potential galaxy-absorber pairs in Appendix~\ref{sec:sightlines}.}
\item{The galaxies identified as candidates for showing CGM absorption in the COS spectra with this method have the following median properties: $z_{\rm phot}=0.13$,
$M_r$=-20.0, $u-r=1.7$, $\log(M_*/M_{\sun}) =9.7$.}
\item{The candidate galaxy-absorber pairs found here show a preference for larger impact parameter, bluer color, and lower stellar mass than non-paired
SDSS galaxies within 500~kpc of the QSO sightlines, but no preference in orientation angle relative to the QSO sightline.}
\item{Among the galaxies paired with absorbers, there is a clear tendency for more luminous, redder galaxies to show absorption at lower impact parameter, within
$4r_{\rm vir}$ and a more
modest tendency for the absorbers in these pairs to show larger equivalent widths.}
\end{itemize}

\begin{acknowledgments}
The authors thank the anonymous referee for helpful comments that improved the manuscript.
This work was supported by National Science Foundation grant AST-0952923, and made extensive use of data from the Sloan Digital Sky Survey, SDSS-III.
Funding for SDSS-III has been provided by the Alfred P. Sloan Foundation, the Participating Institutions, the National Science Foundation, and the U.S. Department of Energy Office of Science. The SDSS-III web site is 
http://www.sdss3.org/.
SDSS-III is managed by the Astrophysical Research Consortium for the Participating Institutions of the SDSS-III Collaboration including the University of Arizona, the Brazilian Participation Group, 
Brookhaven National Laboratory, Carnegie Mellon University, University of Florida, the French Participation Group, the German Participation Group, Harvard University, the Instituto de Astrofisica de Canarias, the 
Michigan State/Notre Dame/JINA Participation Group, 
Johns Hopkins University, Lawrence Berkeley National Laboratory, Max Planck Institute for Astrophysics, Max Planck Institute for Extraterrestrial Physics, 
New Mexico State University, New York University, Ohio State University, Pennsylvania State University, University of Portsmouth, Princeton University, the 
Spanish Participation Group, University of Tokyo, University of Utah, Vanderbilt University, University of Virginia, University of Washington, and Yale University.
This work also used High Level Science Products provided by the Barbara A. Mikulski Archive for Space Telescopes under a Creative Commons Attribution license (CC BY 4.0);
and the NASA/IPAC Extragalactic Database (NED),
which is operated by the Jet Propulsion Laboratory, California Institute of Technology,
under contract with the National Aeronautics and Space Administration.
\end{acknowledgments}

\appendix
\section{Notes on Individual QSO Sightlines}
\label{sec:sightlines}
\subsection{1ES 1028+511} 
\label{sec:1028}
\cite{Stocke2013} found absorbing
galaxies UGC~5740 and SDSS~J103108.88+504708.7 with $(z_{\rm gal}, \rho)=(0.002, 90 $~kpc) and $(0.003, 25 $~kpc),
respectively. 
UGC~5740 is also matched by \cite{Stocke2013} to an absorber in the spectrum of 1SAX~J1032.3+5051.
These galaxies are not in our galaxy catalog due to their very low redshifts and the photometric redshift error cut we applied.
Instead, our formalism combines the $z_{\rm abs}=0.0024$ and $z_{\rm abs}=0.0032$ \lya\ lines into one system and associates them with 
either
SDSS~J103541.06+500601.4, $(z_{\rm phot}, \rho)=(0.021, 255 $~kpc) or
 SDSS~J103001.38+505047.5 
$(z_{\rm phot}, \rho)=(0.021, 50 $~kpc). 
The SDSS spectroscopic redshift measurements of these galaxies are $z_{\rm gal}=0.030$ and $z_{\rm gal}=0.044$, respectively.
Thus the pairings reported by \cite{Stocke2013} are more likely.

As tabulated in Table~\ref{tab:lit}, \cite{Liang2014} associate a \lya\ absorption feature with a
$z = 0.137274$ galaxy SDSS~J103110.35+505211.0 at $\rho = 278.6$~kpc from the QSO sightline.
We recover this pair as the top galaxy match for this absorber, which we associate with additional \lya\ and \ion{O}{6} components.
However, our uniqueness criteria pair this galaxy with a different absorber, listed in Tables~\ref{tab:rvir} and \ref{tab:mstar}, 
for which it is also the top match: a \lya\ line with
$z_{\rm abs}=0.1406$.

Of the other potential galaxy-absorber pairs listed in these tables, we find
a match of a $z_{\rm abs}=0.051$ absorber with SDSS~J103157.08+505753.2, $(z_{\rm phot}, \rho)=(0.047, 447$~kpc).
This galaxy is also matched, with similar single ion probabilities, to an absorption system in the spectrum of 1SAX~J1032.3+5051. It is flagged as
a nearby galaxy for both QSOs by \cite{Keeney2018}.

\subsection{1ES1553+113} 
We report 
27(28)
potential galaxy-absorber matches in Table~\ref{tab:rvir}(\ref{tab:mstar}), including 
one \ion{O}{6} absorber $z_{\rm abs}=0.395$ and galaxy with SDSS~J123952.34+104515.1, $(z_{\rm phot}, \rho)=(0.41, 189 $~kpc). 

\subsection{1SAX J1032.3+5051}
\cite{Stocke2013} reported an absorbing galaxy, UGC~5740, very close to the sightline at $\rho = 65$~kpc, with $z_{gal} = 0.00216$ \kms.
UGC~5740 is also matched by \cite{Stocke2013} to an absorber in the spectrum of 1ES~1028+511. As reported above, UGC~5740 is not in our SDSS galaxy catalog.

\begin{figure}
\centering
\includegraphics[width=0.6\textwidth, viewport=0 0 1000 1000]{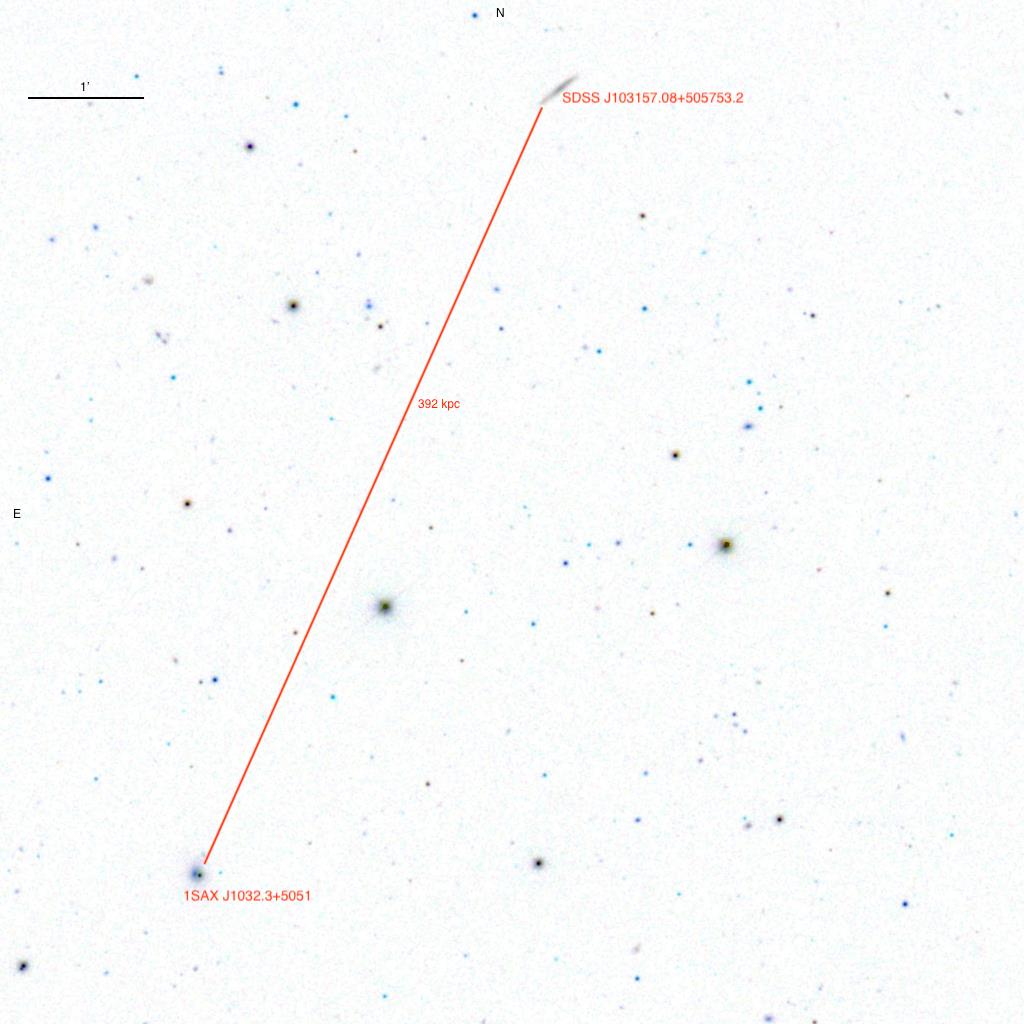} 
\caption{Galaxy SDSS~J103157.08+505753.2, $(z_{\rm phot}, \rho)=(0.047, 392 $~kpc), paired with $z_{\rm abs}=0.045$ in the spectrum of 1SAX~J1032.3+5051.
}~\label{fig:sj1032}
\end{figure}

Instead, we find 
six
other potential galaxy-absorber pairs, including a multicomponent absorption system (two \lya\ components and a \ion{Si}{3} line)
paired with SDSS~J103157.08+505753.2, the galaxy also paired with an absorber in the spectrum of 1ES~1028+511, discussed above. 
It lies at $\rho=392$~kpc from this sightline.
The SDSS image is shown in Figure~\ref{fig:sj1032}. This galaxy is flagged by \cite{Keeney2018} as a nearby galaxy,
but not the closest known,
to this absorber.

\subsection{3C 263} \cite{Liang2014} report that the galaxy SDSS~J114015.59+655551.7 ($\rho=456$~kpc) is not matched with any absorber in the QSO spectrum. 
We do find a three-component $z_{\rm abs}=0.046$ 
\lya\ absorber paired to this galaxy. Its spectroscopic redshift, $z_{\rm gal}=0.04588$, is in even better agreement with $z_{\rm abs}$ than the 
photometric estimate, 
$z_{\rm phot}=0.050$. 

A $z_{\rm abs}=0.063$ system with three \lya\ components and \ion{Si}{2}, two \ion{Si}{3}, \ion{Si}{4}, \ion{C}{2} and two \ion{C}{4} components
is matched with SDSS~J114005.17+654801.3 with $\rho=63$~kpc and $z_{\rm phot}=0.093$ and which is
flagged by \cite{Keeney2018} as the closest known galaxy to this absorber.
The stellar mass method instead pairs this galaxy with a different multicomponent metal line system with $z_{\rm abs}=0.113$, at $\rho=107$~kpc.

We find 23 other possible galaxy-absorber pairs for this QSO sightline from both methods,
including a 
$z_{\rm abs}=0.11$ system with \lya, \ion{O}{6}, \ion{C}{4}, and \ion{Si}{3} components paired with 
SDSS~J113956.96+654459.6 ($z_{\rm phot}, \rho=0.11, 353 $~kpc) by the virial radius method
and with SDSS~J114005.17+654801.3 ($z_{\rm phot}, \rho=0.093, 107 $~kpc) by the stellar mass method. Both galaxies
are flagged by \cite{Keeney2018} as being the closest known galaxies to absorbers.

\subsection{3C 273}  
This sightline passes through the Virgo cluster.
The galaxy HI~1225+01 lies at $\rho = 126$~kpc from the QSO sightline.
\cite{Bowen1996} associate it with $z_{\rm abs}=0.004$ \lya\ absorption in the HST Faint Object Spectrograph data and \cite{Penton2002} note its 
coincidence with $z_{\rm abs}=0.00337$ and $z_{\rm abs}=0.00527$ features in the spectrum taken by the Goddard High Resolution Spectrograph.
This galaxy is not in our SDSS sample due to its low redshift and the photometric redshift error cut we applied.

The following absorption systems are matched to galaxies by various authors:

\begin{itemize}
\item{$z_{\rm abs}=0.0034$:
The galaxies in this region may reside in a group environment \citep{Stocke2013}.
\cite{Bowen1996}, \cite{Penton2002}, and \cite{Stocke2013} note the coincidence of this absorber to SDSS~J123103.89+014034.4, with  
$z_{\rm gal}=0.00368$, at 
$\rho=118$~kpc from the line of sight.
This galaxy is not in our SDSS sample due to the photometric redshift error cut.
\cite{Liang2014} associate this absorber with SDSS~J122815.96+014944.1, $(z_{\rm gal}, \rho)=(0.00303, 69 $~kpc).
We do recover this match with both methods. However, because the photometric redshift for this galaxy, $z_{\rm phot}=0.013$, is not in good agreement with the
spectroscopic value, we find 
$P(r_{\rm vir})=P(M_*)=0.13$, and its unique match is to a different \lya\ absorber: 
$z_{\rm abs}=0.020$, $\rho=472$~kpc, with $P(r_{\rm vir}), P(M_*)=$0.26, 0.29.
Our statistics match the $z_{\rm abs}=0.0034$ absorber instead to SDSS~J122806.95+032056.2, $(z_{\rm phot}, \rho)=(0.012, 377$~kpc), with a probability
close to our 10\% threshold value.
Given the spectroscopic redshift information available for SDSS~J122815.96+014944.1 and used in the analysis of 
\cite{Liang2014}, we consider its association
with the $z_{\rm abs}=0.0034$ absorber to be most plausible.
}

\item{$z_{\rm abs}=0.0053$:
Absorption at this redshift is linked to two different galaxies in this likely group environment by \cite{Stocke2013}, NGC~4420 and SDSS~J122950.57+020153.7.
The former is not in our SDSS catalog due to our magnitude cut. The latter is assigned a photometric redshift of $0.036$, so it is not among the top ten probable
galaxy matches to this absorber using either of our two statistics.
Our two methods find no unique galaxy matches to this absorber.
}

\item{
$z_{\rm abs}=0.0072, 0.0075$:
\cite{Penton2002} and \cite{WakkerSavage2009} match the $z_{\rm abs}=0.0072$ absorber to UGC~7625, at a fairly large impact parameter to the line of sight, 
771 kpc, and 
also to absorption in the Q1230+0115 sightline, with $\rho=339$~kpc. 
This galaxy is not in our SDSS catalog due to our photometric redshift error cut.
\cite{WakkerSavage2009} match the $z_{\rm abs}=0.0075$ absorber to SDSS~J122815.88+024202.9 $(z_{\rm gal}, \rho)=(0.00762, 429 $~kpc).
In the methodology presented here, the two absorbers are combined into one system, and, since UGC~7625 is absent from our catalog,
SDSS~J122815.88+024202.9 is the top match to the absorber
for both methods, even though the photometric redshift of the galaxy, $z_{\rm phot}=0.019$, 
is in poor agreement with $z_{\rm abs}$ and with the spectroscopic value reported by SDSS, $z_{\rm phot}=0.0074$.
}
\end{itemize}

We report 25 other possible galaxy-absorber pairs along this sightline with both methods,
including an \ion{O}{6} absorber with $z_{\rm abs}=0.12$ paired with SDSS~J122910.05+020120.1, ($(z_{\rm phot}, \rho)=(0.125, 24 $~kpc) by both methods.

\subsection{FBQS 1010+3003} 
\cite{Stocke2013} report an association of a $z_{\rm abs}=0.00462$ absorber with the galaxy UGC~5478 $(z_{\rm gal}, \rho)=(0.00459, 48 $~kpc). 
This galaxy is not in our SDSS catalog due to the photometric redshift error cut.
Our best unique match to this absorber, SDSS~J101042.80+293136.7, $(z_{\rm phot}, \rho)=(0.019, 191 $~kpc), has a rather poor redshift agreement
and so  
we would not choose this over UGC~5478 as the most likely galaxy associated with this system.

\cite{Liang2014} find that the $z_{\rm abs}=0.0875$ absorber is associated with SDSS~J101008.85+300252.5 $(z_{\rm gal}, \rho)=(0.0874, 181 $~kpc). 
Because the SDSS photometric redshift estimate for this galaxy is $0.111$, it is the
7$^{\rm th}$ best match for this absorber with our methods.
Both methods find that the galaxy SDSS~J100950.99+300703.0, $(z_{\rm phot}, \rho)=(0.091, 392 $~kpc), is the best match for this absorber.
The galaxy has no spectroscopically measured redshift.

We report 18(17) other unique galaxy-absorber pairs for this sightline from the virial radius(stellar mass) method.
The highest probability pairs are:
$z_{\rm abs}=0.082$ and SDSS~J100940.28+300640.9, $(z_{\rm phot}, \rho)=(0.083, 490 $~kpc); 
$z_{\rm abs}=0.155$  and SDSS~J101006.14+300613.8, $(z_{\rm phot}, \rho)=(0.15, 479 $~kpc);
$z_{\rm abs}=0.25$ and SDSS~J101003.92+300141.6, $(z_{\rm phot}, \rho)=(0.25, 491 $~kpc).
However, the relatively large impact parameters and the SDSS spectroscopic redshift for SDSS~J100940.28+300640.9,
$z_{\rm gal}=0.094$, throw these matches into some question.

\subsection{HS 1102+3441} 
We find 19 possible pairs with the virial radius method, 18 with the stellar mass method.
Of these paired absorbers, five are metal line systems:
\begin{itemize}
\item{$z_{\rm abs}=0.201$, matched with SDSS~J110534.38+342403.5, $(z_{\rm phot}, \rho)=(0.20, 377 $~kpc);}
\item{$z_{\rm abs}=0.205$, matched with SDSS~J110546.32+342721.2, $(z_{\rm phot}, \rho)=(0.25, 457 $~kpc);}
\item{$z_{\rm abs}=0.238$, matched with SDSS~J110539.49+342653.8, $(z_{\rm phot}, \rho)=(0.22, 303 $~kpc);}
\item{$z_{\rm abs}=0.289$, matched with SDSS J110540.90+342514.9, $(z_{\rm phot}, \rho)=(0.26, 104 $~kpc); and}
\item{$z_{\rm abs}=0.408$, matched with SDSS~J110537.33+342535.9, $(z_{\rm phot}, \rho)=(0.38, 168 $~kpc).}
\end{itemize}
None of these galaxies has a spectroscopically confirmed redshift and plausible associations would require better agreement with the
absorber redshifts in most cases. 
None of the galaxy pairings we find for the absorbers in this sightline overlaps with the spectroscopic sample of \cite{Keeney2018}.

\subsection{MRK 106} 
There are two galaxy-absorber pairs for this sightline reported in the literature. \cite{WakkerSavage2009} match the $z_{\rm abs}=0.00811$ absorber with 
SDSS~J091923.29+553137.2 and 
\cite{Liang2014} pair the $z_{\rm abs}=0.0317$ absorber with UGC~4800.
Neither of these galaxies are in our SDSS galaxy catalog due to the photometric redshift error 
cut. Also, UGC~4800 lies at a projected distance of 1030 kpc, larger than the
maximum of 500 kpc that we imposed.
We report 9(7) possible galaxy-absorber pairs for this sightline from  the virial radius(stellar mass) method.
The highest probability matches are to two \lya\ absorbers: 
$z_{\rm abs}=0.008$  paired with SDSS J092000.79+554635.2, $(z_{\rm phot}, \rho)=(0.011, 248 $~kpc); and
$z_{\rm abs}=0.038$ matched to SDSS~J092030.67+553102.2, $(z_{\rm phot}, \rho)=(0.032, 487 $~kpc).
The latter is
is unlikely given the galaxy's SDSS spectroscopic redshift, $z_{\rm gal}=0.046$.

\subsection{MRK 1513} 
We find only three(four) unique galaxy-absorber pairs in this field with the virial radius(stellar mass) method, all with $\rho > 400$ kpc.

\subsection{MRK 478} \cite{WakkerSavage2009} associate an absorber with $z_{\rm abs}=0.00525$  with the galaxy NGC~5727 ($\rho=645$~kpc). The photometric redshift
estimate for this galaxy, $0.198$, is in poor agreement with the spectroscopic value, $z_{\rm gal}=0.00497$, and its impact parameter is larger than our
500 kpc upper limit,
so we do not find this galaxy in our top ten
matches to this absorber. 
Both methods find the top match to this absorber is SDSS~J143800.57+360124.2, $(z_{\rm phot}, \rho)=(0.018, 401 $~kpc), and identify two other
possible galaxy-absorber pairs.

\subsection{PG 0003+158} \cite{Liang2014} report a galaxy-absorber pair for $z_{\rm abs}=0.0909$ and SDSS~J000556.15+160804.1 ($\rho=193$~kpc).
This galaxy is identified as the highest probability unique match to the five-component absorber (three \lya, two \ion{C}{4}) with both methods.

\begin{figure}
\centering
\includegraphics[width=0.4\textwidth, viewport=0 0 450 450]{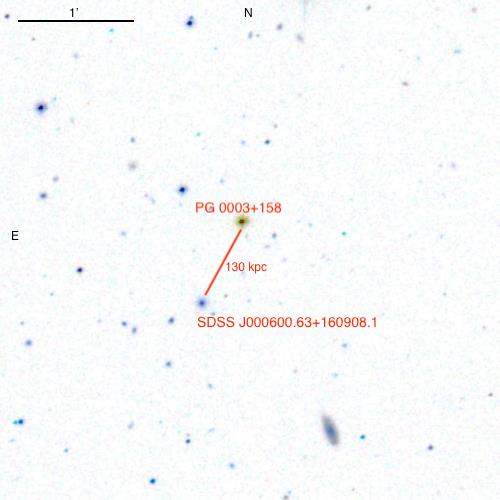}
\caption{Galaxy SDSS~J000600.63+160908.1, $(z_{\rm phot}, \rho)=(0.16, 130 $~kpc), paired with $z_{\rm abs}=0.165$ 
in the PG~0003+158 spectrum.
}~\label{fig:pg0003}
\end{figure}

The virial radius(stellar mass) method reveals 22(25) other possible galaxy-absorber pairs, including
a $z_{\rm abs}=0.165$ system with \lya, \ion{O}{6}, and \ion{C}{3},
paired with SDSS~J000600.63+160908.1, $(z_{\rm phot}, \rho)=(0.16, 130$~kpc). 
The spectroscopic redshift is in excellent agreement, $z_{\rm gal}=0.165$, strenghening the conclusion of a physical association.
This galaxy is shown in Figure~\ref{fig:pg0003}.

We also pair a five-component $z_{\rm abs}=0.347$ system showing three \lya, two \ion{O}{6}, \ion{C}{3}, and \ion{Si}{2} components with
the galaxy SDSS J000601.78+160837.7, $(z_{\rm phot}, \rho)=(0.34, 398$~kpc).

\subsection{PG 0026+129} 
\cite{Bowen1997} match a $z_{\rm abs}=0.0391$ absorber to the galaxy WISEA~J002915.37+132056.5 ($\rho=111$~kpc).
Because $z_{\rm phot}=0.028$, this galaxy is identified as the second best match for this absorber with our methods.
Instead, this galaxy is paired with a $z_{\rm abs}=0.033$ \lya\ absorber after uniqueness is imposed.
The top galaxy match to the $z_{\rm abs}=0.0391$ absorber in both methods is SDSS~J002843.85+131421.4, $(z_{\rm phot}, \rho)=(0.031, 348 $~kpc).
The virial radius(stellar mass) method identifies five(four) additional possible galaxy-absorber pairs listed in Tables~\ref{tab:rvir} and \ref{tab:mstar}.

\subsection{PG 0157+001} This sightline has no previously reported galaxy-absorber pairs.
Both methods find 12 potential pairs, including
a multicomponent absorber with two \lya\ components and one \ion{Si}{3} with $z_{\rm abs}=0.146$ and the galaxy SDSS~J015946.54+002320.5,
$(z_{\rm phot}, \rho)=(0.158, 153 $~kpc). 
However, the spectroscopic redshift of this galaxy,
$z_{\rm gal}=0.161$, suggests that if the absorber is circumgalactic in origin, it likely arises from another galaxy.

\subsection{PG 0832+251} 
\begin{figure}
\centering
\includegraphics[width=0.4\textwidth, viewport=0 0 450 450]{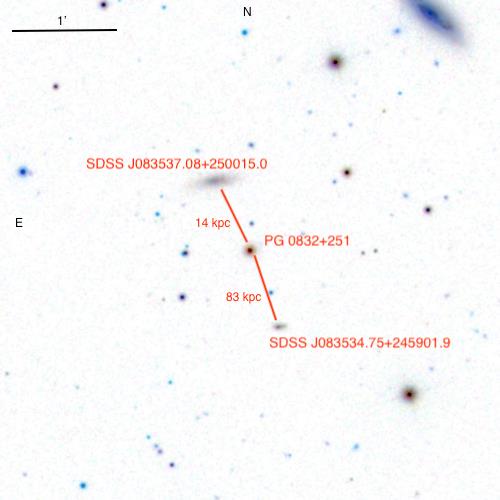}
\caption{Galaxy SDSS~J083537.08+250015.0, $(z_{\rm phot}, \rho)=(0.031, 14$~kpc), paired with $z_{\rm abs}=0.0174$ in the spectrum of PG~0832+251; and
Galaxy SDSS~J083534.75+245901.9, $(z_{\rm phot}, \rho)=(0.098, 83$~kpc), paired with $z_{\rm abs}=0.108$.
}~\label{fig:pg0832}
\end{figure}

\cite{Liang2014} identify a $z_{\rm abs}=0.0073$ system with SDSS~J083335.65+250847.1 ($\rho=263$~kpc). The SDSS photometric redshift cut removes 
this galaxy from our catalog.
\cite{Stocke2013} find an association between $z_{\rm abs}=0.0174$  and NGC~2611, at $\rho=53$~kpc from the QSO line of sight.
We recover this galaxy as the second match for the virial radius method and the top match for the stellar mass method, even though its
SDSS photometric redshift estimate, $z_{\rm phot}=0.026$, is not in good agreement with the absorber redshift.
Interestingly, our top match from the virial radius method yielded a different galaxy, SDSS~J083537.08+250015.0, with $(z_{\rm phot}, \rho)=(0.031, 14 $~kpc). The 
spectroscopic redshift, $z_{\rm gal}=0.017$, places 
it in even better agreement with $z_{\rm abs}$. Its even closer proximity to the QSO sightline merits 
follow-up investigation. This pair is shown in Figure~\ref{fig:pg0832}.

\cite{Liang2014} pair a $z_{\rm abs}=0.0233$ absorber with SDSS~J083607.41+250645.7 ($\rho=283$~kpc), and \cite{Keeney2018} flag this galaxy as the closest known galaxy to the absorber. 
The SDSS photometric redshift of this galaxy is in good agreement with the spectroscopic value and with $z_{\rm abs}$.
Both methods 
find this to be the second best galaxy match. Instead, the top galaxy match is SDSS~J083645.44+250652.0, $(z_{\rm phot}, \rho)=(0.021, 492 $~kpc).
However, the galaxy's spectroscopic redshift, $z_{\rm gal}=0.026$, rules it out. Our methods instead pair SDSS~J083607.41+250645.7 with a double \lya\
component system with $z_{\rm abs}=0.028$ and $\rho=343$~kpc. 
Given the spectroscopic redshift, 
we favor the galaxy-absorber pair for $z_{\rm abs}=0.023$ found by \cite{Liang2014}.

Our methods identify 16 additional galaxy-absorber pairs with the two methods.
We find two \lya\ lines with $z_{\rm abs}=0.101$ paired with SDSS~J083531.17+245548.7, $(z_{\rm phot}, \rho)=(0.10, 453 $~kpc) and
an \ion{O}{6} absorber with $z_{\rm abs}=0.233$ paired with SDSS~J083535.75+250032.9, $(z_{\rm phot}, \rho)=(0.25, 194 $~kpc).
However, \cite{Keeney2018} find no absorber within 1000 \kms\ of these galaxies.

Finally a \lya\ absorber with $z_{\rm abs}=0.108$ is matched to
SDSS~J083534.75+245901.9 with $(z_{\rm phot}, \rho)=(0.098, 83$~kpc). 
The galaxy is flagged as the closest to this absorber by \cite{Keeney2018}.
This pair is also shown in Figure~\ref{fig:pg0832}. 

\subsection{PG 0844+349} \cite{WakkerSavage2009} find associations between absorbers with $z_{\rm abs}=0.00137$ and $z_{\rm abs}=0.00769$ with NGC~2683 and UGC~4621, respectively.
These two galaxies are too bright to be in our SDSS galaxy catalog. 
The virial radius(stellar mass) method yields 3(2) candidate galaxy-absorber pairs with probability greater than our 10\% threshold.

\subsection{PG 0953+414} 
\cite{WakkerSavage2009} pair a $z_{\rm abs}=0.00204$ absorber with NGC~3104 a galaxy with the same redshift and an impact parameter of 296 kpc from the sightline.
This galaxy is too bright to be part of our SDSS galaxy sample. Instead, our methods pair this absorber with SDSS~J094758.66+412057.7 $(z_{\rm phot}, \rho)=(0.015, 296 $~kpc).

\cite{Stocke2013} and \cite{Liang2014} identify a pairing of $z_{\rm abs}=0.142$  with SDSS~J095638.90+411646.0, at $\rho=435$~kpc.
This galaxy's photometric redshift estimate is $z_{\rm phot}=0.131$, and without the reliable spectroscopic redshift information, 
it is the seventh best match for the absorber in our virial radius method. It is the third best in the stellar mass method, and 
our uniqueness criteria do recover this galaxy-absorber pair in Table~\ref{tab:mstar}.
Instead, the galaxy uniquely matched to this absorber by the virial radius method is
SDSS~J095706.09+411707.8, $(z_{\rm phot}, \rho)=(0.16, 473 $~kpc).

We report 16(18) other possible galaxy-absorber pairs from the virial radius(stellar mass) method, 
including a $z_{\rm abs}=0.068$ \ion{C}{4} system matched to SDSS~J095640.12+411107.6 with $(z_{\rm phot}, \rho)=(0.083, 380
$~kpc). However, this galaxy's spectroscopic redshift is reported by SDSS to be $z_{\rm gal}=0.041$.

\subsection{PG 1001+291} 
This sightline has several galaxy-absorber pairs reported in the literature.
The following are not reproduced by our methods because the galaxy is excluded from our SDSS galaxy catalog for the reasons listed in in Table~\ref{tab:lit}:
UGC~5427 and $z_{\rm abs}=0.00165$  \citep{WakkerSavage2009};
SDSS~J100618.16+285641.9 and  $z_{\rm abs}=0.0036$ \citep{Bowen1997, Liang2014};
SDSS~J100402.36+285512.5 and $z_{\rm abs}=0.137$; and 
TON0028:[KSS94]~39 and $z_{\rm abs}=0.214$ \citep{Mathes2014}.

Our methods match the $z_{\rm abs}=0.00165$ absorber with SDSS~J100534.47+273051.7 $(z_{\rm phot}, \rho)=(0.014, 180 $~kpc), but this is ruled out by its
spectroscopic redshift, $z_{\rm gal}=0.0212$.

Two other galaxies paired with absorbers in this sightline by \cite{WakkerSavage2009} are included in our catalog: UGC~5464 and UGC~5461, 
paired with $z_{\rm abs}=0.00357$ and $z_{\rm abs}=0.0153$,
respectively. This is a different galaxy match to the $z_{\rm abs}=0.00357$ absorber mentioned above.
Neither of the these galaxies are recovered as unique matches to the absorberss
with our methods due to the inaccuracies of the photometric redshifts, estimated to be $0.0236$ and $0.035$. Additionally,
UGC~5461 lies at an impact parameter larger than our 500 kpc limit.
Instead our methods match $z_{\rm abs}=0.00357$ with SDSS~J100223.17+294333.3, $(z_{\rm phot}, \rho)=(0.014, 234 $~kpc), which is 
plausible given its reported spectroscopic redshift, $z_{\rm gal}=0.00274$.
The $z_{\rm abs}=0.0153$ absorber is matched with SDSS~J100427.09+284043.9 $(z_{\rm phot}, \rho)=(0.020, 298 $~kpc) by both methods,
but the spectroscopic redshift is in good agreement with the photometric estimate, which is actually in rather poor agreement with the absorber redshift, so
this match is unlikely.

Both methods do find that SDSS~J100403.24+285650.2 is the top match to $z_{\rm abs}=0.134$, $(z_{\rm phot}, \rho)=(0.143, 181 $~kpc), as reported by \cite{Liang2014}. 
Because the photometric redshift is larger than the spectroscopically measured value, 
the probability of this galaxy matching with $z_{\rm abs}=0.137$ ($\rho=185 $~kpc) is found to be higher by both methods.
This absorption system consists of two \lya\ 
and one \ion{O}{6} components. However, given the spectroscopic information, we judge that this galaxy is more likely to be physically associated with $z_{\rm abs}=0.134$.

The methods find 11 additional potential pairs,
including a double \lya\ system with $z_{\rm abs}=0.165$ and the galaxy SDSS~J100352.49+285508.9,
$(z_{\rm phot}, \rho)=(0.162, 385$~kpc).

\subsection{PG 1048+342} \cite{Liang2014} pair a $z_{\rm abs}=0.0593$ absorber with SDSS~J105111.41+335935.6, at $\rho=465$~kpc.
The photometric redshift is in good agreement with the spectroscopic value and this pair is recovered with both methods.

\begin{figure}
\centering
\includegraphics[width=\textwidth, viewport=0 0 2200 1200]{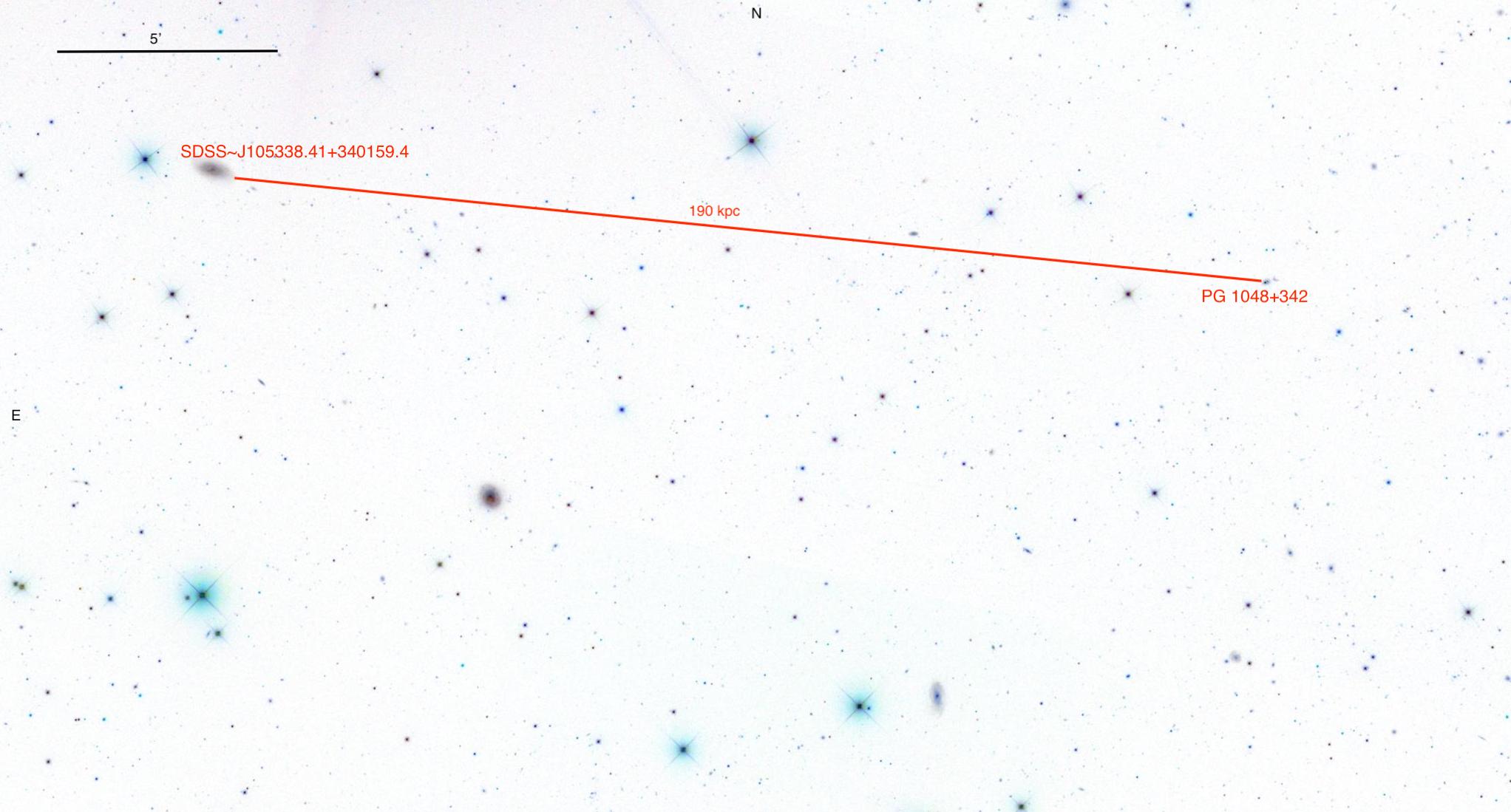}
\caption{Galaxy SDSS~J105338.41+340159.4, $(z_{\rm phot}, \rho)=(0.011, 190 $~kpc), paired with $z_{\rm abs}=0.005$ absorber in the spectrum of PG~1048+342.
}~\label{fig:pg1048}
\end{figure}

Both methods result in seven additional matches, including a multicomponent $z_{\rm abs}=0.005$ absorber with 
three \lya\ and two \ion{Si}{4} components matched with SDSS~J105338.41+340159.4,
$(z_{\rm phot}, \rho)=(0.011,, 190 $~kpc). The spectroscopic redshift, $z_{\rm gal}=0.006$, strengthens the case. This pair is shown in Figure~\ref{fig:pg1048}.

\subsection{PG 1049-005} We find no previously reported specific galaxy-absorber pairs for this sightline.
Our virial radius and stellar mass methods identify 17 and 16 possible pairs respectively.
Two paired absorption systems show \lya\ and \ion{Si}{3} components:
$z_{\rm abs}=0.038$ \citep{Richter2016} is matched with SDSS~J105147.77-005020.3, with $(z_{\rm phot}, \rho)=(0.040, 61 $~kpc) and spectroscopic redshift $z_{\rm gal}=0.038$;
and $z_{\rm abs}=0.171$ is matched with SDSS~J105147.68-005203.4, $(z_{\rm phot}, \rho)=(0.17, 214 $~kpc).

\subsection{PG 1115+407} 
\begin{figure}
\centering
\includegraphics[width=0.4\textwidth, viewport=0 0 600 600]{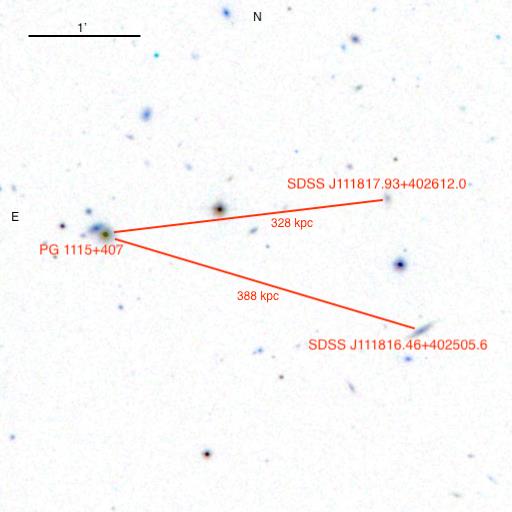}
\caption{Galaxy
SDSS~J111817.93+402612.0, $(z_{\rm phot}, \rho)=(0.12, 328$~kpc), paired with a $z_{\rm abs}=0.12$ \lya\ system; 
and Galaxy SDSS~J111816.46+402505.6, $(z_{\rm phot}, \rho)=(0.13, 388 $~kpc), paired with a $z_{\rm abs}=0.13$ \ion{O}{6} absorber in the spectrum of PG1115+407.
}~\label{fig:pg1115}
\end{figure}

Our methods find 11 possible galaxy-absorber pairs along this sightline.
A two \lya\ and two \ion{O}{6} component system with $z_{\rm abs}=0.12$ is paired with SDSS~J111814.91+402343.3, $(z_{\rm phot}, \rho)=(0.11, 487 $~kpc)  with both methods.
Five other galaxy-absorber pairs found by our methods are flagged by \cite{Keeney2018} as the closest galaxy to an absorber, including
the pairing of a $z_{\rm abs}=0.12$ \lya\ system with SDSS~J111817.93+402612.0, $(z_{\rm phot}, \rho)=(0.12, 328$~kpc) and
the pairing of a $z_{\rm abs}=0.13$ \ion{O}{6} system with SDSS~J111816.46+402505.6, $(z_{\rm phot}, \rho)=(0.13, 388$~kpc).
Both are shown in Figure~\ref{fig:pg1115}.

\subsection{PG 1116+215} This sightline has a plethora of possible galaxy-absorber pairs presented in the literature, several of which we do not recover:

\begin{itemize} 

\item{$z_{\rm abs}= 0.00493$, UGC~6258 \citep{WakkerSavage2009}: We do not recover this pair with either method primarily due to the fact that its projected distance, 543~kpc, is
larger than our threshold of 500~kpc. Instead, our methods match this absorber with SDSS~J111349.11+213101.3, $(z_{\rm phot}, \rho)=(0.010, 465 $~kpc).
}

\item{$z_{\rm abs}=0.0163$, NGC~3649 \citep{WakkerSavage2009}: This galaxy is excluded from our SDSS galaxy catalog due to the photometric redshift error cut.
Instead, we find a match with SDSS~J111844.37+213351.0, with $(z_{\rm phot}, \rho)=(0.017, 313 $~kpc).
The spectroscopic redshift, $z_{\rm gal}=0.0212$, calls this into question, however, so NGC~3649 is a more likely association.
}

\item{$z_{\rm abs}=0.0195$, SDSS~J112045.94+211115.3 \citep{Tripp1998}: We do not recover this pair with either method primarily due to the fact that its projected distance, 557~kpc, is 
larger than our threshold of 500~kpc. Our methods find no unique match for this absorber, i.e. the top galaxy match satisfying $P > 0.10$ is assigned to other absorbers.
In fact, the algorithms find the top match for this absorber to be SDSS~J111844.37+213351.0, and its spectroscopic redshift, $z_{\rm gal}=0.021$, is in better agreement with this absorber than with 
the $z_{\rm abs}=0.0163$ absorber it was uniquely matched with based on its $z_{\rm phot}$, which equals 0.018.
}

\item{$z_{\rm abs}=0.0322$, SDSS~J111843.28+212723.0 \citep{Liang2014}: Both
methods recover this galaxy-absorber pair as the top galaxy match for this absorber. However, after the uniqueness criteria are applied, the galaxy paired with this absorber is
SDSS~J111833.63+211300.4, $(z_{\rm phot}, \rho)=(0.039, 400 $~kpc) and SDSS~J111843.28+212723.0 is instead paired with $z_{\rm abs}=0.041$ ($\rho=495$~kpc). Since the SDSS spectroscopic 
redshifts of SDSS~J111843.28+212723.0 and SDSS~J111833.63+211300.4, are $z_{\rm gal}=0.0323$ and $z_{\rm gal}=0.0601$, respectively, we consider the match reported by
\citep{Liang2014} to be the most plausible.
}

\item{$z_{\rm abs}=0.0412$, SDSS~J111909.56+210243.4 \citep{Tripp1998} : We do not recover this pair with either method primarily due to the fact that its projected distance, 746~kpc, is
larger than our threshold of 500~kpc.
The galaxy found to be the most probable match to this absorber by both methods is SDSS~J111843.28+212723.0, uniquely identified with $z_{\rm abs}=0.0322$ as noted above. 
}

\item{$z_{\rm abs}=0.0590$, SDSS~J111924.26+211029.9 \citep{Tripp1998}: We do not recover this pair with either method primarily due to the fact that its projected distance, 601~kpc, is
larger than our threshold of 500~kpc. 
Instead, our top match for this absorber is SDSS~J111848.70+211441.4, $(z_{\rm phot}, \rho)=(0.058, 450 $~kpc).
This galaxy is flagged by \cite{Keeney2018} as the closest to an absorber but not this absorber, as the galaxy's spectroscopic redshift is reported to be $z_{\rm gal}=0.041$. Thus, it is more likely 
a match to the $z_{\rm abs}=0.0412$ system discussed above.
}

\item{$z_{\rm abs}=0.0590$, SDSS~J111905.34+211537.7, SDSS~J111905.51+211733.0  \citep{Stocke2013}: 
SDSS~J111905.34+211537.7 is the 7$^{\rm th}$(6$^{\rm th}$) best match to this absorber with the virial radius(stellar mass) method.
SDSS~J111905.51+211733.0 is the 10$^{\rm th}$ best match for the stellar mass method and not in the top ten highest probability galaxy matches with the virial radius method.
Both methods pair this absorber instead with
SDSS~J111848.70+211441.4, $(z_{\rm phot}, \rho)=(0.058, 450 $~kpc), as noted above. 
}

\item{$z_{\rm abs}=0.0608$, SDSS~J111942.04+212610.3 \citep{Tripp1998}: We do not recover this pair with either method primarily due to the fact that its projected distance, 677~kpc, is
larger than our threshold of 500~kpc.
}

\item{$z_{\rm abs}=0.138$, SDSS~J111906.68+211828.7 \citep{Liang2014}: This is a multicomponent absorber with \ion{Si}{2}, \ion{Si}{3}, \ion{Si}{4}, \ion{C}{2}, \ion{C}{4}, and \lya.
The galaxy's photometric redshift estimate, $z_{\rm phot}=0.134$, is in reasonable agreement with the spectroscopic redshift so this match is recovered with both methods. 
}
\end{itemize}

\subsection{PG 1121+422} \cite{Liang2014} report two galaxies associated with absorbers in this QSO sightline:
SDSS~J112418.74+420323.1 and   $z_{\rm abs}=0.0245$, with an impact parameter of 123 kpc, and
SDSS~J112457.15+420550.8 and $z_{\rm abs}=0.0338$ with $\rho=213$~kpc.
Our methods both recover the former as the unique galaxy match.
The methods find the latter as the second best galaxy match for the absorber since $z_{\rm phot}=0.042$. The top match for this absorber is also
SDSS~J112418.74+420323.1, and so we report no unique galaxy match for $z_{\rm abs}=0.0338$  with $P>0.10$.

The virial radius(stellar mass) method finds 12(9) other possible galaxy-absorber pairs for this line of sight. 
We find a match between a $z_{\rm abs}=0.192$ metal line system with \ion{C}{2}, \ion{Si}{2}, \ion{C}{3}, \ion{Si}{3} and \ion{O}{6} components and 
the galaxy SDSS~J112433.82+420152.3, $(z_{\rm phot}, \rho)=(0.175, 194$~kpc).

\subsection{PG 1216+069} 
This Virgo cluster sightline has several galaxy-absorber associations reported in the literature, primarily from \cite{Bowen1996} using {\it HST}/Faint Object Spectrograph data.
These authors associate one strong $z_{\rm abs}=0.00550$ absorber with eleven different possible galaxies. In the COS spectrum, this system shows 
four components, \lya, two \ion{Si}{2} lines,
and one \ion{C}{2} line, with $z_{\rm abs}=0.0061-0.0063$:
\begin{itemize}
\item{VCC 381, $z_{\rm gal}=0.00160$: This galaxy is excluded from our SDSS galaxy catalog due to the photometric redshift error cut.
}
\item{VCC 538, $z_{\rm gal}=0.00167$: The SDSS photometric redshift estimate for this galaxy is 0.384, so it is not in the top ten matches 
to this absorber for either method.
}
\item{IC 3115, $z_{\rm gal}=0.00244$: This galaxy is excluded from our SDSS galaxy catalog due to our magnitude cut of $r < 14$.
}
\item{VCC 446, $z_{\rm gal}=0.00283$:  This galaxy, with $z_{\rm phot}=0.0128$, is the 
9$^{\rm th}$
best match to the $z_{\rm abs}=0.0063$ absorber for the virial radius method but is not in the top ten matches with the stellar mass method.
}

\item{UGC 7423, $z_{\rm gal}=0.00419$: The SDSS photometric redshift estimate for this galaxy is 0.0243, and it is not in the top ten matches 
to this absorber for either method.
}

\item{VCC 340, $z_{\rm gal}=0.00504$: This galaxy, with $z_{\rm phot}=0.0116$, 
is the 10$^{\rm th}$ best match to the absorber for the virial radius method but is not in the top ten matches with the stellar mass method.
}

\item{VCC 329, $z_{\rm gal}=0.00541$: The SDSS photometric redshift estimate for this galaxy is 0.0913, and it is not in the top ten matches 
to this absorber for either method.
}

\item{NGC 4260, $z_{\rm gal}=0.00653$:  This galaxy is excluded from our SDSS galaxy catalog due to our magnitude cut of $r < 14$.
}

\item{VCC 223, $z_{\rm gal}=0.00690$: The SDSS photometric redshift for this galaxy is estimated at 0.0159. Thus, it is not in the top ten 
matches 
to this absorber 
with either method.
}

\item{NGC 4241, $z_{\rm gal}=0.00745$:  This galaxy is excluded from our SDSS galaxy catalog due to our magnitude cut of $r < 14$.
}

\item{VCC 415, $z_{\rm gal}=0.00854$: This galaxy is excluded from our SDSS galaxy catalog due to the photometric redshift error cut.
}

\end{itemize}

\begin{figure}
\centering
\includegraphics[width=\textwidth, viewport=0 0 2050 1120]{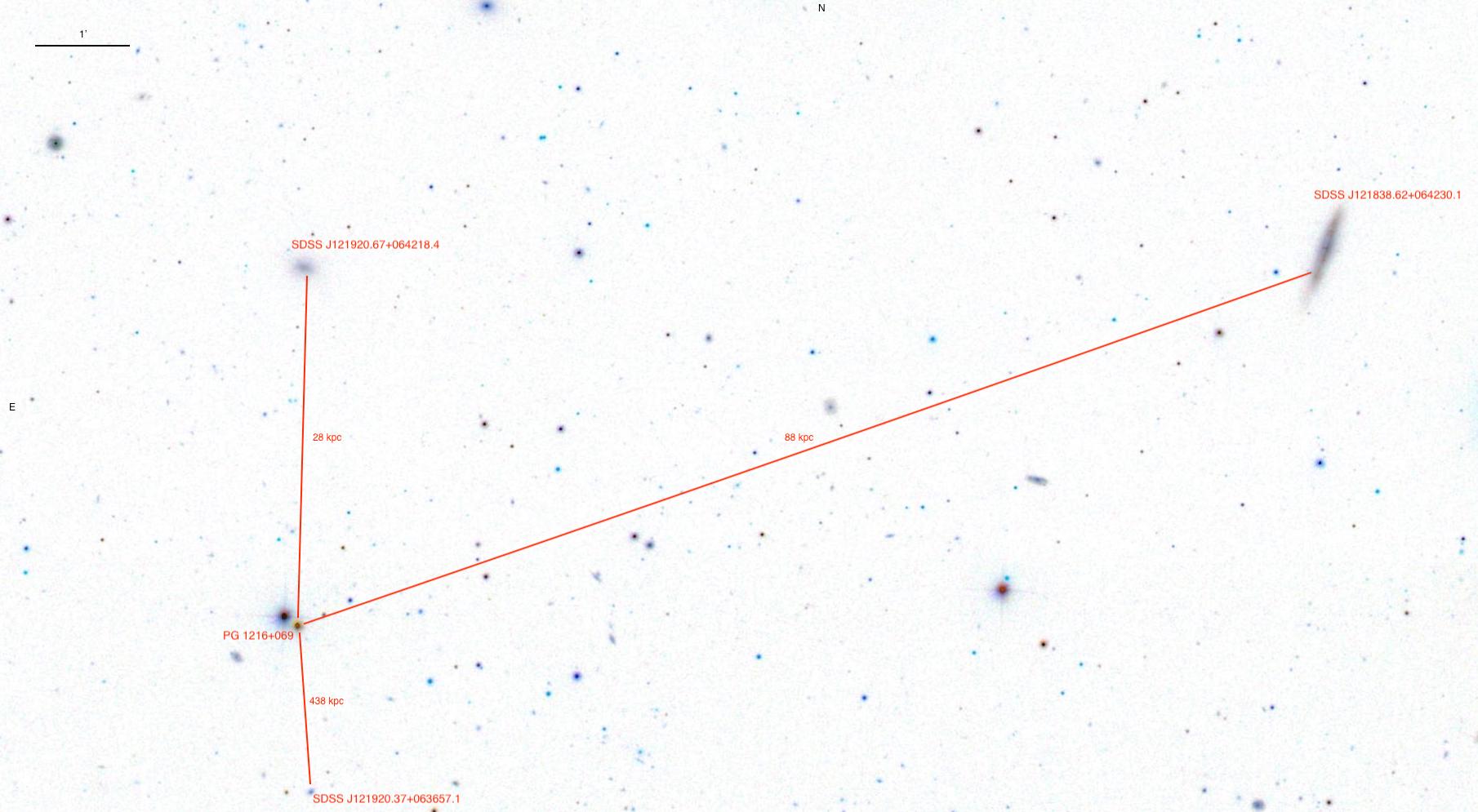}
\caption{
Galaxies SDSS~J121920.67+064218.4, $(z_{\rm phot}, \rho)= (0.03, 28$~kpc) and
SDSS~J121838.62+064230.1, $(z_{\rm phot}, \rho)=(0.019, 88$~kpc), paired with
$z_{\rm abs}=0.006$, and 
SDSS~J121920.37+063657.1, $(z_{\rm phot}, \rho) = (0.30, 438 $~kpc)
paired with $z_{\rm abs}=0.282$
in the spectrum of PG~1216+069.
}~\label{fig:pg1216}
\end{figure}

For this absorber, our virial radius method identifies the highest probability unique match to be 
SDSS~J121920.67+064218.4, $(z_{\rm phot}, \rho)=(0.03, 28 $~kpc). This is also the fourth top match for the 
stellar mass method. 
The spectroscopic redshift of this galaxy is $z_{\rm gal}=0.00741$, within $\sim 400$ \kms\ of the absorber. The very small impact parameter is consistent with the
large, 2.8\AA\ \lya\ equivalent width. 
The second top match from the stellar mass method is SDSS~J121838.62+064230.1, $(z_{\rm phot}, \rho)=(0.019, 88$~kpc). 
It is not reported in Table~\ref{tab:mstar} because $P_{\rm ion}< 0.10$ for the \lya\ component.
(Similarly for the top match to this absorber,
SDSS~J121912.37+062253.6, $(z_{\rm phot}, \rho)=(0.016, 125$~kpc), but $z_{\rm gal}=0.00943$.)
Nevertheless, the spectroscopic redshift of SDSS~J121838.62+064230.1, $z_{\rm gal}=0.00669$,
is in even better agreement with the absorber redshift than SDSS~J121920.67+064218.4. These galaxies are both shown in Figure~\ref{fig:pg1216}.

Other galaxy-absorber pairs reported along this sightline are:

\begin{itemize}
\item{$z_{\rm abs}=0.0784-0.0805$ and SDSS~J121930.87+064334.4 \citep{Bowen1996, Stocke2013}: This galaxy is excluded from our SDSS galaxy catalog due to 
the photometric redshift error cut.}

\item{$z_{\rm abs}=0.1240$ and SDSS~J121923.43+063819.7 \citep{Mathes2014}: This absorption system consists of 13 components, showing multicomponent \lya, \ion{O}{6}, \ion{C}{4}, and
\ion{Si}{3}. This 
galaxy, with $(z_{\rm phot}, \rho)=(0.1318, 94 $~kpc),
is the 
6$^{\rm th}$ 
highest overall probability match to this absorber in the virial radius
method but it is the top match for the stellar mass method. After uniqueness criteria are applied, however, it is the galaxy reported for virial radius method in 
Tables~\ref{tab:rvir}. 
The galaxy's spectroscopic redshift is reported by \cite{Keeney2018} to be $z_{\rm gal}=0.12365$ 
and these authors flag it as the closest galaxy to this absorber.}
\end{itemize}

Our two methods identify 14 other possible galaxy-absorber pairs along this rich sightline. One of particular interest is
a $z_{\rm abs}=0.282$ system with \lya, \ion{O}{6}, \ion{Si}{3}, and \ion{C}{3} components
and SDSS~J121920.37+063657.1, $(z_{\rm phot}, \rho)=(0.30, 438 $~kpc), a unique match reported in Table~\ref{tab:rvir}.
This system is also shown in 
Figure~\ref{fig:pg1216}. 
The stellar mass method uniquely pairs this galaxy with a $z_{\rm abs}=0.267$ \lya\ absorber.
This galaxy is not in the spectroscopic sample of \cite{Keeney2018}. 

\subsection{PG 1222+216} 
Our methods find possible galaxy matches to several multicomponent systems:

\begin{itemize}
\item{$z_{\rm abs}=0.0544$: This absorber shows \ion{Si}{3}, \ion{Si}{4}, \ion{C}{4} in addition to \lya. Both methods pair it with
SDSS~J122445.32+212401.1, with $(z_{\rm phot}, \rho)=
(0.064, 158 $~kpc). 
The galaxy is not in the spectroscopic sample of \cite{Keeney2018}. 
}

\item{$z_{\rm abs}=0.0987$: This \ion{Si}{4}, \ion{C}{4} and \lya\ system is 
is matched with
SDSS~J122456.68+212312.2, $(z_{\rm phot}, \rho)=
(0.091, 74 $~kpc). 
\cite{Keeney2018} report a spectroscopic redshift in good agreement with $z_{\rm phot}$ for this galaxy and it is flagged as the closest to this absorber.
This pair is shown in Figure~\ref{fig:pg1222}.}

\item{$z_{\rm abs}=0.222$: This system shows both \ion{C}{3} and \ion{O}{6} along with two \lya\ components. Both methods choose SDSS~J122457.88+212426.1 as the most likely galaxy responsible
for this absorber, with $(z_{\rm phot}, \rho)=(0.235, 400 $~kpc)
This galaxy is not in the spectroscopic sample of \cite{Keeney2018}. }

\item{$z_{\rm abs}=0.3773$: This \lya\ redshift is one of six in this absorbing complex, along with five \ion{O}{6} components, four \ion{C}{3}, and one
\ion{Si}{3}. 
Both methods pair this with SDSS~J122454.53+212218.3, $(z_{\rm phot}, \rho)=(0.394, 147 $~kpc).
This galaxy is not in the spectroscopic sample of \cite{Keeney2018}. 
}

\item{$z_{\rm abs}=0.4213$: This system consists of three \lya\ lines, two \ion{C}{3}, a \ion{Si}{2}, two \ion{Si}{3} and an \ion{O}{6} component.
It is paired with SDSS~J122451.79+212157.6 by $(z_{\rm phot}, \rho)=(0.41, 342$~kpc) by the virial radius method and with 
SDSS~J122452.68+212337.1, $(z_{\rm phot}, \rho)=(0.39, 316$~kpc) by the stellar mass method, after uniqueness is imposed.
}

\item{$z_{\rm abs}=0.4231$: This system of \lya, \ion{C}{3}, and \ion{O}{6} components is matched with
SDSS~ J122455.68+212346.0, $(z_{\rm phot}, \rho)=(0.43, 348 $~kpc), also not part of the spectroscopic sample of \cite{Keeney2018}. 
This pair is also shown in Figure~\ref{fig:pg1222}.
}

\begin{figure}
\centering
\includegraphics[width=0.4\textwidth, viewport=0 0 600 600]{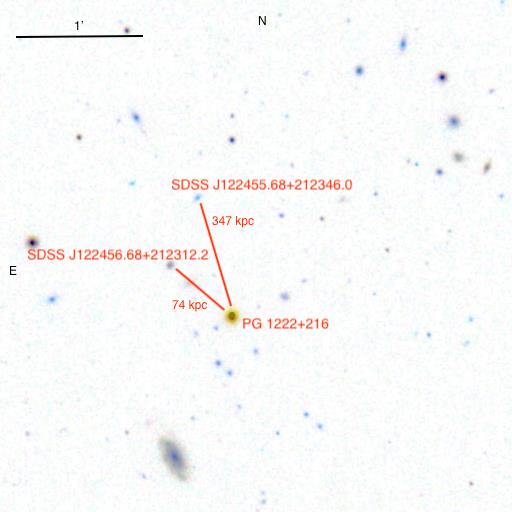}
\caption{
Galaxies SDSS~J122456.68+212312.2, $(z_{\rm phot}, \rho)= (0.091, 74$~kpc) and
SDSS~J122455.68+212346.0, $(z_{\rm phot}, \rho)= (0.43,348$~kpc)
paired with $z_{\rm abs}=0.0987,0.423$, respectively, in the spectrum of PG~1222+216.
}~\label{fig:pg1222}
\end{figure}

\end{itemize}

\subsection{PG 1229+204} \cite{Cote2005} associate UGC~7697 with $z_{\rm abs}=0.00859$ but this galaxy is excluded from our galaxy sample due to the photometric
redshift error cut.
We find five other potential galaxy-absorber pairs, all with $\rho > 300$~kpc.

\subsection{PG 1259+593} 
\begin{figure}
\centering
\includegraphics[width=0.8\textwidth, viewport=0 0 1125 1125]{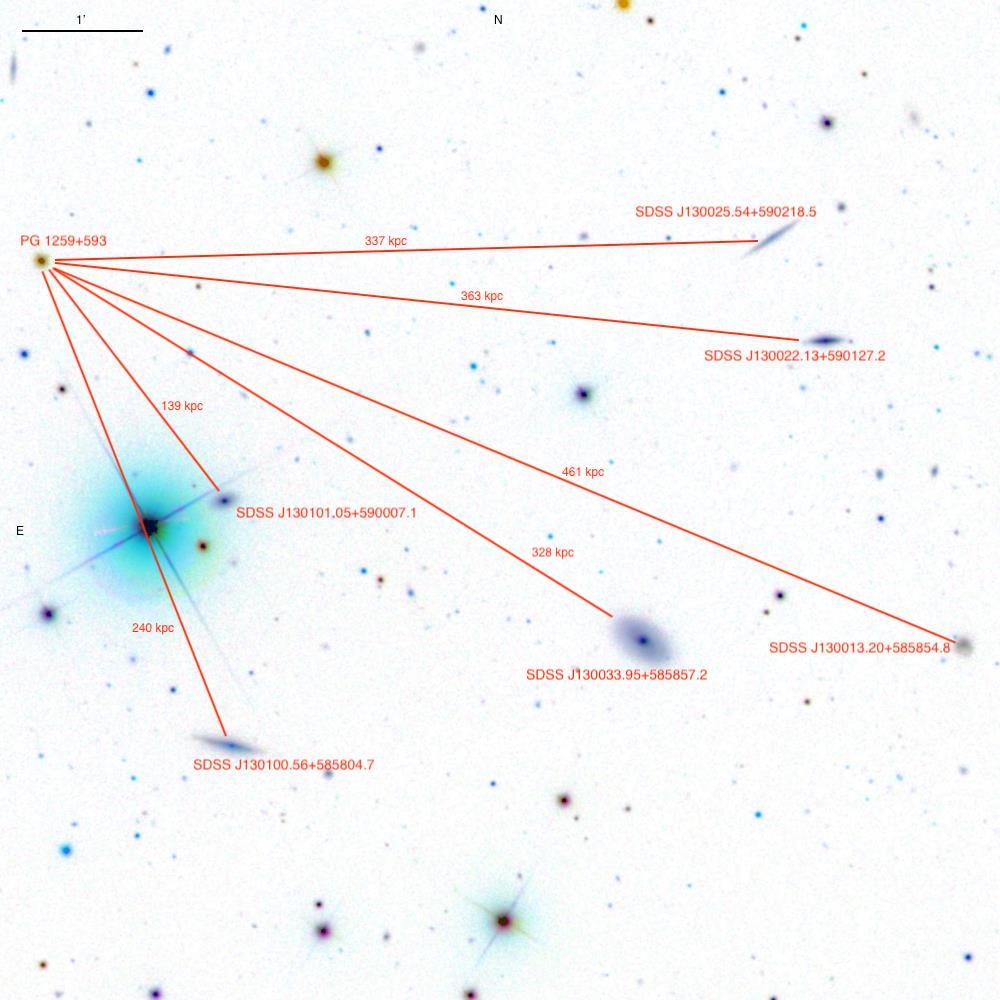}
\caption{
Galaxies
SDSS~J130101.05+590007.1, $(z_{\rm phot}, z_{\rm gal}, \rho)= (0.043, 0.0462, 139 $~kpc);
SDSS~J130100.56+585804.7, $(z_{\rm phot}, z_{\rm gal}, \rho)= (0.042, 0.0459, 240$~kpc);
SDSS~J130033.95+585857.2, $(z_{\rm phot}, z_{\rm gal}, \rho)=(0.033, 0.460,  328 $~kpc)
SDSS~J130022.13+590127.2, $(z_{\rm phot}, z_{\rm gal}, \rho)= (0.053, 0.0464, 363 $~kpc);
SDSS~J130013.20+585854.8, $(z_{\rm phot}, z_{\rm gal}, \rho)= (0.046, 0.04639, 461$~kpc) and
SDSS~J130025.54+590218.5, $(z_{\rm phot}, z_{\rm gal}, \rho)= (0.069, 0.04658, 337$~kpc)
paired with $z_{\rm abs}=0.046$ in the spectrum of PG~1259+593.
}~\label{fig:pg1259}
\end{figure}

This sightline has been studied by several authors, finding the following galaxy-absorber pairs:

\begin{itemize}
\item{$z_{\rm abs}=0.00229$: \cite{Cote2005} associate this absorber with UGC~8146, while \cite{Stocke2013}
report another possible associated galaxy, SDSS~J130207.44+584153.8. Both galaxies are excluded from our catalog due to the photometric redshift error cut 
we employ.
}

\item{$z_{\rm abs}=0.00759$: \cite{WakkerSavage2009} pair this absorber with UGC~8040. This galaxy is the fifth best match to the absorber in both methods, in part due to
its photometric redshift, $z_{\rm phot}=0.019$. 
The unique galaxy match to the absorber found by both methods is SDSS~J125503.96+584728.8, with $(z_{\rm phot}, \rho)=(0.012, 472 $~kpc).
The spectroscopic redshift of this galaxy, $z_{\rm gal}=0.00874$ places it within $\sim 350$ \kms\ of this $z_{\rm abs}=0.00759$ absorber.
}

\item{$z_{\rm abs}=0.046$: This absorber shows \lya, \ion{Si}{2}, \ion{Si}{3}, and \ion{C}{4} components.
\cite{Stocke2013} discuss the association of two absorbers, with $z_{\rm abs}=0.0460$ and $z_{\rm abs}=0.0464$ with SDSS~J130101.05+590007.1, $z_{\rm gal}=0.0462$, and
with other members of 
a possible galaxy group, including SDSS~J130033.95+585857.2 ($z_{\rm gal}=0.0460$), SDSS~J130100.56+585804.7 ($z_{\rm gal}=0.0459$), and SDSS~J130022.13+590127.2 ($z_{\rm gal}=0.0464$).

The galaxy with the highest overall probabilities for both methods is SDSS~J130101.05+590007.1, but, due to
the emphasis the uniqueness algorithm places on the \lya\ $P_{\rm ion}$,
the unique pairing listed in Tables~\ref{tab:rvir} and \ref{tab:mstar} is the second highest probability match,
SDSS~J130100.56+585804.7, $(z_{\rm phot}, \rho)=(0.042, 240 $~kpc). 
Its SDSS spectrum shows it to be star-forming galaxy. 

Other possible group galaxies found in the top ten probable 
galaxies paired with this absorber by the virial radius method are: 
SDSS~J130033.95+585857.2 with $(z_{\rm phot}, \rho)=(0.033, 328 $~kpc),
SDSS~J130022.13+590127.2, with $(z_{\rm phot}, \rho)=(0.053, 363 $~kpc),
SDSS~J130013.20+585854.8, a star-forming galaxy with $z_{\rm gal}=0.04639$; and SDSS~J130025.54+590218.5 ($z_{\rm gal}=0.04658$).
All of these galaxies are shown in Figure~\ref{fig:pg1259}.
}

\item{$z_{\rm abs}=0.196$: \cite{Mathes2014} associate this absorber with SDSS~J130116.43+590135.7, which is only the 7$^{\rm th}$(8$^{\rm th}$) 
best match with the virial radius(stellar mass) method due to its photometric redshift
estimate, $z_{\rm phot}=0.3517$. 
After imposing the uniqueness criteria, the galaxy matched with this absorber in Tables~\ref{tab:rvir} or \ref{tab:mstar} is the second highest probability match,
SDSS~J130131.84+590128.5, $(z_{\rm phot}, \rho)=(0.26, 495$~kpc). However, \cite{Keeney2018} report $z_{\rm gal}=0.24076$, which is in better agreement with the absorber discussed below.
}

\item{$z_{\rm abs}=0.241$: \cite{Mathes2014} find that this absorber is associated with SDSS~J130109.88+590315.3, the fifth best match found by our two methods. 
Instead, our methods
give a top match with SDSS~J130059.79+590059.7, $(z_{\rm phot}, \rho)=(0.36, 466$~kpc). This galaxy is uniquely paired with
$z_{\rm abs}=0.256$ in Tables~\ref{tab:rvir} and \ref{tab:mstar}. \cite{Keeney2018} report the spectroscopic redshift of SDSS~J130059.79+590059.7
to be 0.2929, so it is unlikely to be truly associated with either of these absorbers but instead with $z_{\rm abs}=0.292$.
}
\end{itemize}

One remaining absorption system along this line of sight is a $z_{\rm abs}=0.22$ system showing \lya, \ion{O}{6}, \ion{Si}{3} and \ion{C}{3}.
We do find a galaxy paired with this absorber, SDSS~J130059.79+590059.7, $(z_{\rm phot}, \rho)=(0.35, 435 $~kpc), but its association with the absorber would
depend on its true redshift being substantially different from the photometric estimate. There is no spectroscopic redshift reported by \cite{Keeney2018}.

\subsection{PG 1307+085} 
\begin{figure}
\centering
\includegraphics[width=0.4\textwidth, viewport=0 0 600 600]{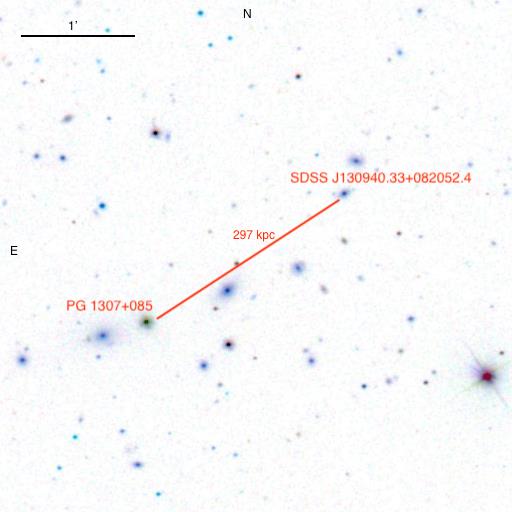}
\caption{
Galaxy SDSS~J130940.33+082052.4, $(z_{\rm phot}, \rho)= (0.145, 297$~kpc) 
paired with $z_{\rm abs}=0.14$ in the spectrum of PG~1307+085.
}~\label{fig:pg1307}
\end{figure}

This sightline has no reported galaxy-absorber pairs reported in the literature.
There are only four possible unique pairs meeting the probability threshold reported by both methods in 
Tables~\ref{tab:rvir} and \ref{tab:mstar}, including
a multicomponent $z_{\rm abs}=0.14$ system with \lya\ and \ion{O}{6} 
matched with SDSS~J130940.33+082052.4, with $(z_{\rm phot}, \rho)=(0.145, 297 $~kpc). This is shown in Figure~\ref{fig:pg1307}.

\subsection{PG 1309+355}  \cite{Cote2005} associate $z_{\rm abs}=0.00292$ absorber with the NGC~5033 group.
We report 7(9) potential galaxy-absorber pairs meeting the 10\% probability threshold found using the virial radius(stellar mass) method,
including one galaxy, SDSS J131220.99+351518.3 ($(z_{\rm phot}=0.15$), paired with
$z_{\rm abs}=0.0877$ ($\rho=65$~kpc) by the virial radius method and
$z_{\rm abs}=0.120$ ($\rho=87$~kpc) by the stellar mass method.

\subsection{PG 1424+240} 
\begin{figure}
\centering
\includegraphics[width=0.4\textwidth, viewport=0 0 600 600]{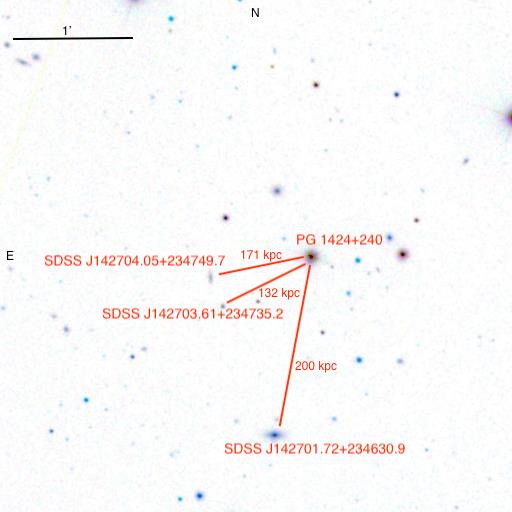}
\caption{
Galaxies SDSS~J142701.72+234630.9, $(z_{\rm phot}, \rho)= (0.12, 200$~kpc) and
SDSS~J142703.61+234735.2, $(z_{\rm phot}, \rho)= (0.15, 132$~kpc); 
SDSS~J142704.05+234749.7, $(z_{\rm phot}, \rho)= (0.19,171$~kpc)
paired with absorbers in the spectrum of PG~1424+240.
}~\label{fig:pg1424}
\end{figure}

We find 24 possible galaxy-absorber pairs for this sightline, including:
\begin{itemize}
\item{$z_{\rm abs}=0.12$: This system showing \lya, \ion{O}{6}, \ion{Si}{3}, and \ion{C}{4} is matched with
SDSS~J142701.72+234630.9, $(z_{\rm phot}, \rho)=(0.12, 200 $~kpc). The photometric redshift is 
in agreement with the spectroscopic measurement $z_{\rm gal}=0.1211$.
This pair is shown in Figure~\ref{fig:pg1424}.}

\item{$z_{\rm abs}=0.14$: This system with \lya, \ion{O}{6}, \ion{Si}{3}, \ion{Si}{4} and \ion{C}{4}, is paired with 
as a unique match with SDSS~J142703.61+234735.2, $(z_{\rm phot}, \rho)=(0.15, 132 $~kpc), which is shown in Figure~\ref{fig:pg1424}.
This galaxy is not in the \cite{Keeney2018} sample, who report the closest galaxy to this absorber having an impact parameter of 493~kpc.
}

\item{$z_{\rm abs}=0.20$: This absorber with one \lya\ and two \ion{C}{3} components is paired with SDSS~J142704.05+234749.7, $(z_{\rm phot}, \rho)=(0.20, 171 $~kpc)
by the virial radius method.
This pair
is shown in Figure~\ref{fig:pg1424} and is marginally favored over the pairing reported by the stellar mass method,
SDSS~J142701.63+234833.0, $(z_{\rm phot}, \rho)=(0.19, 124 $~kpc).
Neither galaxy is part of the spectroscopic
sample of \cite{Keeney2018}, who report the closest galaxy to this absorber having an impact parameter of 683~kpc.
}
\end{itemize}

\subsection{PG 1626+554} Our methods find eight potential galaxies paired with single \lya\ lines along this line of sight. All but one have $\rho \sim 400-500$~kpc.
One pair, $z_{\rm abs}=0.0613$ and SDSS~J162828.16+552648.4, $(z_{\rm phot}, \rho)=(0.060, 446$~kpc), is supported by the spectroscopic
redshift measurement reported by \cite{Keeney2018}, $z_{\rm gal}=0.06168$.

\subsection{PG 2349+014} \cite{Bowen1997} find an association between a $z_{\rm abs}=0.038$ system showing \lya\ and \ion{Si}{3} in the COS spectrum and SDSS~J235142.21-010100.9.
This is recovered as the most favored galaxy matched to this absorber by both methods $(z_{\rm phot}, \rho)=(0.043, 406 $~kpc), and 
reported as the unique match for the virial radius method. But our uniqueness criteria with the stellar mass method
instead pair this galaxy with a $z_{\rm abs}=0.0455$ \lya\ line.
We find 8(9) other possible galaxy- \lya\ absorber pairs with the virial radius(stellar mass) method.

\subsection{QSO 0045+3926} This sightline has no reported galaxy-absorber pairs reported in the literature.
We find 6(7) possible pairs with the virial radius(stellar mass) method, including a $z_{\rm abs}=0.110$ \lya\ and \ion{O}{6} system associated with SDSS~J004833.42+394057.1,
$(z_{\rm phot}, \rho)=(0.12, 339 $~kpc).

\subsection{Q 1230+0115} As for 3C~273 and PG 1216+069, the Virgo cluster lies in the foreground.
\cite{Penton2002} and \cite{WakkerSavage2009}  note a correspondence between $z_{\rm abs}=0.00769$ and UGC~7625. The SDSS photometric redshift estimate for this galaxy
is 0.114, so it is not in the top ten most probable matches we find for this absorber. 
Our algorithms identify SDSS~J122815.96+014944.1, $(z_{\rm phot}, \rho)=(0.013, 495$~kpc), as the unique galaxy match for this absorber. However, the spectroscopic
redshift reported by SDSS,
$z_{\rm gal}=0.0030$, contradicts this result.

Our highest probability unique match from the virial radius method to a $z_{\rm abs}=0.0057$ system with five \lya, two \ion{C}{2}, one \ion{Si}{2}, a \ion{Si}{3}, and two \ion{Si}{4} components 
is SDSS~J122921.63+010325.0, $(z_{\rm phot}, \rho)=(0.018, 194 $~kpc). However, the spectroscopic redshift, $z_{\rm gal}=0.0231$, does not support an association.

\begin{figure}
\centering
\includegraphics[width=0.8\textwidth, viewport=0 0 600 600]{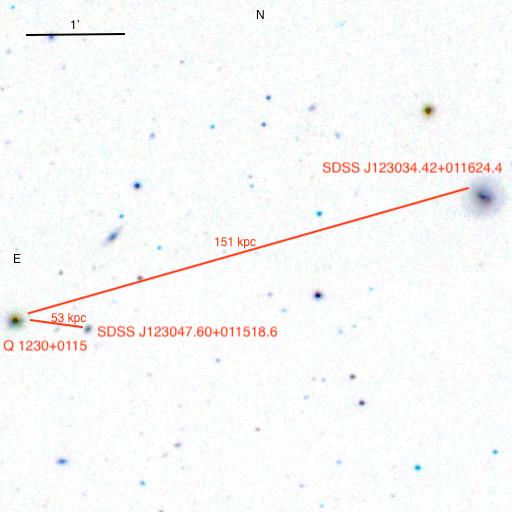}
\caption{
Galaxies 
SDSS~J123034.42+011624.4, $(z_{\rm phot}, \rho)= (0.031, 151$~kpc),
paired with $z_{\rm abs}=0.031$ 
and SDSS~J123047.60+011518.6, $(z_{\rm phot}, \rho)= (0.12, 53$~kpc)
paired with $z_{\rm abs}=0.077$
in the spectrum of Q~1230+0115
}~\label{fig:q1230}
\end{figure}

A more promising association found by the stellar mass method
between a $z_{\rm abs}=0.031$ \lya\ and \ion{C}{4} system  and SDSS~J123034.42+011624.4, with $(z_{\rm phot}, \rho)=(0.031, 151 $~kpc). The spectroscopic redshift is in 
good agreement in this case, $z_{\rm gal}=0.031$, and this pair is shown in Figure~\ref{fig:q1230}.  The unique galaxy match to the absorber found by the virial radius method,
SDSS~J123141.38+011821.5, $(z_{\rm phot}, \rho)=(0.048, 495 $~kpc),
is less plausible, both because of the large impact parameter and because SDSS reports $z_{\rm gal}=0.0234$.

Finally, our methods both find a match of a $z_{\rm abs}=0.077$ \ion{C}{4}, \ion{Si}{4}, and \ion{Si}{3} system with
SDSS~J123047.60+011518.6, $(z_{\rm phot}, \rho)=(0.12, 53 $~kpc). This is supported by the spectroscopic redshift reported by \cite{Keeney2018}, $z_{\rm gal}=0.07782$, 
and is also shown in Figure~\ref{fig:q1230}.

\subsection{SBS 1108+560} 
\cite{Stocke2013} find an association of M~108 with two possible absorbers, $z_{\rm abs}=0.0222$ and $z_{\rm abs}=0.0259$.
This galaxy is excluded from our SDSS galaxy sample due to its brightness and large projected size on the sky.
\cite{Liang2014} report that the galaxy WISEA~J111125.60+554435.4 is a pair with $z_{\rm abs}=0.138$.
This galaxy is our top match to the absorber with both methods. However, the after imposing the uniqueness criteria on the stellar mass method results, the unique galaxy
paired with this absorber and reported in Table~\ref{tab:mstar} is SDSS~J111116.84+554812.2 $(z_{\rm phot}, \rho)=(0.142, 339 $~kpc), and \cite{Keeney2018} report
its spectroscopic redshift to be $z_{\rm gal}=0.13813$, an even better redshift match with $z_{\rm abs}$.

\begin{figure}
\centering
\includegraphics[width=0.4\textwidth, viewport=0 0 560 560]{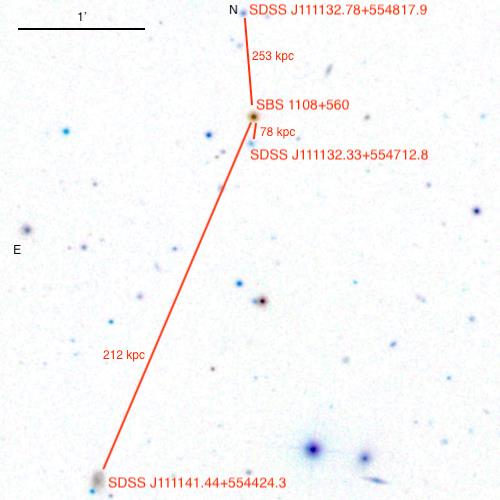}
\caption{
Galaxies
SDSS~J111141.44+554424.3, $(z_{\rm phot}, \rho)= (0.057, 212$~kpc),
SDSS~J111132.78+554817.9, $(z_{\rm phot}, \rho)= (0.33, 253$~kpc), and 
SDSS~J111132.33+554712.8, $(z_{\rm phot}, \rho)= (0.43, 78$~kpc),
paired with $z_{\rm abs}=0.140, 0.32, 0.46$, respectively, in the spectrum of SBS~1108+560.
}~\label{fig:sbs1108}
\end{figure}

Between the two methods, we report 22 other possible galaxy-absorber pairs for this sightline.
One possible pair identified by both methods is $z_{\rm abs}=0.0548$ and SDSS~J111141.44+554424.3, $(z_{\rm phot}, \rho)=   (0.057, 212$~kpc).
The spectroscopic redshift reported by \cite{Keeney2018}, $z_{\rm gal}=0.05493$, supports this association and the pair is shown in Figure~\ref{fig:sbs1108}. 
These authors report that another SDSS galaxy lies closer 
than SDSS~J111141.44+554424.3 to the absorber.

The candidate pairs also include a $z_{\rm abs}=0.33$ \ion{O}{6} absorber, uniquely paired with
SDSS~J111132.78+554817.9, $(z_{\rm phot}, \rho)=(0.33, 253 $~kpc) by the virial radius method. This galaxy is the top stellar mass method match to the absorber also, but
the uniqueness criteria match it with two $z_{\rm abs}=0.285$ \lya\ absorbers, and no unique match is listed for $z_{\rm abs}=0.33$. This pair is shown in Figure~\ref{fig:sbs1108}.

The stellar mass method yields a match of a $z_{\rm abs}=0.46$ system with \ion{O}{6}, \ion{Si}{3}, and \ion{C}{3} and
SDSS~J111132.33+554712.8, $(z_{\rm phot}, \rho)=(0.43, 78 $~kpc). 
The galaxy's spectroscopic redshift, $z_{\rm gal}=0.462$, supports this association.
This pair is also shown in Figure~\ref{fig:sbs1108}.
The virial radius method finds this galaxy as the top match to the $z_{\rm abs}=0.46$ system  also, but the uniqueness criteria match it with an \ion{O}{6} - only
$z_{\rm abs}=0.4358$ absorber.

\subsection{SBS 1122+594} 
In total, our virial radius(stellar mass) method identifies 28(27) possible galaxy-absorber pairs for this long line of sight to the $z=0.852$ QSO.
\cite{Stocke2013} and \cite{Liang2014} determine an association between $z_{\rm abs}=0.04$ and IC~691. This galaxy is excluded from our SDSS galaxy lists because of our
photometric redshift error cut. 
Instead, our methods find SDSS~J112505.99+591804.6 as the reported match for this \lya\ line, with $(z_{\rm phot}, \rho)=(0.06, 477 $~kpc).
The spectroscopic redshift of this galaxy is 0.1125, so this is not considered a more likely association than IC~691.

Other galaxy-absorber pairs found along this sightline include:

\begin{itemize}

\item{$z_{\rm abs}=0.060$: This system shows \lya, \ion{C}{4}, \ion{Si}{4} and \ion{Si}{3} absorption.
The virial radius method finds a match with SDSS~J112517.67+590828.8, $(z_{\rm phot}, \rho)=(0.064, 350 $~kpc). The SDSS spectroscopic redshift
is $z_{\rm gal}=0.058$, a difference of $\sim$565 \kms\ with the absorber redshift. 
}

\item{$z_{\rm abs}=0.138$:  This system consists of four \lya\ lines. 
Both methods yield a match with SDSS~J112618.11+590925.8, $(z_{\rm phot}, \rho)=(0.13, 481 $~kpc).
This galaxy is not part of the \cite{Keeney2018} sample.
}

\item{$z_{\rm abs}=0.194$: For this system showing \ion{O}{6}, \ion{Si}{3}, and \ion{C}{3} absorption, both methods uniquely pair
SDSS~J112549.24+590957.7, $(z_{\rm phot}, \rho)=(0.18, 137 $~kpc). However, given the spectroscopic redshift reported by \cite{Keeney2018}, 
$z_{\rm gal}=0.15527$, this galaxy is more likely associated with $z_{\rm abs}=0.155$, as these authors note it is the closest galaxy known.
}

\item{$z_{\rm abs}=0.420$: This \ion{O}{6} system is matched with 
SDSS~J112548.53+590916.8, $(z_{\rm phot}, \rho)=(0.40, 426 $~kpc).
This galaxy is not in the 
spectroscopic sample of \cite{Keeney2018}.
}

\end{itemize}

\subsection{SDSS~J080908.13+461925.6} For this sightline, we find 18(16) possible galaxy-absorber pairs with the virial radius(stellar mass) method.
\cite{Liang2014} report that the  $z_{\rm abs}=0.0464$ system, with \lya, \ion{Si}{3}, and \ion{Si}{4}, is associated with SDSS~J080913.17+461842.7.  
This is the top match
for the virial radius method and the
second most probable match for the stellar mass method.
The other top match for both methods is 
SDSS~J080914.28+461822.2, $(z_{\rm phot}, \rho)=(0.044, 82 $~kpc), which is reported as the
unique pairing in Tables~\ref{tab:rvir} and \ref{tab:mstar}.

\subsection{SDSS~J092554.43+453544} 
We find 25 possible galaxy-absorber pairs with both methods.
\cite{Liang2014} report two galaxy-absorber pairs for this sightline: 
$z_{\rm abs}=0.0170$ and SDSS~J092721.06+454158.8, the second match to the absorber and the unique pair outcome of both our methods; and
$z_{\rm abs}=0.0261$  and SDSS~J092617.38+452924.9, the top match from the virial radius method and the 
second match to the absorber from the stellar mass method.
For $z_{\rm abs}=0.0261$, the unique galaxy match is instead with the second virial radius method match and the top
stellar mass method match, SDSS~J092530.98+453157.8, $(z_{\rm phot}, \rho)=(0.027, 183 $~kpc). 
However, the SDSS spectroscopic redshift of this galaxy is 0.0142, which does not support this proposed
association.

\begin{figure}
\centering
\includegraphics[width=0.4\textwidth, viewport=0 0 600 600]{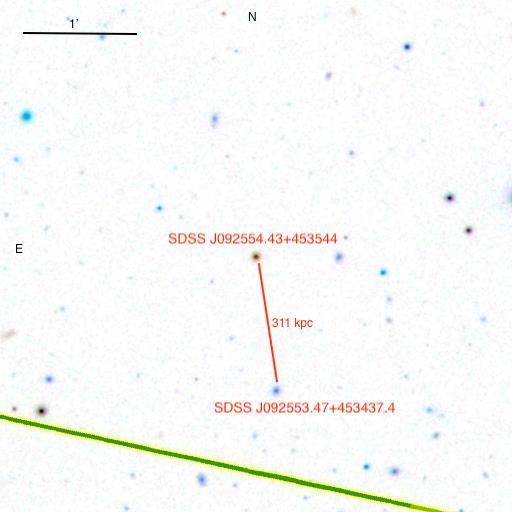}
\caption{
Galaxy
SDSS~J092553.47+453437.4, $(z_{\rm phot}, \rho)= (0.30, 311$~kpc)
paired with $z_{\rm abs}=0.309$ absorber in the spectrum of QSO SDSS~J080908.13+461925.6.
}~\label{fig:sdss0925}
\end{figure}

Our methods also match a 
a $z_{\rm abs}=0.296$ system with \ion{O}{6} and \ion{C}{3} absorption with  SDSS~J092556.40+453653.3, $(z_{\rm phot}, \rho)=(0.28, 320 $~kpc);  and
a $z_{\rm abs}=0.309$ system with \ion{O}{6} and \ion{C}{3} with SDSS~J092553.47+453437.4, $(z_{\rm phot}, \rho)=(0.30, 311 $~kpc), 
shown in Figure~\ref{fig:sdss0925}.
The spectroscopic 
redshift of SDSS~J092553.47+453437.4, $z_{\rm gal}=0.3088$, does support this association. 

\subsection{SDSS~J092909.79+464424} 
\begin{figure}
\centering
\includegraphics[width=0.8\textwidth, viewport=0 0 1150 1150]{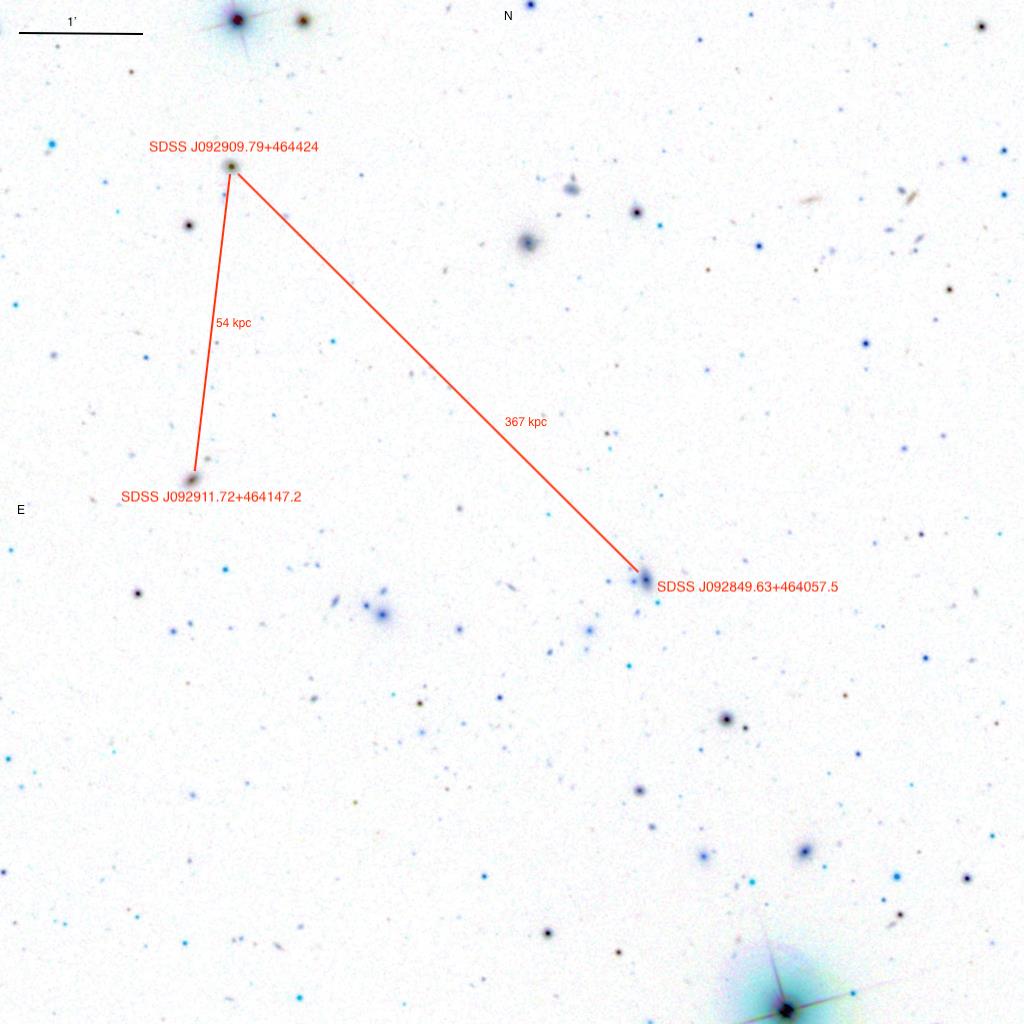}
\caption{
Galaxies
SDSS~J092911.72+464147.2, $(z_{\rm phot}, \rho)= (0.012, 54$~kpc) and
SDSS~J092849.63+464057.5, $(z_{\rm phot}, \rho)= (0.065, 367$~kpc) 
paired with absorbers in the spectrum of QSO SDSS~J092909.79+464424.
}~\label{fig:sdss0929}
\end{figure}

With both methods, we find 18 potential galaxy absorber pairs.
Some candidates of particular interest include:
\begin{itemize}
\item{
$z_{\rm abs}=0.016$: a \ion{C}{4} system matched with SDSS~J092911.72+464147.2, $(z_{\rm phot}, \rho)=(0.012, 54 $~kpc) and spectroscopic 
redshift
$z_{\rm gal}=0.01676$. This pair is shown in Figure~\ref{fig:sdss0929}.
}

\item{$z_{\rm abs}=0.064$: a double \lya\ component system paired with SDSS~J092849.63+464057.5, $(z_{\rm phot}, \rho)=(0.065, 367 $~kpc) and spectroscopic 
redshift
$z_{\rm gal}=0.06402$. This pair is shown in Figure~\ref{fig:sdss0929}.
}

\item{$z_{\rm abs}=0.145$: an \ion{O}{6} absorber 
and SDSS~J092853.27+464412.6 by $(z_{\rm phot}, \rho)=(0.10, 435 $~kpc).
}

\end{itemize}

\subsection{SDSS~J094952.91+390203} We find 14 galaxy-absorber pairs for this sightline with the both methods.
\cite{Liang2014} report a pairing of $z_{\rm abs}=0.0669$ with SDSS~J095002.76+390308.7.
This is our top ranked galaxy match  to this absorber, with $(z_{\rm phot}, \rho)=(0.058, 159 $~kpc).

\begin{figure}
\centering
\includegraphics[width=0.4\textwidth, viewport=0 0 600 600]{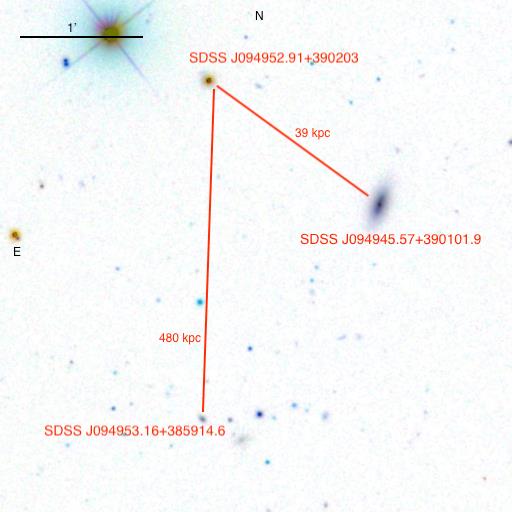}
\caption{
Galaxies
SDSS~J094945.57+390101.9, $(z_{\rm phot}, \rho)= (0.027, 39$~kpc) and
SDSS~J094953.16+385914.6, $(z_{\rm phot}, \rho)= (0.164, 480$~kpc)
paired with absorbers in the spectrum of QSO SDSS~J094952.91+390203.
}~\label{fig:sdss0949}
\end{figure}

We also find a $z_{\rm abs}=0.018$ system showing \lya, \ion{Si}{3}, and \ion{C}{4} components with SDSS~J094945.57+390101.9, $(z_{\rm phot}, \rho)=(0.027, 39 $~kpc), shown 
in Figure~\ref{fig:sdss0949}.
The spectroscopic redshift of the galaxy, $z_{\rm gal}=0.0180$, is in excellent agreement with $z_{\rm abs}$.
Finally, we report the match of a $z_{\rm abs}=0.164$ double \lya\ component system with SDSS~J094953.16+385914.6,
$(z_{\rm phot}, \rho)=(0.164, 480 $~kpc). This pair is also shown in Figure~\ref{fig:sdss0949}.

\subsection{SDSS~J135712.61+170444} We report 7 possible galaxy-absorber pairs for this QSO sightline. One is a rich metal line system:
a $z_{\rm abs}=0.097$ system with \ion{Si}{3}, \ion{Si}{4}, \ion{C}{2} and \ion{C}{4} paired with SDSS J135719.68+170410.3, $(z_{\rm phot}, \rho)=(0.10, 195 $~kpc).
However, the galaxy's SDSS spectroscopic redshift, $z_{\rm gal}=0.12605$, does not support this pairing.

\begin{figure}
\centering
\includegraphics[width=0.4\textwidth, viewport=0 0 600 600]{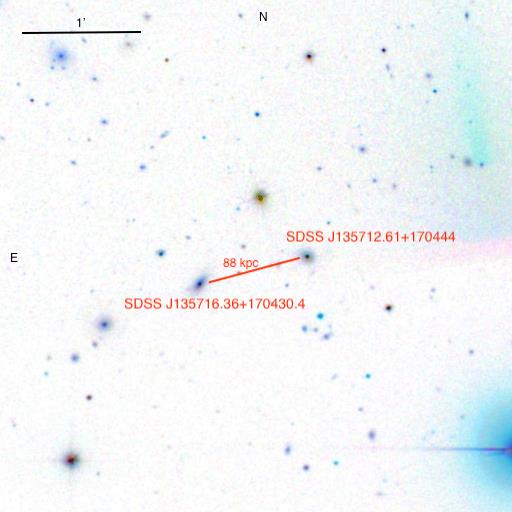}
\caption{
Galaxy
SDSS~J135716.36+170430.4, $(z_{\rm phot}, \rho)= (0.087, 88$~kpc)
paired with $z_{\rm abs}=0.083$ absorber in the spectrum of QSO SDSS~J135712.61+170444.
}~\label{fig:sdss1357}
\end{figure}

On the other hand, a $z_{\rm abs}=0.083$ system which shows absorption from multiple components of \ion{Si}{3}, \ion{Si}{4} and \ion{C}{4}, is matched with
SDSS~J135716.36+170430.4, $(z_{\rm phot}, \rho)=(0.087, 88 $~kpc). Its spectroscopic redshift, $z_{\rm gal}=0.08334$, does support the association.
This pair is shown in Figure~\ref{fig:sdss1357}.

\subsection{TON 1187} We find eight candidate galaxy-absorber pairs for this line of sight with the two methods, all consisting of \lya\ only. 
One absorber, $z_{\rm abs}=0.0678$ is paired with a galaxy
with $\rho=117$~kpc, SDSS~J101255.81+355120.4. 
A true association depends on a better redshift agreement between $z_{\rm abs}$ and the spectroscopic redshift than with the
photometric estimate, $z_{\rm phot}=0.076$. 

Another pair, a three-component $z_{\rm abs}=0.035$ system and SDSS J101222.19+355523.0, $(z_{\rm phot}, \rho)=(0.039, 394$~kpc),
is unlikely given the fact that $z_{\rm gal}=0.0393$. Instead, this galaxy is likely a better match with
$z_{\rm abs}=0.04022$ ($\rho=443$~kpc), for which it is the top ranked pairing before the uniqueness criteria are imposed.

\subsection{TON 236} 
\cite{Liang2014} report an association of with $z_{\rm abs}=0.0451$ with  SDSS~J152827.39+282738.6. This galaxy is excluded from our SDSS galaxy catalog
due to the photometric redshift error cut we employ.

\begin{figure}
\centering
\includegraphics[width=0.4\textwidth, viewport=0 0 560 560]{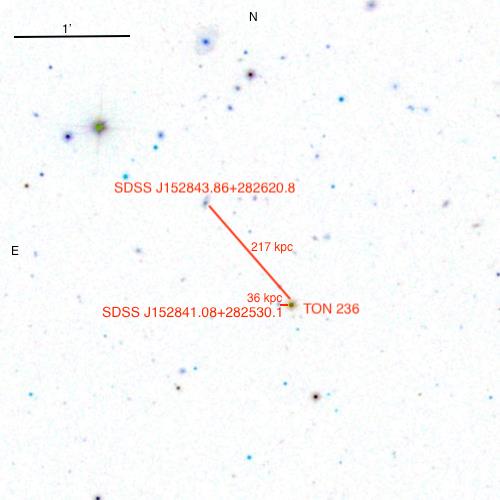}
\caption{
Galaxies
SDSS~J152843.86+282620.8, $(z_{\rm phot}, \rho)= (0.14, 217$~kpc) and
SDSS~J152841.08+282530.1, $(z_{\rm phot}, \rho)= (0.47, 36$~kpc),
paired with absorbers in the spectrum of TON~236.
}~\label{fig:ton236}
\end{figure}

Our methods yield 20 possible galaxy-absorber pairs for this QSO sightline.
An \ion{O}{6} system
with $z_{\rm abs}=0.194$
is matched with SDSS~J152843.86+282620.8, $(z_{\rm phot}, \rho)=(0.14, 217 $~kpc) with both methods. This galaxy has no measured spectroscopic redshift. 
The
pair is shown in Figure~\ref{fig:ton236}.

A $z_{\rm abs}=0.259$ \ion{O}{6} absorber is paired with SDSS~J152845.56+282654.6, $(z_{\rm phot}, \rho)=(0.25, 434 $~kpc) by both methods. 
This galaxy is not in the 
\cite{Keeney2018} sample. 

Finally, we note that our methods each pair 
a galaxy very close to the sightline, SDSS~J152841.08+282530.1, with two different absorption systems, as defined by our algorithm.
The virial radius method pairs the galaxy with 
a triple component $z_{\rm abs}=0.433$ \lya\ absorber
and the stellar mass method pairs it with
a double component $z_{\rm abs}=0.439$ \lya\ absorber.
This galaxy is also not in the spectroscopic sample of \cite{Keeney2018}.
It is shown in Figure~\ref{fig:ton236}.

\subsection{TON 580} \cite{Liang2014} find that a $z_{\rm abs}=0.0744$ \ion{C}{4} system is associated with  SDSS~J113056.11+311445.6 at $\rho=250$~kpc.
We recover this pairing as the top match with both of our methods.
We report 14 other galaxy-\lya\ absorber pairs for this sightline.
One match, $z_{\rm abs}=0.1023$ paired with SDSS~J113121.67+311055.8, $(z_{\rm phot}, \rho)=(0.125, 464 $~kpc), is ruled out by the spectroscopic redshift
reported by \cite{Keeney2018}, $z_{\rm gal}=0.12434$.
A $z_{\rm abs}=0.2026$ metal system with \ion{O}{6}, \ion{C}{3}, and \ion{Si}{3}  components is matched with
SDSS~J113108.50+311332.6, $(z_{\rm phot}, \rho)=(0.224, 116 $~kpc).

\bibliography{cgm}{}

\begin{thebibliography}{}
\expandafter\ifx\csname natexlab\endcsname\relax\def\natexlab#1{#1}\fi
\providecommand{\url}[1]{\href{#1}{#1}}
\providecommand{\dodoi}[1]{doi:~\href{http://doi.org/#1}{\nolinkurl{#1}}}
\providecommand{\doeprint}[1]{\href{http://ascl.net/#1}{\nolinkurl{http://ascl.net/#1}}}
\providecommand{\doarXiv}[1]{\href{https://arxiv.org/abs/#1}{\nolinkurl{https://arxiv.org/abs/#1}}}

\bibitem[{{Alam} {et~al.}(2015){Alam}, {Albareti}, {Allende Prieto}, {Anders},
  {Anderson}, {Anderton}, {Andrews}, {Armengaud}, {Aubourg}, {Bailey}, {Basu},
  {Bautista}, {Beaton}, {Beers}, {Bender}, {Berlind}, {Beutler}, {Bhardwaj},
  {Bird}, {Bizyaev}, {Blake}, {Blanton}, {Blomqvist}, {Bochanski}, {Bolton},
  {Bovy}, {Shelden Bradley}, {Brandt}, {Brauer}, {Brinkmann}, {Brown},
  {Brownstein}, {Burden}, {Burtin}, {Busca}, {Cai}, {Capozzi}, {Carnero
  Rosell}, {Carr}, {Carrera}, {Chambers}, {Chaplin}, {Chen}, {Chiappini},
  {Chojnowski}, {Chuang}, {Clerc}, {Comparat}, {Covey}, {Croft}, {Cuesta},
  {Cunha}, {da Costa}, {Da Rio}, {Davenport}, {Dawson}, {De Lee}, {Delubac},
  {Deshpande}, {Dhital}, {Dutra-Ferreira}, {Dwelly}, {Ealet}, {Ebelke},
  {Edmondson}, {Eisenstein}, {Ellsworth}, {Elsworth}, {Epstein}, {Eracleous},
  {Escoffier}, {Esposito}, {Evans}, {Fan}, {Fern{\'a}ndez-Alvar}, {Feuillet},
  {Filiz Ak}, {Finley}, {Finoguenov}, {Flaherty}, {Fleming}, {Font-Ribera},
  {Foster}, {Frinchaboy}, {Galbraith-Frew}, {Garc{\'\i}a},
  {Garc{\'\i}a-Hern{\'a}ndez}, {Garc{\'\i}a P{\'e}rez}, {Gaulme}, {Ge},
  {G{\'e}nova-Santos}, {Georgakakis}, {Ghezzi}, {Gillespie}, {Girardi},
  {Goddard}, {Gontcho}, {Gonz{\'a}lez Hern{\'a}ndez}, {Grebel}, {Green},
  {Grieb}, {Grieves}, {Gunn}, {Guo}, {Harding}, {Hasselquist}, {Hawley},
  {Hayden}, {Hearty}, {Hekker}, {Ho}, {Hogg}, {Holley-Bockelmann}, {Holtzman},
  {Honscheid}, {Huber}, {Huehnerhoff}, {Ivans}, {Jiang}, {Johnson},
  {Kinemuchi}, {Kirkby}, {Kitaura}, {Klaene}, {Knapp}, {Kneib}, {Koenig},
  {Lam}, {Lan}, {Lang}, {Laurent}, {Le Goff}, {Leauthaud}, {Lee}, {Lee},
  {Licquia}, {Liu}, {Long}, {L{\'o}pez-Corredoira}, {Lorenzo-Oliveira},
  {Lucatello}, {Lundgren}, {Lupton}, {Mack}, {Mahadevan}, {Maia}, {Majewski},
  {Malanushenko}, {Malanushenko}, {Manchado}, {Manera}, {Mao}, {Maraston},
  {Marchwinski}, {Margala}, {Martell}, {Martig}, {Masters}, {Mathur},
  {McBride}, {McGehee}, {McGreer}, {McMahon}, {M{\'e}nard}, {Menzel},
  {Merloni}, {M{\'e}sz{\'a}ros}, {Miller}, {Miralda-Escud{\'e}}, {Miyatake},
  {Montero-Dorta}, {More}, {Morganson}, {Morice-Atkinson}, {Morrison},
  {Mosser}, {Muna}, {Myers}, {Nandra}, {Newman}, {Neyrinck}, {Nguyen},
  {Nichol}, {Nidever}, {Noterdaeme}, {Nuza}, {O'Connell}, {O'Connell},
  {O'Connell}, {Ogando}, {Olmstead}, {Oravetz}, {Oravetz}, {Osumi}, {Owen},
  {Padgett}, {Padmanabhan}, {Paegert}, {Palanque-Delabrouille}, {Pan},
  {Parejko}, {P{\^a}ris}, {Park}, {Pattarakijwanich}, {Pellejero-Ibanez},
  {Pepper}, {Percival}, {P{\'e}rez-Fournon}, {P{\'e}rez-Ra`fols}, {Petitjean},
  {Pieri}, {Pinsonneault}, {Porto de Mello}, {Prada}, {Prakash},
  {Price-Whelan}, {Protopapas}, {Raddick}, {Rahman}, {Reid}, {Rich}, {Rix},
  {Robin}, {Rockosi}, {Rodrigues}, {Rodr{\'\i}guez-Torres}, {Roe}, {Ross},
  {Ross}, {Rossi}, {Ruan}, {Rubi{\~n}o-Mart{\'\i}n}, {Rykoff},
  {Salazar-Albornoz}, {Salvato}, {Samushia}, {S{\'a}nchez}, {Santiago},
  {Sayres}, {Schiavon}, {Schlegel}, {Schmidt}, {Schneider}, {Schultheis},
  {Schwope}, {Sc{\'o}ccola}, {Scott}, {Sellgren}, {Seo}, {Serenelli}, {Shane},
  {Shen}, {Shetrone}, {Shu}, {Silva Aguirre}, {Sivarani}, {Skrutskie},
  {Slosar}, {Smith}, {Sobreira}, {Souto}, {Stassun}, {Steinmetz}, {Stello},
  {Strauss}, {Streblyanska}, {Suzuki}, {Swanson}, {Tan}, {Tayar}, {Terrien},
  {Thakar}, {Thomas}, {Thomas}, {Thompson}, {Tinker}, {Tojeiro}, {Troup},
  {Vargas-Maga{\~n}a}, {Vazquez}, {Verde}, {Viel}, {Vogt}, {Wake}, {Wang},
  {Weaver}, {Weinberg}, {Weiner}, {White}, {Wilson}, {Wisniewski},
  {Wood-Vasey}, {Ye`che}, {York}, {Zakamska}, {Zamora}, {Zasowski}, {Zehavi},
  {Zhao}, {Zheng}, {Zhou}, {Zhou}, {Zou}, \& {Zhu}}]{Alam2015}
{Alam}, S., {Albareti}, F.~D., {Allende Prieto}, C., {et~al.} 2015, \apjs, 219,
  12, \dodoi{10.1088/0067-0049/219/1/12}

\bibitem[{{Bahcall} \& {Salpeter}(1965)}]{BahcallSalpeter1965}
{Bahcall}, J.~N., \& {Salpeter}, E.~E. 1965, \apj, 142, 1677,
  \dodoi{10.1086/148460}

\bibitem[{{Bahcall} \& {Spitzer}(1969)}]{BahcallSpitzer1969}
{Bahcall}, J.~N., \& {Spitzer}, Lyman, J. 1969, \apjl, 156, L63,
  \dodoi{10.1086/180350}

\bibitem[{{Beck} {et~al.}(2016){Beck}, {Dobos}, {Budav{\'a}ri}, {Szalay}, \&
  {Csabai}}]{Beck2016}
{Beck}, R., {Dobos}, L., {Budav{\'a}ri}, T., {Szalay}, A.~S., \& {Csabai}, I.
  2016, \mnras, 460, 1371, \dodoi{10.1093/mnras/stw1009}

\bibitem[{{Bennett} {et~al.}(2014){Bennett}, {Larson}, {Weiland}, \&
  {Hinshaw}}]{Bennet2014}
{Bennett}, C.~L., {Larson}, D., {Weiland}, J.~L., \& {Hinshaw}, G. 2014, \apj,
  794, 135, \dodoi{10.1088/0004-637X/794/2/135}

\bibitem[{{Bielby} {et~al.}(2019){Bielby}, {Stott}, {Cullen}, {Tripp},
  {Burchett}, {Fumagalli}, {Morris}, {Tejos}, {Crain}, {Bower}, \&
  {Prochaska}}]{Bielby2019}
{Bielby}, R.~M., {Stott}, J.~P., {Cullen}, F., {et~al.} 2019, \mnras, 486, 21,
  \dodoi{10.1093/mnras/stz774}

\bibitem[{{Bordoloi} {et~al.}(2018){Bordoloi}, {Prochaska}, {Tumlinson},
  {Werk}, {Tripp}, \& {Burchett}}]{Bordoloi2018}
{Bordoloi}, R., {Prochaska}, J.~X., {Tumlinson}, J., {et~al.} 2018, \apj, 864,
  132, \dodoi{10.3847/1538-4357/aad8ac}

\bibitem[{{Bordoloi} {et~al.}(2014){Bordoloi}, {Tumlinson}, {Werk},
  {Oppenheimer}, {Peeples}, {Prochaska}, {Tripp}, {Katz}, {Dav{\'e}}, {Fox},
  {Thom}, {Ford}, {Weinberg}, {Burchett}, \& {Kollmeier}}]{Bordoloi2014}
{Bordoloi}, R., {Tumlinson}, J., {Werk}, J.~K., {et~al.} 2014, \apj, 796, 136,
  \dodoi{10.1088/0004-637X/796/2/136}

\bibitem[{{Borthakur} {et~al.}(2013){Borthakur}, {Heckman}, {Strickland},
  {Wild}, \& {Schiminovich}}]{Borthakur2013}
{Borthakur}, S., {Heckman}, T., {Strickland}, D., {Wild}, V., \&
  {Schiminovich}, D. 2013, \apj, 768, 18, \dodoi{10.1088/0004-637X/768/1/18}

\bibitem[{{Borthakur} {et~al.}(2015){Borthakur}, {Heckman}, {Tumlinson},
  {Bordoloi}, {Thom}, {Catinella}, {Schiminovich}, {Dav{\'e}}, {Kauffmann},
  {Moran}, \& {Saintonge}}]{Borthakur2015}
{Borthakur}, S., {Heckman}, T., {Tumlinson}, J., {et~al.} 2015, \apj, 813, 46,
  \dodoi{10.1088/0004-637X/813/1/46}

\bibitem[{{Borthakur} {et~al.}(2016){Borthakur}, {Heckman}, {Tumlinson},
  {Bordoloi}, {Kauffmann}, {Catinella}, {Schiminovich}, {Dav{\'e}}, {Moran}, \&
  {Saintonge}}]{Borthakur2016}
---. 2016, \apj, 833, 259, \dodoi{10.3847/1538-4357/833/2/259}

\bibitem[{{Bowen} {et~al.}(1995){Bowen}, {Blades}, \& {Pettini}}]{Bowen1995}
{Bowen}, D.~V., {Blades}, J.~C., \& {Pettini}, M. 1995, \apj, 448, 634,
  \dodoi{10.1086/175993}

\bibitem[{{Bowen} {et~al.}(1996){Bowen}, {Blades}, \& {Pettini}}]{Bowen1996}
---. 1996, \apj, 464, 141, \dodoi{10.1086/177306}

\bibitem[{{Bowen} {et~al.}(1997){Bowen}, {Osmer}, {Blades}, \&
  {Tytler}}]{Bowen1997}
{Bowen}, D.~V., {Osmer}, S.~J., {Blades}, J.~C., \& {Tytler}, D. 1997, \mnras,
  284, 599, \dodoi{10.1093/mnras/284.3.599}

\bibitem[{{Bowen} {et~al.}(2002){Bowen}, {Pettini}, \& {Blades}}]{Bowen2002}
{Bowen}, D.~V., {Pettini}, M., \& {Blades}, J.~C. 2002, \apj, 580, 169,
  \dodoi{10.1086/343106}

\bibitem[{{Burchett} {et~al.}(2020){Burchett}, {Elek}, {Tejos}, {Prochaska},
  {Tripp}, {Bordoloi}, \& {Forbes}}]{Burchett2020}
{Burchett}, J.~N., {Elek}, O., {Tejos}, N., {et~al.} 2020, \apjl, 891, L35,
  \dodoi{10.3847/2041-8213/ab700c}

\bibitem[{{Burchett} {et~al.}(2015){Burchett}, {Tripp}, {Prochaska}, {Werk},
  {Tumlinson}, {O'Meara}, {Bordoloi}, {Katz}, \& {Willmer}}]{Burchett2015}
{Burchett}, J.~N., {Tripp}, T.~M., {Prochaska}, J.~X., {et~al.} 2015, \apj,
  815, 91, \dodoi{10.1088/0004-637X/815/2/91}

\bibitem[{{Burchett} {et~al.}(2016){Burchett}, {Tripp}, {Bordoloi}, {Werk},
  {Prochaska}, {Tumlinson}, {Willmer}, {O'Meara}, \& {Katz}}]{Burchett2016}
{Burchett}, J.~N., {Tripp}, T.~M., {Bordoloi}, R., {et~al.} 2016, \apj, 832,
  124, \dodoi{10.3847/0004-637X/832/2/124}

\bibitem[{{Chen} \& {Lanzetta}(2003)}]{Chen2003}
{Chen}, H.-W., \& {Lanzetta}, K.~M. 2003, \apj, 597, 706,
  \dodoi{10.1086/378635}

\bibitem[{{Chen} {et~al.}(2001{\natexlab{a}}){Chen}, {Lanzetta}, \&
  {Webb}}]{Chen2001b}
{Chen}, H.-W., {Lanzetta}, K.~M., \& {Webb}, J.~K. 2001{\natexlab{a}}, \apj,
  556, 158, \dodoi{10.1086/321537}

\bibitem[{{Chen} {et~al.}(1998){Chen}, {Lanzetta}, {Webb}, \&
  {Barcons}}]{Chen1998}
{Chen}, H.-W., {Lanzetta}, K.~M., {Webb}, J.~K., \& {Barcons}, X. 1998, \apj,
  498, 77, \dodoi{10.1086/305554}

\bibitem[{{Chen} {et~al.}(2001{\natexlab{b}}){Chen}, {Lanzetta}, {Webb}, \&
  {Barcons}}]{Chen2001a}
---. 2001{\natexlab{b}}, \apj, 559, 654, \dodoi{10.1086/322414}

\bibitem[{{Chen} \& {Mulchaey}(2009)}]{ChenMulchaey2009}
{Chen}, H.-W., \& {Mulchaey}, J.~S. 2009, \apj, 701, 1219,
  \dodoi{10.1088/0004-637X/701/2/1219}

\bibitem[{{Chen} {et~al.}(2005){Chen}, {Prochaska}, {Weiner}, {Mulchaey}, \&
  {Williger}}]{Chen2005}
{Chen}, H.-W., {Prochaska}, J.~X., {Weiner}, B.~J., {Mulchaey}, J.~S., \&
  {Williger}, G.~M. 2005, \apjl, 629, L25, \dodoi{10.1086/444377}

\bibitem[{{Chilingarian} {et~al.}(2010){Chilingarian}, {Melchior}, \&
  {Zolotukhin}}]{Chilingarian2010}
{Chilingarian}, I.~V., {Melchior}, A.-L., \& {Zolotukhin}, I.~Y. 2010, \mnras,
  405, 1409, \dodoi{10.1111/j.1365-2966.2010.16506.x}

\bibitem[{{Chilingarian} \& {Zolotukhin}(2012)}]{Chilingarian2012}
{Chilingarian}, I.~V., \& {Zolotukhin}, I.~Y. 2012, \mnras, 419, 1727,
  \dodoi{10.1111/j.1365-2966.2011.19837.x}

\bibitem[{{C{\^o}t{\'e}} {et~al.}(2005){C{\^o}t{\'e}}, {Wyse}, {Carignan},
  {Freeman}, \& {Broadhurst}}]{Cote2005}
{C{\^o}t{\'e}}, S., {Wyse}, R.~F.~G., {Carignan}, C., {Freeman}, K.~C., \&
  {Broadhurst}, T. 2005, \apj, 618, 178, \dodoi{10.1086/425853}

\bibitem[{{Danforth} {et~al.}(2016){Danforth}, {Keeney}, {Tilton}, {Shull},
  {Stocke}, {Stevans}, {Pieri}, {Savage}, {France}, {Syphers}, {Smith},
  {Green}, {Froning}, {Penton}, \& {Osterman}}]{Danforth2016}
{Danforth}, C.~W., {Keeney}, B.~A., {Tilton}, E.~M., {et~al.} 2016, \apj, 817,
  111, \dodoi{10.3847/0004-637X/817/2/111}

\bibitem[{{Gunn} \& {Peterson}(1965)}]{GunnPeterson1965}
{Gunn}, J.~E., \& {Peterson}, B.~A. 1965, \apj, 142, 1633,
  \dodoi{10.1086/148444}

\bibitem[{{Heckman} {et~al.}(2017){Heckman}, {Borthakur}, {Wild},
  {Schiminovich}, \& {Bordoloi}}]{Heckman2017}
{Heckman}, T., {Borthakur}, S., {Wild}, V., {Schiminovich}, D., \& {Bordoloi},
  R. 2017, \apj, 846, 151, \dodoi{10.3847/1538-4357/aa80dc}

\bibitem[{{Johnson} {et~al.}(2015){Johnson}, {Chen}, \&
  {Mulchaey}}]{Johnson2015}
{Johnson}, S.~D., {Chen}, H.-W., \& {Mulchaey}, J.~S. 2015, \mnras, 452, 2553,
  \dodoi{10.1093/mnras/stv1481}

\bibitem[{{Johnson} {et~al.}(2017){Johnson}, {Chen}, {Mulchaey}, {Schaye}, \&
  {Straka}}]{Johnson2017}
{Johnson}, S.~D., {Chen}, H.-W., {Mulchaey}, J.~S., {Schaye}, J., \& {Straka},
  L.~A. 2017, \apjl, 850, L10, \dodoi{10.3847/2041-8213/aa9370}

\bibitem[{{Keeney} {et~al.}(2017){Keeney}, {Stocke}, {Danforth}, {Shull},
  {Pratt}, {Froning}, {Green}, {Penton}, \& {Savage}}]{Keeney2017}
{Keeney}, B.~A., {Stocke}, J.~T., {Danforth}, C.~W., {et~al.} 2017, \apjs, 230,
  6, \dodoi{10.3847/1538-4365/aa6b59}

\bibitem[{{Keeney} {et~al.}(2018){Keeney}, {Stocke}, {Pratt}, {Davis},
  {Syphers}, {Danforth}, {Shull}, {Froning}, {Green}, {Penton}, \&
  {Savage}}]{Keeney2018}
{Keeney}, B.~A., {Stocke}, J.~T., {Pratt}, C.~T., {et~al.} 2018, \apjs, 237,
  11, \dodoi{10.3847/1538-4365/aac727}

\bibitem[{{Lanzetta} {et~al.}(1995){Lanzetta}, {Bowen}, {Tytler}, \&
  {Webb}}]{Lanzetta1995}
{Lanzetta}, K.~M., {Bowen}, D.~V., {Tytler}, D., \& {Webb}, J.~K. 1995, \apj,
  442, 538, \dodoi{10.1086/175459}

\bibitem[{{Liang} \& {Chen}(2014)}]{Liang2014}
{Liang}, C.~J., \& {Chen}, H.-W. 2014, \mnras, 445, 2061,
  \dodoi{10.1093/mnras/stu1901}

\bibitem[{{Mathes} {et~al.}(2014){Mathes}, {Churchill}, {Kacprzak}, {Nielsen},
  {Trujillo-Gomez}, {Charlton}, \& {Muzahid}}]{Mathes2014}
{Mathes}, N.~L., {Churchill}, C.~W., {Kacprzak}, G.~G., {et~al.} 2014, \apj,
  792, 128, \dodoi{10.1088/0004-637X/792/2/128}

\bibitem[{{McQuinn}(2016)}]{McQuinn2016}
{McQuinn}, M. 2016, \araa, 54, 313, \dodoi{10.1146/annurev-astro-082214-122355}

\bibitem[{{Meiksin}(2009)}]{Meiksin2009}
{Meiksin}, A.~A. 2009, Reviews of Modern Physics, 81, 1405,
  \dodoi{10.1103/RevModPhys.81.1405}

\bibitem[{{Morris} {et~al.}(1993){Morris}, {Weymann}, {Dressler}, {McCarthy},
  {Smith}, {Terrile}, {Giovanelli}, \& {Irwin}}]{Morris1993}
{Morris}, S.~L., {Weymann}, R.~J., {Dressler}, A., {et~al.} 1993, \apj, 419,
  524, \dodoi{10.1086/173505}

\bibitem[{{Muzahid}(2014)}]{Muzahid2014}
{Muzahid}, S. 2014, \apj, 784, 5, \dodoi{10.1088/0004-637X/784/1/5}

\bibitem[{{Nelson} {et~al.}(2019){Nelson}, {Pillepich}, {Springel}, {Pakmor},
  {Weinberger}, {Genel}, {Torrey}, {Vogelsberger}, {Marinacci}, \&
  {Hernquist}}]{Nelson2019}
{Nelson}, D., {Pillepich}, A., {Springel}, V., {et~al.} 2019, \mnras, 490,
  3234, \dodoi{10.1093/mnras/stz2306}

\bibitem[{{Oppenheimer} {et~al.}(2018){Oppenheimer}, {Schaye}, {Crain}, {Werk},
  \& {Richings}}]{Oppenheimer2018}
{Oppenheimer}, B.~D., {Schaye}, J., {Crain}, R.~A., {Werk}, J.~K., \&
  {Richings}, A.~J. 2018, \mnras, 481, 835, \dodoi{10.1093/mnras/sty2281}

\bibitem[{{Peeples} {et~al.}(2019){Peeples}, {Corlies}, {Tumlinson}, {O'Shea},
  {Lehner}, {O'Meara}, {Howk}, {Earl}, {Smith}, {Wise}, \&
  {Hummels}}]{Peeples2019}
{Peeples}, M.~S., {Corlies}, L., {Tumlinson}, J., {et~al.} 2019, \apj, 873,
  129, \dodoi{10.3847/1538-4357/ab0654}

\bibitem[{{Penton} {et~al.}(2002){Penton}, {Stocke}, \& {Shull}}]{Penton2002}
{Penton}, S.~V., {Stocke}, J.~T., \& {Shull}, J.~M. 2002, \apj, 565, 720,
  \dodoi{10.1086/324483}

\bibitem[{{Pointon} {et~al.}(2019){Pointon}, {Kacprzak}, {Nielsen}, {Muzahid},
  {Murphy}, {Churchill}, \& {Charlton}}]{Pointon2019}
{Pointon}, S.~K., {Kacprzak}, G.~G., {Nielsen}, N.~M., {et~al.} 2019, \apj,
  883, 78, \dodoi{10.3847/1538-4357/ab3b0e}

\bibitem[{{Prochaska} {et~al.}(2011{\natexlab{a}}){Prochaska}, {Weiner},
  {Chen}, {Cooksey}, \& {Mulchaey}}]{Prochaska2011a}
{Prochaska}, J.~X., {Weiner}, B., {Chen}, H.-W., {Cooksey}, K.~L., \&
  {Mulchaey}, J.~S. 2011{\natexlab{a}}, \apjs, 193, 28,
  \dodoi{10.1088/0067-0049/193/2/28}

\bibitem[{{Prochaska} {et~al.}(2011{\natexlab{b}}){Prochaska}, {Weiner},
  {Chen}, {Mulchaey}, \& {Cooksey}}]{Prochaska2011b}
{Prochaska}, J.~X., {Weiner}, B., {Chen}, H.-W., {Mulchaey}, J., \& {Cooksey},
  K. 2011{\natexlab{b}}, \apj, 740, 91, \dodoi{10.1088/0004-637X/740/2/91}

\bibitem[{{Prochaska} {et~al.}(2019){Prochaska}, {Burchett}, {Tripp}, {Werk},
  {Willmer}, {Howk}, {Lange}, {Tejos}, {Meiring}, {Tumlinson}, {Lehner},
  {Ford}, \& {Dav{\'e}}}]{Prochaska2019}
{Prochaska}, J.~X., {Burchett}, J.~N., {Tripp}, T.~M., {et~al.} 2019, \apjs,
  243, 24, \dodoi{10.3847/1538-4365/ab2b9a}

\bibitem[{{Rao} {et~al.}(2011){Rao}, {Belfort-Mihalyi}, {Turnshek}, {Monier},
  {Nestor}, \& {Quider}}]{Rao2011}
{Rao}, S.~M., {Belfort-Mihalyi}, M., {Turnshek}, D.~A., {et~al.} 2011, \mnras,
  416, 1215, \dodoi{10.1111/j.1365-2966.2011.19119.x}

\bibitem[{{Richter} {et~al.}(2016){Richter}, {Wakker}, {Fechner}, {Herenz},
  {Tepper-Garc{\'\i}a}, \& {Fox}}]{Richter2016}
{Richter}, P., {Wakker}, B.~P., {Fechner}, C., {et~al.} 2016, \aap, 590, A68,
  \dodoi{10.1051/0004-6361/201527038}

\bibitem[{{Rudie} {et~al.}(2019){Rudie}, {Steidel}, {Pettini}, {Trainor},
  {Strom}, {Hummels}, {Reddy}, \& {Shapley}}]{Rudie2019}
{Rudie}, G.~C., {Steidel}, C.~C., {Pettini}, M., {et~al.} 2019, \apj, 885, 61,
  \dodoi{10.3847/1538-4357/ab4255}

\bibitem[{{Rudie} {et~al.}(2013){Rudie}, {Steidel}, {Shapley}, \&
  {Pettini}}]{Rudie2013}
{Rudie}, G.~C., {Steidel}, C.~C., {Shapley}, A.~E., \& {Pettini}, M. 2013,
  \apj, 769, 146, \dodoi{10.1088/0004-637X/769/2/146}

\bibitem[{{Rudie} {et~al.}(2012){Rudie}, {Steidel}, {Trainor}, {Rakic},
  {Bogosavljevi{\'c}}, {Pettini}, {Reddy}, {Shapley}, {Erb}, \&
  {Law}}]{Rudie2012}
{Rudie}, G.~C., {Steidel}, C.~C., {Trainor}, R.~F., {et~al.} 2012, \apj, 750,
  67, \dodoi{10.1088/0004-637X/750/1/67}

\bibitem[{{Rykoff} {et~al.}(2015){Rykoff}, {Rozo}, \& {Keisler}}]{Rykoff2015}
{Rykoff}, E.~S., {Rozo}, E., \& {Keisler}, R. 2015, arXiv e-prints,
  arXiv:1509.00870.
\newblock \doarXiv{1509.00870}

\bibitem[{{Smailagi{\'c}} {et~al.}(2018){Smailagi{\'c}}, {Prochaska},
  {Burchett}, {Zhu}, \& {M{\'e}nard}}]{Smailagic2018}
{Smailagi{\'c}}, M., {Prochaska}, J.~X., {Burchett}, J., {Zhu}, G., \&
  {M{\'e}nard}, B. 2018, \apj, 867, 106, \dodoi{10.3847/1538-4357/aae384}

\bibitem[{{Steidel} {et~al.}(2011){Steidel}, {Bogosavljevi{\'c}}, {Shapley},
  {Kollmeier}, {Reddy}, {Erb}, \& {Pettini}}]{Steidel2011}
{Steidel}, C.~C., {Bogosavljevi{\'c}}, M., {Shapley}, A.~E., {et~al.} 2011,
  \apj, 736, 160, \dodoi{10.1088/0004-637X/736/2/160}

\bibitem[{{Stocke} {et~al.}(2013){Stocke}, {Keeney}, {Danforth}, {Shull},
  {Froning}, {Green}, {Penton}, \& {Savage}}]{Stocke2013}
{Stocke}, J.~T., {Keeney}, B.~A., {Danforth}, C.~W., {et~al.} 2013, \apj, 763,
  148, \dodoi{10.1088/0004-637X/763/2/148}

\bibitem[{{Strateva} {et~al.}(2001){Strateva}, {Ivezi{\'c}}, {Knapp},
  {Narayanan}, {Strauss}, {Gunn}, {Lupton}, {Schlegel}, {Bahcall}, {Brinkmann},
  {Brunner}, {Budav{\'a}ri}, {Csabai}, {Castander}, {Doi}, {Fukugita},
  {Gy{\H{o}}ry}, {Hamabe}, {Hennessy}, {Ichikawa}, {Kunszt}, {Lamb}, {McKay},
  {Okamura}, {Racusin}, {Sekiguchi}, {Schneider}, {Shimasaku}, \&
  {York}}]{Strateva2001}
{Strateva}, I., {Ivezi{\'c}}, {\v{Z}}., {Knapp}, G.~R., {et~al.} 2001, \aj,
  122, 1861, \dodoi{10.1086/323301}

\bibitem[{{Taylor} {et~al.}(2011){Taylor}, {Hopkins}, {Baldry}, {Brown},
  {Driver}, {Kelvin}, {Hill}, {Robotham}, {Bland-Hawthorn}, {Jones}, {Sharp},
  {Thomas}, {Liske}, {Loveday}, {Norberg}, {Peacock}, {Bamford}, {Brough},
  {Colless}, {Cameron}, {Conselice}, {Croom}, {Frenk}, {Gunawardhana},
  {Kuijken}, {Nichol}, {Parkinson}, {Phillipps}, {Pimbblet}, {Popescu},
  {Prescott}, {Sutherland}, {Tuffs}, {van Kampen}, \&
  {Wijesinghe}}]{Taylor2011}
{Taylor}, E.~N., {Hopkins}, A.~M., {Baldry}, I.~K., {et~al.} 2011, \mnras, 418,
  1587, \dodoi{10.1111/j.1365-2966.2011.19536.x}

\bibitem[{{Tejos} {et~al.}(2014){Tejos}, {Morris}, {Finn}, {Crighton},
  {Bechtold}, {Jannuzi}, {Schaye}, {Theuns}, {Altay}, {Le F{\`e}vre},
  {Ryan-Weber}, \& {Dav{\'e}}}]{Tejos2014}
{Tejos}, N., {Morris}, S.~L., {Finn}, C.~W., {et~al.} 2014, \mnras, 437, 2017,
  \dodoi{10.1093/mnras/stt1844}

\bibitem[{{Thom} {et~al.}(2012){Thom}, {Tumlinson}, {Werk}, {Prochaska},
  {Oppenheimer}, {Peeples}, {Tripp}, {Katz}, {O'Meara}, {Ford}, {Dav{\'e}},
  {Sembach}, \& {Weinberg}}]{Thom2012}
{Thom}, C., {Tumlinson}, J., {Werk}, J.~K., {et~al.} 2012, \apjl, 758, L41,
  \dodoi{10.1088/2041-8205/758/2/L41}

\bibitem[{{Tripp} {et~al.}(1998){Tripp}, {Lu}, \& {Savage}}]{Tripp1998}
{Tripp}, T.~M., {Lu}, L., \& {Savage}, B.~D. 1998, \apj, 508, 200,
  \dodoi{10.1086/306397}

\bibitem[{{Trotta}(2017)}]{Trotta2017}
{Trotta}, R. 2017, arXiv e-prints, arXiv:1701.01467.
\newblock \doarXiv{1701.01467}

\bibitem[{{Tumlinson} {et~al.}(2017){Tumlinson}, {Peeples}, \&
  {Werk}}]{Tumlinson2017}
{Tumlinson}, J., {Peeples}, M.~S., \& {Werk}, J.~K. 2017, \araa, 55, 389,
  \dodoi{10.1146/annurev-astro-091916-055240}

\bibitem[{{Tumlinson} {et~al.}(2011){Tumlinson}, {Thom}, {Werk}, {Prochaska},
  {Tripp}, {Weinberg}, {Peeples}, {O'Meara}, {Oppenheimer}, {Meiring}, {Katz},
  {Dav{\'e}}, {Ford}, \& {Sembach}}]{Tumlinson2011}
{Tumlinson}, J., {Thom}, C., {Werk}, J.~K., {et~al.} 2011, Science, 334, 948,
  \dodoi{10.1126/science.1209840}

\bibitem[{{Tumlinson} {et~al.}(2013){Tumlinson}, {Thom}, {Werk}, {Prochaska},
  {Tripp}, {Katz}, {Dav{\'e}}, {Oppenheimer}, {Meiring}, {Ford}, {O'Meara},
  {Peeples}, {Sembach}, \& {Weinberg}}]{Tumlinson2013}
---. 2013, \apj, 777, 59, \dodoi{10.1088/0004-637X/777/1/59}

\bibitem[{{Turner} {et~al.}(2014){Turner}, {Schaye}, {Steidel}, {Rudie}, \&
  {Strom}}]{Turner2014}
{Turner}, M.~L., {Schaye}, J., {Steidel}, C.~C., {Rudie}, G.~C., \& {Strom},
  A.~L. 2014, \mnras, 445, 794, \dodoi{10.1093/mnras/stu1801}

\bibitem[{{Turner} {et~al.}(2015){Turner}, {Schaye}, {Steidel}, {Rudie}, \&
  {Strom}}]{Turner2015}
---. 2015, \mnras, 450, 2067, \dodoi{10.1093/mnras/stv750}

\bibitem[{{Wakker} \& {Savage}(2009)}]{WakkerSavage2009}
{Wakker}, B.~P., \& {Savage}, B.~D. 2009, The Astrophysical Journal Supplement
  Series, 182, 378, \dodoi{10.1088/0067-0049/182/1/378}

\bibitem[{{Werk} {et~al.}(2012){Werk}, {Prochaska}, {Thom}, {Tumlinson},
  {Tripp}, {O'Meara}, \& {Meiring}}]{Werk2012}
{Werk}, J.~K., {Prochaska}, J.~X., {Thom}, C., {et~al.} 2012, \apjs, 198, 3,
  \dodoi{10.1088/0067-0049/198/1/3}

\bibitem[{{Werk} {et~al.}(2013){Werk}, {Prochaska}, {Thom}, {Tumlinson},
  {Tripp}, {O'Meara}, \& {Peeples}}]{Werk2013}
---. 2013, \apjs, 204, 17, \dodoi{10.1088/0067-0049/204/2/17}

\bibitem[{{Werk} {et~al.}(2014){Werk}, {Prochaska}, {Tumlinson}, {Peeples},
  {Tripp}, {Fox}, {Lehner}, {Thom}, {O'Meara}, {Ford}, {Bordoloi}, {Katz},
  {Tejos}, {Oppenheimer}, {Dav{\'e}}, \& {Weinberg}}]{Werk2014}
{Werk}, J.~K., {Prochaska}, J.~X., {Tumlinson}, J., {et~al.} 2014, \apj, 792,
  8, \dodoi{10.1088/0004-637X/792/1/8}

\bibitem[{{Werk} {et~al.}(2016){Werk}, {Prochaska}, {Cantalupo}, {Fox},
  {Oppenheimer}, {Tumlinson}, {Tripp}, {Lehner}, \& {McQuinn}}]{Werk2016}
{Werk}, J.~K., {Prochaska}, J.~X., {Cantalupo}, S., {et~al.} 2016, \apj, 833,
  54, \dodoi{10.3847/1538-4357/833/1/54}

\bibitem[{{Wilde} {et~al.}(2021){Wilde}, {Werk}, {Burchett}, {Prochaska},
  {Tchernyshyov}, {Tripp}, {Tejos}, {Lehner}, {Bordoloi}, {O'Meara}, \&
  {Tumlinson}}]{Wilde2020}
{Wilde}, M.~C., {Werk}, J.~K., {Burchett}, J.~N., {et~al.} 2021, \apj, 912, 9,
  \dodoi{10.3847/1538-4357/abea14}

\bibitem[{{Wolfe} {et~al.}(2005){Wolfe}, {Gawiser}, \& {Prochaska}}]{Wolfe2005}
{Wolfe}, A.~M., {Gawiser}, E., \& {Prochaska}, J.~X. 2005, \araa, 43, 861,
  \dodoi{10.1146/annurev.astro.42.053102.133950}

\bibitem[{{Zahedy} {et~al.}(2019){Zahedy}, {Chen}, {Johnson}, {Pierce},
  {Rauch}, {Huang}, {Weiner}, \& {Gauthier}}]{Zahedy2019}
{Zahedy}, F.~S., {Chen}, H.-W., {Johnson}, S.~D., {et~al.} 2019, \mnras, 484,
  2257, \dodoi{10.1093/mnras/sty3482}

\end{thebibliography}
\bibliographystyle{aasjournal}

\end{document}